\documentclass[aps,preprint,nofootinbib,floatfix]{revtex4}
\usepackage{color}
\usepackage{enumerate}
\usepackage[dvipsnames]{xcolor}

\pdfoutput=1
\usepackage{graphicx}
\usepackage[latin1]{inputenc}
\usepackage{amsmath,amssymb}
\usepackage{hyperref}
\usepackage{epstopdf}

\newcommand{\lsim}{\mathrel{\mathop{\kern 0pt \rlap
  {\raise.2ex\hbox{$<$}}}
  \lower.9ex\hbox{\kern-.190em $\sim$}}}
\newcommand{\gsim}{\mathrel{\mathop{\kern 0pt \rlap
  {\raise.2ex\hbox{$>$}}}
  \lower.9ex\hbox{\kern-.190em $\sim$}}}

\newcommand{\gev}{\ensuremath{\,\mathrm{GeV}}}

\def  \bcen   {\begin{center}}
\def  \ecen   {\end{center}}
\def  \beq    {\begin{equation}}
\def  \eeq    {\end{equation}}
\def  \bpm    {\begin{pmatrix}}
\def  \epm    {\end{pmatrix}}
\def  \beqa   {\begin{eqnarray}}
\def  \eeqa   {\end{eqnarray}}
\def  \nn     {\nonumber }
\def\bea{\begin{eqnarray}}
\def\eea{\end{eqnarray}}

\def\ga   {\gamma}
\def\Ga   {\Gamma}

\def\la   {\lambda}

\def\sig   {\sigma}

\def\nn{\nonumber}
\def\lee { \left( }
\def\rii { \right) }
\def\lan   {\langle}
\def\ran   {\rangle}

\def\De {\Delta}

\def\to {\rightarrow}
\def\lphp {\la^\prime_{H\Phi}}

\begin{document}

{\small
\begin{flushright}
CP3-Origins-2018-021 DNRF90
\end{flushright} }

\title{
Consistency of a gauged two-Higgs-doublet model: Scalar sector}
\author{
Abdesslam Arhrib$^1$, Wei-Chih Huang$^2$, \\
Raymundo Ramos$^3$, Yue-Lin Sming Tsai$^3$ and Tzu-Chiang Yuan$^3$
}

\affiliation{
\small{
$^1$D${\it \acute{e}}$partement de Math${\it \acute{e}}$matique, 
Facult${\it \acute{e}}$ des Sciences et Techniques,\\}
{\it Universit${\it \acute{e}}$ Abdelmalek Essaadi, B. 416, Tangier, Morocco\\
$^2$ CP$^{\, 3}$-Origins, University of Southern Denmark, Campusvej 55, DK-5230 Odense M, Denmark \\
$^3$Institute of Physics, Academia Sinica, Nangang, Taipei 11529, Taiwan
}
}

\date{\today}

\begin{abstract}

We study the theoretical and phenomenological constraints imposed on the scalar sector of
the gauged two Higgs doublet model proposed recently as a variant of 
the popular inert Higgs doublet model of dark matter.
The requirements of tree-level vacuum stability and perturbative unitarity 
in the scalar sector are analyzed in detail. 
Furthermore, taking into account the constraints from the 125 GeV Higgs boson measurements 
at the Large Hadron Collider, we map out the allowed ranges 
for the fundamental parameters of the scalar potential in the model.

\end{abstract}

\maketitle

\section{Introduction} \label{section:intro}

Despite the plentiful and cogent evidences of dark matter~(DM) from astrophysical and cosmological observations 
over several decades, the identity of DM is still unknown. Whether the solution should reside on the particle physics side or  belong to the pure gravity doctrine is not settled. Perhaps this dilemma will not be resolved anytime soon and will 
continue baffling us for a while.
On the other hand, after the monumental discovery of the 125 GeV 
Higgs boson at the Large Hadron Collider (LHC), more data and statistics are still needed to determine 
whether the observed Higgs boson is the one predicted in the standard model (SM)
or it is part of an extended scalar sector proposed in many models beyond the SM (BSM). 
Among these BSM models with additional scalars, the general two Higgs doublet model (2HDM) 
is possibly the most popular and well-studied class of models~\cite{Branco:2011iw}. 
One of the intriguing 2HDM variants is the inert Higgs doublet model (IHDM)~\cite{Deshpande:1977rw,Ma:2006km,Barbieri:2006dq,LopezHonorez:2006gr}, 
where the neutral component of the
second Higgs doublet, in light of an imposed discrete $Z_2$ symmetry on the scalar potential,
is stable and hence can be a DM candidate. 
Over the years, many thorough phenomenological analysis of the IHDM had been
carried out in the
literature~\cite{Arhrib:2013ela,Ilnicka:2015jba,Belyaev:2016lok,Eiteneuer:2017hoh},
while the origin of multiple inert Higgs 
doublets in the context of grand unification was explored in~\cite{Kephart:2015oaa}.

The discrete $Z_2$ symmetry, however, is introduced by hand without proper justification. 
Besides, it is commonly believed that global symmetry (regardless of being discrete or continuous) is deemed 
to be strongly violated by gravitational effects~\cite{Krauss:1988zc,Kallosh:1995hi}.
To remedy these unappealing features, we elevate the IHDM into a more theoretically elegant 
setting~\cite{Huang:2015wts}, in which the two Higgs doublets $H_1$ and $H_2$ comprise 
a doublet $H=(H_1,H_2)^{\rm T}$ of a new non-Abelian $SU(2)_H$ gauge group. For various phenomenological reasons,
$H$  is also charged under an additional Abelian gauge group $U(1)_X$. 
Gauge invariance ensures that only $\langle H_1 \rangle$ can develop a nonzero 
vacuum expectation value (VEV) while $\langle H_2 \rangle$ remains zero. Thus $H_1$ and $H_2$ play the role 
of the SM and inert Higgs doublets respectively. There is no need to impose the discrete $Z_2$ symmetry in the scalar potential 
as it can emerge as a low-energy symmetry after spontaneously symmetry breaking. 
We dubbed this model as 
gauged two-Higgs-doublet model~(G2HDM) in~\cite{Huang:2015wts}.
In IHDM, either the neutral scalar or pseudoscalar 
component of $H_2$ can be a DM candidate, while in G2HDM the whole neutral 
component of $H_2$ does the job. 

Phenomenology of G2HDM at the LHC have been explored previously in~\cite{Huang:2015wts,Huang:2015rkj} 
for Higgs physics and in~\cite{Huang:2017bto} for the new gauge bosons.
In this work, we will focus on some theoretical issues that were not addressed before. 
We will analyze the scalar sector in more detail.
Our main goal is to identify the allowed regions for the fundamental parameters/couplings in the scalar potential by imposing the 
tree-level theoretical constraints from vacuum stability and perturbative unitarity, 
as well as phenomenological constraints from the 125 GeV Higgs boson measurements at the LHC.
Similar studies have been presented
for the general 2HDM~\cite{Kanemura:1999xf,Arhrib:2000is,Akeroyd:2000wc} and
for the IHDM~\cite{Arhrib:2013ela,Arhrib:2012ia}.

Vacuum stability comes from the requirement that the potential 
has to be bounded from below, while  perturbative unitarity demands
the amplitudes of all scattering processes in the scalar sector to lie within the unitarity circle.

In G2HDM, the VEVs of various scalars in the extended Higgs sector will induce 
complicated mixings among the scalars in the flavor basis.
In particular, the discovered 125 GeV boson at the LHC is in general a linear combination of 
the three neutral components of $H_1$, $\Phi_H$ and $\Delta_H$,
where $\Phi_H$ and $\Delta_H$ are the $SU(2)_H$ doublet and triplet respectively. 
These extra scalars are necessarily introduced in G2HDM for 
various phenomenological reasons, as explained in detail in~\cite{Huang:2015wts}.
Only in the absence of mixings, would the neutral component of $H_1$ 
be identified as the observed 125 GeV Higgs boson. 
As a result, we have to include these mixing effects when imposing the constraints
from the measurements of the Higgs boson mass and the diphoton signal strength.
The contributions to the diphoton final state from the charged Higgs bosons and 
new heavy fermions in G2HDM will be carefully scrutinized.
There are also mixings between $H^0_2$, $G_H^m$ and $\Delta_m$, the lower components of $H_2$,  
$\Phi_H$ and $\Delta_H$ respectively, rendering the analysis of DM physics more involved.
We will discuss the extended scalar sector of G2HDM in more detail in the next section.
However, the thorough study of DM phenomenology in G2HDM will be carried out 
in a separate work. 

The content of the paper is laid out as follows. 
In Sec.~\ref{section:model}, we review some crucial features of the G2HDM, 
with a focus on the particle content and the scalar potential of the model,
followed by discussions on the minimization conditions of the scalar potential and the scalar mass spectra. 
We will take this opportunity to investigate the impacts of
two new quartic terms,  
namely $- H_1^\dagger H_1  H_2^\dagger H_2 + H_1^\dagger H_2  H_2^\dagger H_1$
and $H^\dagger \Phi_H \Phi_H^\dagger H$,
in the scalar potential which were missing in our previous works.
In Sec.~\ref{section:th_constraints}, we study the theoretical constraints on the scalar potential parameters 
from  tree-level vacuum stability and perturbative unitarity of the scalar sector.
In Sec.~\ref{section:HiggsPheno}, we further impose the LHC constraints associated with the 125 GeV Higgs boson 
and its diphoton signal strength on the
parameter space.
Finally, we summarize our findings in Sec.~\ref{section:conclusions}.

\section{G2HDM Set Up \label{section:model}}

In this section, we start with a brief review on the setup of G2HDM~\cite{Huang:2015wts} 
by specifying its matter content and writing down the scalar potential including the two new terms 
$- H_1^\dagger H_1  H_2^\dagger H_2 + H_1^\dagger H_2  H_2^\dagger H_1$
and $H^\dagger \Phi_H \Phi_H^\dagger H$ missing in previous work.
Then we will discuss minimization of the potential and the scalar mass spectrum.

\subsection{Matter content \label{subsection:particlecontent}}

Besides the two Higgs doublets $H_1$ and $H_2$ combining to form $H=(H_1,
H_2)^{\rm T}$ in the fundamental 
representation of an extra $SU(2)_H$, we introduced a triplet $\De_H$ and a doublet $\Phi_H$ of this new gauge group.
However $\De_H$ and $\Phi_H$ are singlets under the SM gauge group. 
Only $H$ carries both quantum numbers of the two $SU(2)$s.

There are different ways of introducing new heavy fermions in the model but we choose a
simplest realization: the heavy fermions together with the SM right-handed fermions comprise $SU(2)_H$ doublets,
while the SM left-handed doublets are singlets under $SU(2)_H$. 
We note that heavy right-handed neutrinos paired up with 
a mirror charged leptons forming $SU(2)_L$ doublets was suggested before in \cite{Hung:2006ap}.
To render the model anomaly-free, four additional chiral (left-handed) fermions for each generation, 
all singlets under both $SU(2)_L$ and $SU(2)_H$, are included.
For the Yukawa interactions that couple the fermions to scalars in G2HDM, we refer our readers to~\cite{Huang:2015wts}
 for more details.

To avoid some unwanted pieces in the scalar potential and construct gauge invariant Yukawa couplings, 
we  require the matter fields to  carry extra $U(1)_X$ charges. Thus the complete 
gauge groups in G2HDM consist of $SU(3)_C \times SU(2)_L \times U(1)_Y  \times SU(2)_H \times U(1)_X$.
Apart from the matter content of G2HDM
summarized in Table~\ref{tab:quantumnos}~\footnote{We note that 
$u^H_L,d^H_L,\nu^H_L,e^H_L$ were denoted as $\chi_u,\chi_d,\chi_\nu,\chi_e$ respectively 
in~\cite{Huang:2015wts}.}, there also exist the gauge bosons corresponding to the SM and the extra 
gauge groups.

The salient features of G2HDM are:
(i) it is free of gauge and gravitational anomalies; 
(ii) renormalizable;
(iii) without resorting to the previous ad-hoc $Z_2$ symmetry, 
an inert Higgs doublet $H_2$ can be naturally realized, providing a DM candidate;
(iv) due to the gauge symmetries~\footnote{For using Abelian $U(1)$ symmetry to avoid flavor-changing neutral currents in 2HDM 
at tree level, see for example~\cite{Ko:2012hd,Campos:2017dgc}.}, 
dangerous flavor-changing neutral currents vanish at tree level for the SM sector;
(v) the VEV of the triplet can trigger $SU(2)_L$ symmetry breaking while that of $\Phi_H$ provides 
a mass to the new fermions through $SU(2)_H$-invariant Yukawa couplings;
{\it etc.}

\begin{table}[htbp!]
\begin{tabular}{|c|c|c|c|c|c|}
\hline
Matter Fields & $SU(3)_C$ & $SU(2)_L$ & $SU(2)_H$ & $U(1)_Y$ & $U(1)_X$ \\
\hline \hline
$Q_L=\left( u_L \;\; d_L \right)^{\rm T}$ & 3 & 2 & 1 & 1/6 & 0\\
$U_R=\left( u_R \;\; u^H_R \right)^{\rm T}$ & 3 & 1 & 2 & 2/3 & $1$ \\
$D_R=\left( d^H_R \;\; d_R \right)^{\rm T}$ & 3 & 1 & 2 & $-1/3$ & $-1$ \\
\hline
$u_L^H$ & 3 & 1 & 1 & 2/3 & 0 \\
$d_L^H$ & 3 & 1 & 1 & $-1/3$ & 0 \\
\hline
$L_L=\left( \nu_L \;\; e_L \right)^{\rm T}$ & 1 & 2 & 1 & $-1/2$ & 0 \\
$N_R=\left( \nu_R \;\; \nu^H_R \right)^{\rm T}$ & 1 & 1 & 2 & 0 & $1$ \\
$E_R=\left( e^H_R \;\; e_R \right)^{\rm T}$ & 1 & 1 & 2 &  $-1$  &  $-1$ \\
\hline
$\nu_L^H$ & 1 & 1 & 1 & 0 & 0 \\
$e_L^H$ & 1 & 1 & 1 & $-1$ & 0 \\
\hline\hline
$H=\left( H_1 \;\; H_2 \right)^{\rm T}$ & 1 & 2 & 2 & 1/2 & $1$ \\
$\Delta_H=\left( \begin{array}{cc} \Delta_3/2 & \Delta_p/\sqrt{2}  \\ \Delta_m/\sqrt{2} & - \Delta_3/2 \end{array} \right)$ & 1 & 1 & 3 & 0 & 0 \\
$\Phi_H=\left( \Phi_1 \;\; \Phi_2 \right)^{\rm T}$ & 1 & 1 & 2 & 0 & $1$ \\
\hline
\end{tabular}
\caption{Matter field contents and their quantum number assignments in G2HDM. 
}
\label{tab:quantumnos}
\end{table}

\subsection{Higgs potential \label{subsection:higgspotential}}

The Higgs potential invariant under both $SU(2)_L\times U(1)_Y$ and $SU(2)_H \times U(1)_X$
can be decomposed into four different terms as~\footnote{Here, we consider renormalizable terms only. 
In addition, while the $SU(2)_H$ multiplication is explicitly shown, the $SU(2)_L$ multiplication 
is implicit and suppressed.}
\begin{align}
V_T = V (H) + V (\Phi_H ) + V ( \De_H ) + V_{\rm mix} \left( H , \Delta_H, \Phi_H \right) \; , 
\label{eq:higgs_pot} 
\end{align}
with~\footnote{We should point out that the $\lambda^\prime_H$ term in $V(H)$ was missing in earlier studies~\cite{Huang:2015wts,Huang:2015rkj} 
and $V(H)$ contains just three terms (1 mass term and 2 quartic terms) 
as compared to 8 terms (3 mass terms and 5 quartic terms) in general 2HDM~\cite{Branco:2011iw}.}
\begin{align}
\label{VH1H2}
V (H) 
=& \; \mu^2_H   \left( H^{\alpha i}  H_{\alpha i} \right)
+  \la_H \left( H^{\alpha i}  H_{\alpha i} \right)^2  
+ \frac{1}{2} \la'_H \epsilon_{\alpha \beta} \epsilon^{\gamma \delta}
\left( H^{ \alpha i}  H_{\gamma  i} \right)  \left( H^{ \beta j}  H_{\delta j} \right)  \; , \nn \\
=&  \;  \mu^2_H   \left( H^\dag_1 H_1 + H^\dag_2 H_2 \right) 
+ \la_H   \left( H^\dag_1 H_1 + H^\dag_2 H_2 \right)^2 
+ \la'_H \left( - H^\dag_1 H_1 H^\dag_2 H_2 
                  + H^\dag_1 H_2 H^\dag_2 H_1 \right)  \; , 
\end{align}
where ($\alpha$, $\beta$, $\gamma$, $\delta$) and ($i$, $j$) refer to the $SU(2)_H$ and $SU(2)_L$ indices respectively, 
all of which run from one to two, and $H^{\alpha i} = H^*_{\alpha i}$;
\begin{align}
\label{VPhi}
V ( \Phi_H ) =& \;  \mu^2_{\Phi}   \Phi_H^\dag \Phi_H  + \la_\Phi \lee \Phi_H^\dag \Phi_H  \rii^2  \; , \nn \\
 =& \;  \mu^2_{\Phi} \lee \Phi^*_1\Phi_1 + \Phi^*_2\Phi_2 \rii 
 +  \la_\Phi \lee \Phi^*_1\Phi_1 + \Phi^*_2\Phi_2 \rii^2 \; , 
 \end{align}
with $\Phi_H = (\Phi_1 \; \Phi_2)^{\rm T}$; 
\begin{align}
 \label{VDelta}
V ( \De_H ) =& \; - \mu^2_{\De} {\rm Tr} \lee \De^2_H  \rii  \;  + \la_\De \lee {\rm Tr} \lee \De^2_H  \rii \rii^2 \; , \nn \\
= & \; - \mu^2_{\De} \lee \frac{1}{2} \De^2_3 + \De_p \De_m  \rii +  \la_{\De} \lee \frac{1}{2} \De^2_3 + \De_p \De_m  \rii^2 \; , 
\end{align}
where
 \begin{align}
\De_H=
  \begin{pmatrix}
    \De_3/2   &  \De_p / \sqrt{2}  \\
    \De_m / \sqrt{2} & - \De_3/2   \\
  \end{pmatrix} = \De_H^\dagger \; {\rm with}
  \;\; \Delta_m = \left( \Delta_p \right)^* \; {\rm and} \; \left( \Delta_3 \right)^* = \Delta_3 \;    ;
 \end{align}
and finally the mixed term~\footnote{
We note that the $\lambda^{\prime}_{H\Phi}$ term in (\ref{VMix}) was not included in 
the original work~\cite{Huang:2015wts}. Also another invariant operator 
$\lee \Phi_H^{\rm T}  \epsilon H \rii^\dag \lee \Phi_H^{\rm T} \epsilon H \rii$,
where $\epsilon$ is the second-rank totally antisymmetric tensor acting on the $SU(2)_H$ space,
can be written down. But this term can be expressed as 
$\lee H^\dag H  \rii  \lee \Phi_H^\dag \Phi_H \rii  - \lee H^\dag \Phi_H  \rii  \lee \Phi_H^\dag H \rii$, and therefore
is not linearly independent.
}
\begin{align}
\label{VMix}
V_{\rm{mix}} \left( H , \Delta_H, \Phi_H \right) = 
& \; + M_{H\De}  \lee H^\dag \De_H H \rii -  M_{\Phi\De}  \lee \Phi_H^\dag \De_H \Phi_H \rii  \nn \\
& \; + \la_{H\Phi} \lee H^\dag H  \rii  \lee \Phi_H^\dag \Phi_H \rii  
 + \la^\prime_{H\Phi} \lee H^\dag \Phi_H  \rii  \lee \Phi_H^\dag H \rii
\nn\\
& \;  +  \la_{H\De} \lee H^\dag H  \rii    {\rm Tr} \lee \De^2_H  \rii  
+ \la_{\Phi\De} \lee \Phi_H^\dag \Phi_H \rii {\rm Tr} \lee \De^2_H \rii  \; . 
\end{align}
In terms of the component fields of $H$, $\Delta_H$ and $\Phi_H$, the mixed potential term
$V_{\rm{mix}} \left( H , \Delta_H, \Phi_H \right) $ reads
\begin{align}
V_{\rm{mix}} \left( H , \Delta_H, \Phi_H \right) = 
& \; + M_{H\De} \lee \frac{1}{\sqrt{2}}H^\dag_1 H_2 \De_p  
+  \frac{1}{2} H^\dag_1 H_1\De_3 + \frac{1}{\sqrt{2}}  H^\dag_2 H_1 \De_m  
- \frac{1}{2} H^\dag_2 H_2 \De_3   \rii   \nn \\
& \; - M_{\Phi\De} \lee  \frac{1}{\sqrt{2}} \Phi^*_1 \Phi_2 \De_p  
+  \frac{1}{2} \Phi^*_1 \Phi_1\De_3 + \frac{1}{\sqrt{2}} \Phi^*_2 \Phi_1 \De_m  
- \frac{1}{2} \Phi^*_2 \Phi_2 \De_3   \rii  \nn \\
& \; +  \la_{H\Phi} \lee H^\dag_1 H_1 + H^\dag_2 H_2 \rii  \lee \Phi^*_1\Phi_1 + \Phi^*_2\Phi_2 \rii \nn\\
& \; +  \la^\prime_{H\Phi} \lee H^\dag_1 H_1 \Phi^*_1\Phi_1 + H^\dag_2 H_2  \Phi^*_2\Phi_2 
+ H^\dag_1 H_2 \Phi_2^*\Phi_1 + H^\dag_2 H_1  \Phi^*_1\Phi_2  \rii \nn\\
& \; + \la_{H\De} \lee H^\dag_1 H_1 + H^\dag_2 H_2 \rii   \lee \frac{1}{2} \De^2_3 + \De_p \De_m  \rii \nn\\
& \; + \la_{\Phi\De}  
  \lee  \Phi^*_1\Phi_1 + \Phi^*_2\Phi_2 \rii  \lee \frac{1}{2} \De^2_3 + \De_p \De_m  \rii \; .
  \label{eq:vmix}
\end{align}
One can further substitute the component fields of the two doublets $H_1$ and $H_2$ into Eq.~\eqref{eq:vmix}.
However the resulting 
expression is tedious and not illuminating, and so will not be shown here.

Before moving to the next subsection on spontaneous symmetry breaking 
and minimization of the potential, 
we would like to make some general comments regarding the G2HDM potential.
\begin{enumerate}[(i)]
\item
$U(1)_X$ is introduced to simplify the Higgs potential $V_T $ in Eq.~\eqref{eq:higgs_pot}.
For example, a term $ \Phi^{\rm T}_H \De_H \Phi_H $ obeying the $SU(2)_H$ symmetry would be allowed 
in the absence of $U(1)_X$. Note that as far as the scalar potential is concerned, treating $U(1)_X$ as a global symmetry is sufficient to 
kill this and other unwanted terms. However having a global symmetry lying around in the model defeats 
our original purpose of gauging the discrete global $Z_2$ symmetry. 
Therefore we prefer to treat $U(1)_X$ as a local symmetry.
\item
Note that the $\mu^2_H$ and $\mu_\Phi^2$ terms in
$V(H)$ and $V(\Phi_H)$ have the ``right" signs while the $\mu_\De^2$ in $V(\De_H)$
has the ``wrong" sign.
\item
The quadratic terms for $H_1$ and $H_2$ have the following coefficients
\begin{eqnarray}
\label{H1massterm}
&&\mu^2_H - \frac{1}{2} M_{H\De} \cdot v_\De + \frac{1}{2} \lambda_{H \De} \cdot v_\De^2 
+  \frac{1}{2} \lambda_{H\Phi} 
 \cdot v_\Phi^2 \; , \\
&&\mu^2_H + \frac{1}{2} M_{H\De} \cdot v_\De + \frac{1}{2} \lambda_{H \De} \cdot v_\De^2 
+  \frac{1}{2} (   \lambda_{H\Phi} +  \lambda^\prime_{H\Phi} ) 
 \cdot v_\Phi^2 \; , 
 \label{H2massterm}
\end{eqnarray} 
respectively.
Since the four parameters $M_{H\De}$, $\lambda_{H\De}$, $\lambda_{H\Phi}$
and $\lambda^\prime_{H\Phi}$ can take on either positive or negative values, 
even with a positive $\mu_H^2$, Eqs.~\eqref{H1massterm} and \eqref{H2massterm} 
can be negative and positive respectively 
so that one can achieve $\langle H_1 \rangle \neq 0$ and $\langle H_2 \rangle = 0$ 
to break $SU(2)_L$.
Because the doublet $H_2$ does not obtain a VEV, 
its neutral component, if lighter than the charged Higgs,
can potentially be a DM candidate whose stability is protected by the $SU(2)_H$ gauge symmetry.
\item
Similarly, the quadratic terms for two fields $\Phi_1$ and $\Phi_2$ have the coefficients
\begin{eqnarray}
\label{Phi1massterm}
&&\mu^2_\Phi + \frac{1}{2} M_{\Phi\De} \cdot v_\De + \frac{1}{2} \lambda_{\Phi \De} \cdot v_\De^2 
+  \frac{1}{2} ( \lambda_{H\Phi} + \lambda^\prime_{H\Phi} ) 
\cdot v^2 \; , \\
&&\mu^2_\Phi - \frac{1}{2} M_{\Phi\De} \cdot v_\De + \frac{1}{2} \lambda_{\Phi \De} \cdot v_\De^2 
+  \frac{1}{2} \lambda_{H\Phi} 
\cdot v^2 \; , 
\label{Phi2massterm}
\end{eqnarray}
respectively. 
As in the above cases of $H_1$ and $H_2$,
even with a positive $\mu_\Phi^2$, 
one can achieve $\langle \Phi_1 \rangle = 0$ and $\langle \Phi_2 \rangle \neq 0$ 
by judicious choices of the parameters.
\item
In~\eqref{VDelta}, if $-\mu^2_\De <0$, $SU(2)_H$ is spontaneously broken 
by the VEV $\lan \De_3 \ran =  - v_\De \neq 0$ with $\lan \De_{p,m} \ran=0$ by applying an $SU(2)_H$ rotation. 
In fact, this also triggers the symmetry breaking of the other gauge symmetries as discussed
in the next subsection.
\item
All terms in $V(H)$, $V(\Phi_H)$, $V(\De_H)$ and $V_{\rm mix}(H, \De_H, \Phi_H)$ are Hermitian, 
implying all the coefficients are necessarily real. Thus the scalar potential
in G2HDM is $CP$-conserving.
\item
Classification of all the symmetry breaking patterns of G2HDM like the analysis made 
in \cite{Deshpande:1977rw} for 2HDM would be very interesting but nevertheless beyond the scope of this work. 
We would like to address this issue in the future.
\end{enumerate}

\subsection{Spontaneous symmetry breaking}\label{section:spon_spon}
 
To facilitate spontaneous symmetry breaking, let us shift the fields as follows 
\begin{eqnarray}
\label{HiggsComponents}
H_1 = 
\begin{pmatrix}
G^+ \\ \frac{v + h}{\sqrt 2} + i \frac{G^0}{\sqrt 2}
\end{pmatrix}
, \;
H_2 = 
\begin{pmatrix}
H^+ \\ H_2^0
\end{pmatrix}
, \;
\Phi_H = 
\begin{pmatrix}
G_H^p \\ \frac{v_\Phi + \phi_2}{\sqrt 2} + i \frac{G_H^0}{\sqrt 2}
\end{pmatrix}
, \;
\Delta_H =
\begin{pmatrix}
\frac{-v_\De + \delta_3}{2} & \frac{1}{\sqrt 2}\De_p \\ 
\frac{1}{\sqrt 2}\De_m & \frac{v_\De - \delta_3}{2}
\end{pmatrix} . \nn \\
\end{eqnarray}
Here $v$, $v_\Phi$ and $v_\De$ are VEVs to be determined
by minimization of the potential. The set 
$\Psi_G \equiv \{ G^+, G^0, G^p_H,  G^0_H\}$ are Goldstone bosons, to be
absorbed by the longitudinal components of $W^+$, $W^3$, $W^p$, $W^{\prime 3}$ respectively.

Substituting the VEVs in the potential $V_T$ in Eq.~\eqref{eq:higgs_pot}  leads to
\begin{eqnarray}
\label{Vvevs}
V_T(v,  v_\De , v_\Phi) & = & 
\frac{1}{4} \left[ 
\lambda_H v^4 + \lambda_\Phi v_\Phi^4 + \lambda_\De v_\De^4 + 2 \left( \mu_H^2 v^2 + \mu_\Phi^2 v_\Phi^2 - \mu_\De^2 v_\De^2 \right) \right. \nonumber \\
&& \left. - \left( M_{H\De} v^2 + M_{\Phi\De} v_\Phi^2 \right) v_\De + \lambda_{H\Phi} 
 v^2 v_\Phi^2 + \lambda_{H\De} v^2 v_\De^2 + \lambda_{\Phi\De} v_\Phi^2 v_\De^2
\right] \; .
\end{eqnarray}
Note that the two new couplings $\lambda^\prime_H$ and $\lambda^\prime_{H\Phi}$
do not appear in Eq.~\eqref{Vvevs}.
Thus minimization of the potential in Eq.~\eqref{Vvevs}
yield the same set of VEV equations as in~~\cite{Huang:2015wts}.
For convenience, we here list them again
\begin{eqnarray}
\label{vevv}
\left( 2\lambda_H v^2 + 2 \mu_H^2 - M_{H\De} v_\De  + \lambda_{H\Phi} 
 v_\Phi^2 + \lambda_{H\De} v_\De^2 \right)
& = & 0 \; , \\
\label{vevphi}
 \left( 2\lambda_\Phi v_\Phi^2 + 2 \mu_\Phi^2 -  M_{\Phi\De} v_\De + \lambda_{H\Phi} 
 v^2 + \lambda_{\Phi\De} v_\De^2 \right)
& = &  0 \; , \\
\label{vevdelta}
4\lambda_\De v_\De^3 - 4 \mu_\De^2 v_\De - M_{H \De} v^2 - M_{\Phi \De} v_\Phi^2 
+ 2 v_\De \left( \lambda_{H\De} v^2 + \lambda_{\Phi\De} v_\Phi^2 \right) & = & 0 \; .
\end{eqnarray}
One can solve for nontrivial $v^2$ and $v_\Phi^2$ in terms
of $v_\De$ and other parameters using Eqs.~\eqref{vevv} and \eqref{vevphi}.
Substituting these solutions of $v^2$ and $v_\Phi^2$ into Eq.~\eqref{vevdelta}
gives rise to a cubic equation of $v_\De$ which can be solved either analytically or
numerically. Once $v_\De$ is known, one can substitute its value back to
Eqs.~\eqref{vevv} and \eqref{vevphi} to determine $v$ and $v_\Phi$ respectively.
In this way, one can explicitly see the effects of the triplet's VEV $v_\De$ on
the breaking of the SM $SU(2)_L \times U(1)_Y$ and the $U(1)_X$, after it first triggers
 $SU(2)_H$ symmetry breaking.
In Sec.~\ref{section:HiggsPheno}, we will study numerically 
if this trigger mechanism can indeed happen for $\mu_\De^2 > 0$ and $\mu_H^2 > 0$.

As pointed out before, the scalar potential of G2HDM is
$CP$-conserving, since
all the coefficients in the scalar potential are real. Since the triplet VEV
$v_\De$ satisfies a cubic equation, it may have one real and two complex
solutions. The complex solutions are nevertheless unphysical and we will
discard such solutions in our numerical scans described in later
sections. In the cases where there are three real solutions, we will
pick the one that has the lowest energy as long as it is consistent with the
SM Higgs VEV $v=246$ GeV.


\subsection{Scalar mass spectrum}\label{section:scalar_mass}
 
The scalar boson mass spectrum can be obtained from taking the second derivatives of 
the potential with respect to the various fields and evaluating the results
at the minimum of the potential.
The mass matrix thus obtained contains three diagonal blocks. The first block is $3 \times  3$.
In the basis of $S=\{h, \phi_2, \delta_3\}$ it is given by 
\begin{align}
{\mathcal M}_0^2 =
\begin{pmatrix}
2 \lambda_H v^2 & \lambda_{H\Phi} v v_\Phi 
& \frac{v}{2} \left( M_{H\De} - 2 \lambda_{H \De} v_\De \right)  \\
\lambda_{H\Phi} v v_\Phi
& 
2 \lambda_\Phi v_\Phi^2
&  \frac{ v_\Phi}{2} \left( M_{\Phi\De} - 2 \lambda_{\Phi \De} v_\De \right) \\
\frac{v}{2} \left( M_{H\De} - 2 \lambda_{H \De} v_\De \right)  & \frac{ v_\Phi}{2} \left( M_{\Phi\De} - 2 \lambda_{\Phi \De} v_\De \right) & \frac{1}{4 v_\De} \left( 8 \lambda_\De v_\De^3 + M_{H\Delta} v^2 + M_{\Phi \De} v_\Phi^2 \right)   
\end{pmatrix} \; .
\label{eq:scalarbosonmassmatrix}
\end{align}
This matrix can be diagonalized by a similarity transformation with an orthogonal matrix  $O$,  defined as
$ \vert f \rangle_i \equiv O_{i j}  \vert m \rangle_j $ with $\vert f \rangle_i$ and $\vert m \rangle_j$ referring to the flavor and mass eigenstates respectively,  
\begin{equation}
O^{\rm T} \cdot {\mathcal M}_0^2 \cdot O = {\rm Diag}(m^2_{h_1}, m^2_{h_2}, m^2_{h_3}) \; ,
\label{eq:OTM0sqO}
\end{equation}
where the three eigenvalues are in ascending order: $m_{h_1} \leq m_{h_2} \leq m_{h_3}$.
The lightest eigenstate $h_1$ will be identified as the 125 GeV Higgs observed at the LHC 
while $h_2$ and $h_3$ are  the heavier Higgs bosons. The physical Higgs $h_i$ is 
a linear combination of  the three components of $S$: $h_i = O_{ji}S_j$. 
Thus the 125 GeV scalar boson could be a mixture of the neutral components of 
$H_1$ and the $SU(2)_H$ doublet $\Phi_H$, 
as well as the real component $\delta_3$ of the $SU(2)_H$ triplet $\Delta_H$.

The SM Higgs $h_1$ tree-level couplings to $f \bar f$, $W^+W^-$, $ZZ$ and $H^+H^-$ 
pairs will all be decreased by an overall factor of $O_{11}$, 
implying a reduction by $ | O_{11} |^2 $ on the $h_1$ decay branching ratios
into these channels. On the other hand, as we shall see later, $h_1 \to \gamma\gamma$ 
(and also $h_1 \to Z \gamma$) involves extra contributions from the $\delta_3$
and $\phi_2$ components, which could lead to either enhancement or suppression with respect to the SM prediction.

The second block is also $3 \times  3$. In the basis of 
$G=\{ G^p_H , H^{0*}_2, \De_p  \}$, it is given by 
\begin{align}
{\mathcal M}_0^{\prime 2} =
\begin{pmatrix}
M_{\Phi \De} v_\De  +\frac{1}{2}\lambda^\prime_{H\Phi}v^2 & \frac{1}{2}\lambda^\prime_{H\Phi}  v v_\Phi & - \frac{1}{2} M_{\Phi \De} v_\Phi  \\
\frac{1}{2}\lambda^\prime_{H\Phi} v v_\Phi &  M_{H \De} v_\De 
+\frac{1}{2}\lambda^\prime_{H\Phi} v_\Phi^2
 &  
\frac{1}{2} M_{H \De} v\\
- \frac{1}{2} M_{\Phi \De} v_\Phi & \frac{1}{2} M_{H \De} v & 
\frac{1}{4 v_\De} \left( M_{H\De} v^2 + M_{\Phi \De} v_\Phi^2 \right)\end{pmatrix} \; .
\label{goldstonemassmatrix}
\end{align}
It is straightforward to show that Eq.~\eqref{goldstonemassmatrix} has a massless eigenstate corresponding to 
the physical Goldstone boson $\widetilde G^{p}$, 
which is a mixture of $G^p_H$, $H^{0*}_2$ and $\Delta_p$~\footnote{
The two physical Goldstone fields are $\widetilde G^{p,m} \sim v_\Phi G^{p,m}_H - v H^{0*,0}_2 + 2 v_\De \De_{p,m}$
that can be verified by ${\mathcal M}_0^{\prime 2} \cdot (v_\Phi, \;  - \, v, \; 2 v_\De )^{\rm T} =0$.}.
The other two eigenvalues are the masses of two physical fields $D$ and $\widetilde \Delta$:
\begin{eqnarray}
\label{darkmattermass}
M^2_{D, {\widetilde \Delta}} &=& \frac{-B \mp \sqrt{B^2 - 4 A C}}{2A} \; ,
\end{eqnarray}
with 
\begin{eqnarray}
\label{ABC}
A & =& 8 v_\Delta \; , \nonumber \\
B & = & - 2 \left( M_{H\Delta} \left( v^2 + 4 v_\Delta^2 \right) + M_{\Phi \Delta} \left( 4 v_\Delta^2 + v_\Phi^2 \right)
+ 2 \lambda^\prime_{H\Phi} v_\Delta \left( v^2 + v_\Phi^2 \right) \right) \;, \\
C & = & \left( v^2 + v_\Phi^2 + 4 v_\Delta^2 \right) 
\left( M_{H \Delta} \left( \lambda^\prime_{H\Phi} v^2 + 
2 M_{\Phi \Delta} v_\Delta \right) + \lambda^\prime_{H\Phi} M_{\Phi \Delta} v_\Phi^2  \right) \; .\nonumber
\end{eqnarray}
The field $D$ can be a DM candidate in G2HDM. Note that in the parameter space where the quantity inside the square root 
of  Eq.~\eqref{darkmattermass} is very small, $\widetilde \Delta$ would be nearly degenerate with $D$. Under this circumstance, we need to include coannihilation processes for relic density calculation.
We note that, besides the scalar $D$, it is possible in G2HDM to have 
fermionic $\nu_R$, $\nu_L^H$, $\nu_R^H$ or vector $W^{\prime (p,m)}$
to be a DM candidate as well, depending on which one is the lightest.
In this work, we will assume $D$ is the DM candidate.
In our numerical scan, detailed in later sections, 
we will check to make sure $D$ must be lighter than $W^{\prime (p,m)}$, 
$H^\pm$ and all heavy fermions.

The final  block is $4 \times 4$ and diagonal with 
\begin{equation}
m^2_{H^\pm} = M_{H \De} v_\De  - \frac{1}{2}\la^\prime_H v^2 +\frac{1}{2}\lambda^\prime_{H\Phi}v_\Phi^2 \; ,
\label{chargedHiggsmass}
\end{equation}
for the physical charged Higgs $H^\pm$, and 
\begin{equation}
m^2_{G^\pm} = m^2_{G^0} = m^2_{G^0_H} = 0 \; ,
\end{equation}
for the four Goldstone boson fields $G^\pm$, $G^0$ and $G^0_H$.
Note that we have used the minimization conditions Eqs.~\eqref{vevv}, \eqref{vevphi} and \eqref{vevdelta} 
to simplify various matrix elements of the above mass matrices.
If the charged Higgs mass is close to the DM mass, we should include the corresponding
coannihilation contributions as well.

The six Goldstone particles $G^\pm$, $G^0$, $G^0_H$ and $\widetilde G^{p,m}$ will be absorbed 
by the longitudinal components of the massive gauge bosons 
$W^\pm$, $Z$, $Z^\prime$ and $W^{\prime (p,m)}$. 
It implies that there are two unbroken generators and thus two massless gauge particles
left over after spontaneous symmetry breaking. 
One is naturally identified as the photon while the other one could be interpreted 
as a dark photon $\gamma_D$ or another neutral gauge boson $Z^{\prime\prime}$.
To give a mass to the $\gamma_D$ or $Z^{\prime\prime}$, one can either use the Stueckelberg 
mechanism~\cite{Kors:2004dx,Kors:2005uz,Feldman:2007wj,Cheung:2007ut} 
or introduce yet another Higgs field $\Phi_X$ solely charged under $U(1)_X$ to 
break one of the remaining two unbroken generators. 
Depending on the magnitude of the Stueckelberg mass $M_X$ or the VEV $\langle \Phi_X \rangle$,
one can identify the extra neutral gauge boson as either $\gamma_D$ or $Z^{\prime\prime}$.
Only one unbroken generator for the massless photon should remain at the end of this game. 
The physical neutral gauge bosons $\gamma$, $Z$, $Z^\prime$ and $\gamma_D / Z^{\prime\prime}$
are in principle mixtures of the gauge field components $W^3$, $B$, $W^{\prime \, 3}$ and $X$~\cite{Huang:2015wts}.

After symmetry breaking, by scrutinizing the whole Lagrangian, 
one can discover that an effective $Z_2$ parity can be assigned consistently
to the physical particle spectrum of the model:
All the SM particles (with $h_1$ identified as the 125 GeV Higgs observed at the LHC), 
$Z^\prime$, $\gamma_D/Z^{\prime\prime}$, $h_2$ and $h_3$
are even, while $D$, $\widetilde \Delta$, $H^\pm$, $W^{\prime (p,m)}$ as well as all heavy fermions $f^H$
are odd under this accidental discrete symmetry. 
As mentioned above, we will assume $D$ is the lightest odd particle and can serve as a 
DM candidate in this work.

\section{Theoretical Constraints}\label{section:th_constraints}

In this section, we will discuss the theoretical constraints arising from tree-level vacuum stability~(VS)
and perturbative unitarity~(PU) on the scalar sector in G2HDM.

\subsection{Vacuum stability} \label{section:vac_stab}

For the stability of the vacuum we have to examine the scalar potential at large-field values 
and make sure it is bounded from below. Therefore it is sufficient to consider all the quartic terms in the scalar 
potential which are
\begin{eqnarray}
V_4 & = & 
+ \la_H   \left( H^\dag H \right)^2 
+ \la'_H \left( - H^\dag_1 H_1 H^\dag_2 H_2 
                  + H^\dag_1 H_2 H^\dag_2 H_1 \right)  \nn
 \\
&& +
\la_\Phi \lee \Phi_H^\dag \Phi_H  \rii^2  + 
 \la_\De \lee {\rm Tr} \lee \De^2_H  \rii \rii^2  \nn \\
&& + 
\la_{H\Phi} \lee H^\dag H  \rii  \lee \Phi_H^\dag \Phi_H \rii  
 + \la^\prime_{H\Phi} \lee H^\dag \Phi_H  \rii  \lee \Phi_H^\dag H \rii \nn \\
 && 
 +   \la_{H\De} \lee H^\dag H  \rii    {\rm Tr} \lee \De^2_H  \rii  
+ \la_{\Phi\De} \lee \Phi_H^\dag \Phi_H \rii {\rm Tr} \lee \De^2_H \rii  \; .
 \label{V4UnExpanded}
\end{eqnarray}

 Following the methods in \cite{ElKaffas:2006gdt,Arhrib:2011uy},
 we introduce the following basis ($x, y, z$) and two ratios $\xi$ and $\eta$, defined as
 \begin{eqnarray}
 x & \equiv & H^\dagger H \; , \\
 y & \equiv & \Phi_H^\dagger \Phi_H \; , \\
 z & \equiv & {\rm Tr} \left( \Delta_H^\dagger \Delta_H \right) \; ,
 \end{eqnarray}
and
\begin{eqnarray}
\label{ratioxi}
\xi & \equiv & \frac{\left( H^\dagger \Phi_H \right) \left(\Phi_H^\dagger H \right)}{ \left( H^\dagger H \right) 
\left( \Phi^\dagger_H \Phi_H \right)} \;  , \\
\label{ratioeta}
\eta & \equiv & \frac{\left( - H^\dagger_1 H_1 H_2^\dagger H_2 +  H^\dagger_1 H_2 H_2^\dagger H_1 \right)}
{\left( H^\dagger H  \right)^2} \; .
\end{eqnarray} 
One can show that the ratios satisfy $0 \leq \xi \leq 1$ and $-1/4 \leq \eta
\leq 0$.  While $\eta$ is always above $-1/4$, the actual lower bound and its
effects will be discussed in more detail below.  To deduce the conditions for
the potential to be bounded from below, we rewrite the quartic terms $V_4$ in
terms of $x$, $y$ and $z$ with the ratio parameters $\xi$ and $\eta$. That is
\begin{align}
V_4 = \left( x \; y \; z \right) \cdot {\bf Q}(\xi , \eta)
 \cdot 
\begin{pmatrix}
x\\
y\\
z 
\end{pmatrix} \; ,
\label{quadraticformQ}
\end{align}
with
\begin{align}
{\bf Q}(\xi , \eta) =  
\left( 
\begin{array}{ccc}
 \widetilde \lambda_H (\eta) & \frac{1}{2}\widetilde \lambda_{H\Phi} (\xi) & \frac{1}{2}\lambda_{H \Delta} \\
\frac{1}{2}\widetilde \lambda_{H\Phi} (\xi) & \lambda_{\Phi} & \frac{1}{2}\lambda_{\Phi\Delta} \\
\frac{1}{2}\lambda_{H \Delta}  & \frac{1}{2}\lambda_{\Phi\Delta} & \lambda_{\Delta} 
\end{array}
\right) \; ,
\label{MatQ}
\end{align}
and $\widetilde \lambda_H (\eta) \equiv \lambda_H + \eta \lambda^\prime_H$, ${\widetilde  \lambda_{H\Phi}}(\xi) \equiv \lambda_{H\Phi} + \xi \lphp $.

According to \cite{Kannike:2012pe,Kannike:2016fmd}, using the Sylvester's criterion or requiring
semipositive definite eigenvalues of the quadratic form ${\bf Q}(\xi , \eta)$,
albeit mathematically rigorous, overly constrain the parameter space in this case. Instead, the notion of copositivity 
(conditionally positive) criteria was proposed for vacuum stability conditions. This  is because  
the Sylvester's criterion or positive semidefinite requirement applies to the case that 
the basis ($x,y,z$) can be either positive or negative, whereas copositivity is applicable to the situation of positive~(or non-negative) 
($x,y,z$) only. As our $x$, $y$ and $z$ are the square of the scalar fields and thus non-negative,
we will use the copositivity criteria accordingly~\footnote{For other rigorous method to deduce the necessary and sufficient conditions 
for various scalar potentials in BSM to be bounded from below, see for example \cite{Klimenko:1984qx}.}.
The copositivity criteria for the $3 \times 3$ symmetric matrix ${\bf Q}(\xi , \eta)$ are~\cite{Kannike:2012pe,Kannike:2016fmd}
\begin{itemize}

\item[(A)]
\begin{equation}
\widetilde \lambda_H (\eta) \geq 0 \; , \;\;\; \lambda_\Phi \geq 0 \; , \;\;\; \lambda_\Delta \geq 0 \; ,
\label{coposA}
\end{equation}

\item[(B)]
\begin{eqnarray}
\label{coposB}
\Lambda_{H\Phi}(\xi , \eta) & \equiv & \widetilde \lambda_{H\Phi}(\xi) + 2 \sqrt{\widetilde \lambda_H (\eta) \lambda_\Phi}  \geq  0 \; , \nonumber \\
\Lambda_{H\Delta} (\eta) & \equiv & \lambda_{H\Delta} + 2 \sqrt{\widetilde \lambda_H (\eta) \lambda_\Delta}  \geq  0 \; , \\
\Lambda_{\Phi \Delta} & \equiv & \lambda_{\Phi \Delta} + 2 \sqrt{\lambda_\Phi \lambda_\Delta}  \geq  0 \; , \nonumber
\end{eqnarray}

\item[(C)]
\begin{eqnarray}
\Lambda_{H\Phi\Delta} (\xi , \eta) \equiv \sqrt{\widetilde \lambda_H (\eta) \lambda_\Phi \lambda_\Delta} 
& + &  \frac{1}{2} \left( 
\widetilde \lambda_{H \Phi}(\xi) \sqrt{\lambda_\Delta} 
+ \lambda_{H \Delta} \sqrt{\lambda_\Phi} 
+ \lambda_{\Phi \Delta} \sqrt{\widetilde \lambda_H (\eta)} 
\right) \nonumber \\
& & \;\;\; + \; \frac{1}{2} \sqrt {\Lambda_{H\Phi}(\xi , \eta) \Lambda_{H\Delta} (\eta) \Lambda_{\Phi \Delta} } 
\geq 0 \; ,
\label{coposC}
\end{eqnarray}

\end{itemize}
with $0 \leq \xi \leq 1$ and $-1/4 \leq \eta \leq 0$ as stated before. 
As shown in~\cite{Kannike:2012pe},
these copositivity criteria are necessary and sufficient conditions for the quartic potential to be bounded from below.

It is easy to see that for a given set of $\lambda$ parameters, the conditions
above are monotonic functions of the ratios $\xi$ and $\eta$. If this last two
ratios were perfectly independent, it would be enough to check
Eqs.~\eqref{coposA}, \eqref{coposB} and \eqref{coposC} only at the boundaries
of said ratios.  However, in our case there is a correlation between the lower
bound of $\eta$ and the value of $\xi$ given by $\eta_{\text{min}} = \xi(\xi -
1)$. Since this constraint limits the value of $\eta$ to be well inside the
rectangle defined by the original boundaries mentioned above, checking
the corners of the rectangle would actually yield stronger constraints on the
scalar potential parameters.  Instead, we should check the boundary
defined by $\eta_{\text{min}} = \xi(\xi - 1)$ for $\Lambda_{H \Phi}(\xi, \eta_{\text{min}})$ and 
$\Lambda_{H \Phi \Delta}(\xi, \eta_{\text{min}})$.

We apply the method outlined in \cite{Bonilla:2015eha} to $\Lambda_{H \Phi}(\xi, \eta_{\text{min}})$
and $\Lambda_{H \Phi \Delta}(\xi, \eta_{\text{min}})$ to find the actual conditions to apply. In particular, we
need to check the minimum values of $\Lambda_{H\Phi}(\xi, \eta_{\text{min}})$ and $\Lambda_{H \Phi \Delta}(\xi, \eta_{\text{min}})$,
when necessary, in the boundary mentioned above. Consider the first derivative of
$\Lambda_{H\Phi}(\xi, \eta_{\text{min}})$ with respect to $\xi$ 
at the boundary defined by $\eta_{\text{min}} = \xi(\xi - 1)$,
\begin{equation}
\frac{d \Lambda_{H\Phi}(\xi, \eta_{\text{min}})}{d \xi} = \lambda^\prime_{H\Phi} + \frac{\lambda^\prime_H \left(2 \xi -
1\right) \sqrt{\lambda_\Phi \left(\lambda^\prime_H
    \left(\xi^{2} - \xi\right) + \lambda_H\right)} }{\lambda^\prime_H \left(\xi^{2} - \xi\right) + \lambda_H} \; .
\end{equation}
By solving $d\Lambda_{H\Phi} (\xi, \eta_{\text{min}} ) / d \xi = 0$, we find
\footnote{There are actually two solutions, differing only in the sign in the
second term. However, the solution with positive sign is a solution only when
$\Lambda_{H\Phi}(\xi,\eta_{\text{min}})$ is concave and $\xi_0$ is actually a
maximum.}
the minimum $\xi_0$
\begin{equation}
\xi_0= \frac{1}{2} - \frac{\lambda^\prime_{H\Phi}}{2} \sqrt{\frac{- \lambda^\prime_H + 4
            \lambda_H}{\lambda^\prime_H \left(4 \lambda^\prime_H \lambda_\Phi
- \left(\lambda^\prime_{H\Phi}\right)^{2}\right)}} \; .
\end{equation}
For this $\xi_0$ to be inside the (0,1) range, the absolute value of the
second term must be smaller than 1/2. This results in the condition
$\left(\lambda^\prime_{H\Phi}\right)^{2} \lambda_H <
\left(\lambda^\prime_H\right)^{2} \lambda_\Phi$. If $\xi_0$ is not in the
(0,1) range we can skip the following checks.

If $\xi_0$ is inside the
(0,1) range we can check that it is actually a minimum by finding the sign of
the second derivative, given by
\begin{equation}
\frac{d^2 \Lambda_{H\Phi}(\xi , \eta_{\text{min}}) } {d \xi^2} = \frac{\lambda^\prime_H \sqrt{\lambda_\Phi \left(\lambda^\prime_H \xi
            \left(\xi - 1\right) + \lambda_H\right)} \left(-
            \frac{\lambda^\prime_H \left(2 \xi - 1\right)^{2}}{2
                \left(\lambda^\prime_H \xi \left(\xi - 1\right) +
            \lambda_H\right)} + 2\right)}{\lambda^\prime_H \xi \left(\xi -
    1\right) + \lambda_H} \; .
\end{equation}
By substituting $\xi=\xi_0$ we find that the sign of the second derivative is
given by the sign of $\lambda'_H$. If $\lambda'_H$ is negative, then
$\Lambda_{H\Phi}(\xi,\xi^2-\xi)$ is concave and it is enough to check the
boundary points $(\xi , \eta_{\rm min}) = (0,0)$ and $(1, 0)$. 
But if $\lambda'_H$ is positive then we have a minimum and we
have to check if $\Lambda_{H\Phi}(\xi_0 , \xi_0^2 - \xi_0) \geq 0$.

In summary, to make sure that $\Lambda_{H\Phi} (\xi , \eta) \geq 0$ we have to check:
\begin{equation}
\Lambda_{H\Phi}(\xi=0,\eta=0) \geq 0 \:\:\:\text{ and
}\:\:\:\Lambda_{H\Phi}(\xi=1,\eta=0) \geq 0 \; ,
\end{equation}
and only in the cases that both conditions $\lambda^\prime_H > 0 $ and
$\left(\lambda^\prime_{H\Phi}\right)^{2} \lambda_H <
\left(\lambda^\prime_H\right)^{2} \lambda_\Phi$ apply we also need
to check:
\begin{equation}
\Lambda_{H\Phi}(\xi_0 , \xi_0^2 - \xi_0 ) = 
    \lambda_{H\Phi} + \frac{\lambda^\prime_{H\Phi}}{2}
    + \frac{1}{2} \sqrt{\left(
        4\lambda^\prime_H \lambda_\Phi - \left(\lambda^\prime_{H\Phi}\right)^{2} \right)
    \left(4 \frac{\lambda_H}{\lambda^\prime_H} -1\right)} \geq 0 \; .
\end{equation}

This same procedure can be applied to $\Lambda_{H \Phi \Delta}(\xi, \eta_{\text{min}})$ defined by
Eq.~\eqref{coposC} but the resulting
expressions are far more complicated and add little to nothing to the
discussion. However we follow the same procedure and, in additional to the
boundary points at $(\xi, \eta) = (0,0)$ and $(1,0)$, we also check the
corresponding minimum when needed.

For the rest of the conditions, if such a minimum exists, it corresponds to
the point $(\xi, \eta) = (1/2,-1/4)$, which we also check together with the
$(0,0)$ and $(1,0)$ boundary points.

\subsection{Perturbative unitarity} \label{section:pert_u}

To infer the tree-level perturbative unitarity constraints on the scalar potential parameters, 
similar to the discussion of vacuum stability above
it is sufficient to focus on the quartic couplings. The reason is that at high energies
the $2 \to 2$ scatterings induced by either the scalar cubic couplings or gauge interactions
are  suppressed by the propagator~(which is inversely proportional to momentum transfer squared) compared to 
those induced by the quartic couplings which do not have such a momentum dependence. 
We will follow the same procedures adopted in~\cite{Arhrib:2000is,Akeroyd:2000wc,Arhrib:2012ia,Arhrib:2015dez}
that deduce the perturbative unitarity constraints for the SM, 2HDM and IHDM.

In terms of the physical and Goldstone bosons defined in Eq.~\eqref{HiggsComponents}, the quartic terms 
$V_4$ can be written as
\begin{eqnarray}
V_4 = & + & \frac{1}{4} \lambda_H \left( h^2 + (G^0)^2 + 2 G^+ G^- + 2 H^{0*}_2 H^0_2 + 2 H^+ H^- \right)^2 \nonumber\\
& + &
\frac{1}{2} \la^\prime_H \biggl( 
- (h^2 + G_0^2) H^+ H^- - 2 G^+G^-H^{0 *}_2H^0_2  \biggr.  \nonumber\\
&& \;\;\;\;\;\;\;\;\; \;
 + \; \biggl. \sqrt 2 G^+ H^- H^0_2 ( h - i G_0) + \sqrt 2 G^- H^+ H^{0*}_2 ( h + i G_0)
\biggr) \nonumber\\
& + & \frac{1}{4} \lambda_{\Phi} \left( \phi_2^2 + (G^0_H)^2 + 2 G^p_H G^m_H  \right)^2 \nonumber\\
& + & \frac{1}{4} \lambda_{\Delta} \left( \delta_3^2 + 2 \Delta_p \Delta_m  \right)^2 \nonumber\\
& + & \frac{1}{4} \lambda_{H \Phi} \left( h^2 + (G^0)^2 + 2 G^+ G^- + 2 H^{0*}_2 H^0_2 + 2 H^+ H^- \right) 
\left( \phi_2^2 + (G^0_H)^2 + 2 G^p_H G^m_H  \right) \nonumber \\
& + & \lambda^\prime_{H \Phi} \left\{  \left(\frac{1}{2}(h^2 + (G^0)^2) + G^+G^-\right) G^p_HG^m_H 
+  (H^+H^- + H^{0*}_2 H^0_2) \frac{1}{2} (\phi_2^2 + (G_H^0)^2) \right.  \nonumber \\
&& \; \; \; \;  \; \; \; \; + \left( G^-H^+ + \frac{1}{\sqrt 2} \left( h - i G^0 \right) H^0_2 \right) G_H^p \frac{1}{\sqrt 2} 
\left( \phi_2 - i G^0_H \right) \nonumber \\
&&  \; \; \; \;  \; \; \; \;   \left.  + \left( G^+H^- + \frac{1}{\sqrt 2} \left( h + i G^0 \right) H^{0*}_2 \right) G_H^m \frac{1}{\sqrt 2} 
\left( \phi_2 + i G^0_H \right) \right\}  \nonumber \\ 
& + & \frac{1}{4} \lambda_{H \Delta} \left( h^2 + (G^0)^2 + 2 G^+ G^- + 2 H^{0*}_2 H^0_2 + 2 H^+ H^- \right) 
\left( \delta_3^2 + 2 \Delta_p \Delta_m  \right) \nonumber \\
& + & \frac{1}{4} \lambda_{\Phi \Delta} \left( \phi_2^2 + (G^0_H)^2 + 2 G^p_H G^m_H  \right) 
\left( \delta_3^2 + 2 \Delta_p \Delta_m  \right) \; .
\label{V4Expanded}
\end{eqnarray}

We now analyze all the $2 \to 2$ scalar scatterings  induced by the above quartic potential.

\begin{itemize}
\item[(I)]
The scattering amplitudes with the initial and final states containing one of the 
following states
\begin{equation}
\left\{ \frac{hh}{\sqrt 2}, \frac{G^0G^0}{\sqrt 2}, G^+G^-, H^{0*}_2H^0_2, 
H^+H^-, \frac{\phi_2\phi_2}{\sqrt 2}, \frac{G^0_HG^0_H}{\sqrt 2}, G^p_HG^m_H,
\frac{\delta_3\delta_3}{\sqrt 2},
\Delta_p\Delta_m \right\}
\label{eq:sca_basis}
\end{equation}
can be straightforwardly computed based on
Eq.~\eqref{V4Expanded}~\footnote{We have included in the amplitude a symmetry factor of 
$1/\sqrt2$ for identical particles in the initial and final states.}.
The corresponding amplitude matrix $\mathcal M_1$ in the basis of Eq.~\eqref{eq:sca_basis} is
\begin{equation}
\mathcal M_1 = 
\left( 
\begin{array}{cccccccccc}
3 \lambda_H &  \lambda_H & \frac{2}{\sqrt 2} \lambda_H & \frac{2}{\sqrt 2} \lambda_H  & \frac{2}{\sqrt 2} \widetilde \lambda_H & \frac{1}{2} \lambda_{H\Phi} & \frac{1}{2} \lambda_{H\Phi}  & \frac{1}{\sqrt 2} \widetilde \lambda_{H\Phi} & \frac{1}{2} \lambda_{H\Delta}  & \frac{1}{\sqrt 2} \lambda_{H\Delta} \\

\lambda_H & 3 \lambda_H & \frac{2}{\sqrt 2} \lambda_H & \frac{2}{\sqrt 2} \lambda_H  & \frac{2}{\sqrt 2} \widetilde\lambda_H & \frac{1}{2} \lambda_{H\Phi} & \frac{1}{2} \lambda_{H\Phi}  & \frac{1}{\sqrt 2} \widetilde \lambda_{H\Phi} & \frac{1}{2} \lambda_{H\Delta}  & \frac{1}{\sqrt 2} \lambda_{H\Delta} \\

\frac{2}{\sqrt 2}  \lambda_H & \frac{2}{\sqrt 2} \lambda_H & 4 \lambda_H & 2 \widetilde \lambda_H & 2 \lambda_H  & \frac{1}{\sqrt 2} \lambda_{H\Phi} & \frac{1}{\sqrt 2} \lambda_{H\Phi}  & \widetilde \lambda_{H\Phi} & \frac{1}{\sqrt 2} \lambda_{H\Delta}  & \lambda_{H\Delta} \\

\frac{2}{\sqrt 2} \lambda_H & \frac{2}{\sqrt 2} \lambda_H & 2 \widetilde \lambda_H & 4 \lambda_H  & 2 \lambda_H & \frac{1}{\sqrt 2} \widetilde \lambda_{H\Phi} & \frac{1}{\sqrt 2} \widetilde \lambda_{H\Phi}  & \lambda_{H\Phi} & \frac{1}{\sqrt 2}\lambda_{H\Delta}  & \lambda_{H\Delta} \\

\frac{2}{\sqrt 2} \widetilde \lambda_H & \frac{2}{\sqrt 2} \widetilde \lambda_H & 2  \lambda_H & 2 \lambda_H  & 4 \lambda_H & \frac{1}{\sqrt 2} \widetilde \lambda_{H\Phi} & \frac{1}{\sqrt 2} \widetilde \lambda_{H\Phi}  & \lambda_{H\Phi} & \frac{1}{\sqrt 2}\lambda_{H\Delta}  & \lambda_{H\Delta} \\

\frac{1}{2} \lambda_{H\Phi}  & \frac{1}{2} \lambda_{H\Phi} & \frac{1}{\sqrt 2} \lambda_{H\Phi} & \frac{1}{\sqrt 2} \widetilde \lambda_{H\Phi} & \frac{1}{\sqrt 2} \widetilde \lambda_{H\Phi}  & 3 \lambda_\Phi &  \lambda_\Phi  & \frac{2}{\sqrt 2} \lambda_\Phi & \frac{1}{2} \lambda_{\Phi\Delta}  & \frac{1}{\sqrt 2} \lambda_{\Phi\Delta} \\

\frac{1}{2} \lambda_{H\Phi} & \frac{1}{2} \lambda_{H\Phi} & \frac{1}{\sqrt 2} \lambda_{H\Phi} & \frac{1}{\sqrt 2} \widetilde \lambda_{H\Phi}  & \frac{1}{\sqrt 2} \widetilde \lambda_{H\Phi} & \lambda_\Phi & 3 \lambda_\Phi  & \frac{2}{\sqrt 2} \lambda_\Phi & \frac{1}{2} \lambda_{\Phi\Delta}  & \frac{1}{\sqrt 2} \lambda_{\Phi\Delta} \\

\frac{1}{\sqrt 2} \widetilde \lambda_{H\Phi} & \frac{1}{\sqrt 2} \widetilde \lambda_{H\Phi} & \widetilde \lambda_{H\Phi} & \lambda_{H\Phi}  & \lambda_{H\Phi} & \frac{2}{\sqrt 2}  \lambda_\Phi & \frac{2}{\sqrt 2}  \lambda_\Phi  & 4 \lambda_\Phi & \frac{1}{\sqrt 2} \lambda_{\Phi\Delta}  & \lambda_{\Phi\Delta} \\

\frac{1}{2} \lambda_{H\Delta} & \frac{1}{2} \lambda_{H\Delta}  & \frac{1}{\sqrt 2} \lambda_{H\Delta} & \frac{1}{\sqrt 2} \lambda_{H\Delta} & \frac{1}{\sqrt 2} \lambda_{H\Delta} & \frac{1}{2} \lambda_{\Phi\Delta} & \frac{1}{2} \lambda_{\Phi\Delta}  & \frac{1}{\sqrt 2} \lambda_{\Phi\Delta} & 3 \lambda_\Delta  & \frac{2}{\sqrt 2}  \lambda_\Delta \\

\frac{1}{\sqrt 2} \lambda_{H\Delta} & \frac{1}{\sqrt 2} \lambda_{H\Delta}  & \lambda_{H\Delta} & \lambda_{H\Delta} & \lambda_{H\Delta} & \frac{1}{\sqrt 2} \lambda_{\Phi\Delta} & \frac{1}{\sqrt 2} \lambda_{\Phi\Delta}  & \lambda_{\Phi\Delta} & \frac{2}{\sqrt 2}  \lambda_\Delta  & 4 \lambda_\Delta
\end{array}
\right)\; ,
\label{Mscattering1}
\end{equation}
where the $(i,j)$ element corresponds to the amplitude between the states of
the $i$th and $j$th element of the basis; for instance,
the $(1,1)$ element represents the amplitude of the  process $h \, h \to h \, h$. 
Here $\widetilde \lambda_H \equiv \lambda_H - \lambda_H^\prime/2$, $\widetilde \lambda_{H \Phi}  \equiv \lambda_{H \Phi} + \lambda_{H \Phi}^\prime$.

There are 10 eigenvalues in total and seven of them  are given by:
\begin{eqnarray}
	&\lambda_1 = 2\lambda_H, \:\:\:\:\: \lambda_2 = 2\lambda_\Phi, \:\:\:\:\:
	\lambda_3 = 2\lambda_\Delta, \:\:\:\:\: \lambda_{4,5} = 2\lambda_H \pm
	\lambda^\prime_H\nonumber\\
	&\lambda_{6,7} = \widetilde{\lambda}^+_H +
	\lambda_\Phi \pm \sqrt{2\lambda^{\prime 2}_{H\Phi} +
	(\widetilde{\lambda}^+_H - \lambda_\Phi)^2}
\end{eqnarray}
where $\widetilde{\lambda}^+_H \equiv  \lambda_H + \lambda_H^\prime/2$. The rest of
the eigenvalues are the three roots of  the equation $\lambda^3 + a\lambda^2 +
b\lambda + c = 0$ with
\begin{align}
	a = {}& - 5\lambda_\Delta - 6\lambda_\Phi - 10\lambda_H
		+ \lambda^\prime_H \; , \nonumber \\
	b = {}& - 6\lambda_{H\Delta}^2 - 3\lambda_{\Phi\Delta}^2 
		+ 5\lambda_\Delta(10\lambda_H - \lambda^\prime_H + 6\lambda_\Phi)
		+ 6\lambda_\Phi(10\lambda_H - \lambda^\prime_H)
		\nonumber \\
		& - 8(\lambda_{H\Phi} + \lambda^\prime_{H\Phi}/2)^2 \; , \nonumber \\
	c  = {}& 36\lambda_\Phi\lambda_{H\Delta}^2
		- 24\lambda_{H\Delta}\lambda_{\Phi\Delta}(\lambda_{H\Phi}+ \lambda^\prime_{H\Phi}/2)
		+ 40\lambda_\Delta(\lambda_{H\Phi} + \lambda^\prime_{H\Phi}/2)^2
		\nonumber \\
		&
		+ (3\lambda_{\Phi\Delta}^2 
			- 30\lambda_\Delta\lambda_\Phi)(10\lambda_H - \lambda^\prime_H)
	  \; , \nonumber
\end{align}
which can be solved either numerically or analytically.
Since at very high center-of-mass energies all masses can be ignored, one can take $\sqrt s \to \infty$. 
The $2 \leftrightarrow 2$ scatterings among the scalars are then governed by the quartic couplings. The amplitudes 
are all constants without energy dependence and we can 
focus on the $s$-wave~\cite{Arhrib:2000is}.
Unitarity constraints then require all the 10 real eigenvalues $\lambda_i$ 
of the Hermitian $\mathcal M_1$ to satisfy~\cite{Arhrib:2000is,Akeroyd:2000wc,Arhrib:2012ia,Arhrib:2015dez}
\beq
\vert \lambda_i ( \mathcal M_1 ) \vert \leq 8 \pi \; , \;  \forall i =1,\cdots , 10.
\label{pertunitarity1}
\eeq
\end{itemize}
There also exist processes involving different particles in the initial and final states that
can be divided into the following groups: 
\begin{itemize}

\item[(II)]

For the basis $\left\{ h H^{0*}_2, G^0 H^{0*}_2, G^+ H^- \right\}$ as initial and final states,
we have the following scattering matrix
\begin{equation}
\mathcal M_2 = 
\left( 
\begin{array}{ccc}
2 \lambda_H & 0 & \frac{\sqrt 2}{2} \lambda_H^\prime \\
0 & 2 \lambda_H &  + \frac{i \sqrt 2}{2} \lambda_H^\prime\\
\frac{\sqrt 2}{2} \lambda_H^\prime &  - \frac{i \sqrt 2}{2} \lambda_H^\prime & 2 \lambda_H
\end{array}
\right) \; , 
\label{Mscattering2}
\end{equation}
with eigenvalues $2\lambda_H$ and $2 \lambda_H \pm \lambda_H^\prime$.

\item[(III)]

For the basis $\left\{ h G^+, H^{0*}_2 H^+, G^0 G^+
\right\}$, 
we find the scattering matrix is the same as that  in Eq.~\eqref{Mscattering2}.

\item[(IV)]
\begin{equation}
\mathcal M ( h G^0 \longleftrightarrow h G^0 ) \; = \; \mathcal M (G^+ H^+\longleftrightarrow G^+ H^+) \; =
2\lambda_{H} \; .
\end{equation}

\item[(V)]

\begin{equation}
\mathcal M (h H^+ \longleftrightarrow h H^+)  \,  = \,
\mathcal M (G^0 H^+\longleftrightarrow G^0 H^+) \, =  \,
\mathcal M (H^{0*}_2 G^+ \longleftrightarrow H^{0*}_2 G^+ )\, = \, 
2\widetilde{\lambda}_{H} \; .
\end{equation}

\item[(VI)]

\begin{equation}
\mathcal M \left(G^+ H_2^0 \longleftrightarrow h H^+  \, , G^0 H^+ \right) =
	\frac{1}{\sqrt{2}} (1, -i) \lambda_H^\prime  \;  .
\end{equation}

\item[(VII)]
\begin{equation}
\mathcal M (\phi_2 G_H^0 \longleftrightarrow \phi_2 G_H^0) \, = \,
\mathcal M (\phi_2 G_H^p \longleftrightarrow \phi_2 G_H^p) \, = \, 
\mathcal M (G_H^0 G_H^p \longleftrightarrow G_H^0 G_H^p) \, = \, 2\lambda_{\Phi} \; .
\end{equation}

\item[(VIII)]
\begin{equation}
\mathcal M (\delta_3 \De_p \longleftrightarrow \delta_3 \De_p ) \, = \, 2\lambda_{\De} \; .
\end{equation}

\item[(IX)]
The scattering amplitudes of the following processes
\begin{equation}
\begin{array}{rr}
h\phi_2 \longleftrightarrow  h \phi_2 \; , & h G^0_H \longleftrightarrow  h G^0_H \;,   \\
G^0\phi_2 \longleftrightarrow  G^0 \phi_2 \;, & G^0 G^0_H \longleftrightarrow  G^0 G^0_H \;,   \\ 
G^+\phi_2 \longleftrightarrow  G^+ \phi_2 \;, & G^+ G^0_H \longleftrightarrow  G^+ G^0_H \;,  
\end{array}
\end{equation}
and
\begin{equation}
H^{0*}_2 G^p_H \longleftrightarrow  H^{0*}_2 G^p_H \; , \;
H^+ G^p_H \longleftrightarrow  H^+ G^p_H \; ,
\end{equation}
are all equal to $\lambda_{H \Phi}$.

\item[(X)]
The scattering amplitudes of the following processes
\begin{equation}
h G^p_H \longleftrightarrow  h G^p_H \; ,
G^0 G^p_H \longleftrightarrow  G^0 G^p_H \; ,
G^+ G^p_H \longleftrightarrow  G^+ G^p_H \; ,
\end{equation}
and
\begin{equation}
\begin{array}{rr}
H^{0*}_2\phi_2 \longleftrightarrow  H^{0*}_2 \phi_2 \;, & 
H^{0*}_2 G^0_H \longleftrightarrow  H^{0*}_2 G^0_H \;, \\
H^+\phi_2 \longleftrightarrow  H^+ \phi_2 \;, & H^+ G^0_H \longleftrightarrow  H^+ G^0_H \;, 
\end{array}
\end{equation}
are all equal to $\widetilde \lambda_{H \Phi}$.

\item[(XI)]

\begin{eqnarray}
\mathcal M (G^p_H G^-  \longleftrightarrow   H^- \phi_2 ) & = & \frac{1}{\sqrt 2} \lambda^\prime_{H \Phi} \; , \\
\mathcal M (G^p_H H_2^0 \longleftrightarrow   h \phi_2 ) & = &  
- \mathcal M (G^p_H H_2^0 \longleftrightarrow   G^0 G_H^0 )  =  \frac{1}{2} \lambda^\prime_{H \Phi} \; , \\
\mathcal M (G^p_H G^-  \longleftrightarrow   H^- G_H^0 ) & = & 
\mathcal M (G^p_H H_2^0 \longleftrightarrow   h G_H^0 ) \; = \;
\mathcal M (G^p_H H_2^0 \longleftrightarrow   G^0 \phi_2 ) \nn \\
 & = &   - \frac{i}{\sqrt 2} \lambda^\prime_{H \Phi} \; . 
\end{eqnarray}

\item[(XII)]
The scattering amplitudes of the following processes
\begin{equation}
\begin{array}{rr}
h \delta_3 \longleftrightarrow h \delta_3 \; , & h \Delta_p \longleftrightarrow  h \Delta_p \;  ,\\
G^0 \delta_3 \longleftrightarrow G^0 \delta_3 \; , & G^0 \Delta_p \longleftrightarrow  G^0 \Delta_p \; ,\\
G^+ \delta_3 \longleftrightarrow G^+ \delta_3 \; , & G^+ \Delta_p \longleftrightarrow  G^+ \Delta_p \; ,\\
H^{0*}_2 \delta_3 \longleftrightarrow H^{0*}_2 \delta_3 \; , & H^{0*}_2 \Delta_p \longleftrightarrow  H^{0*}_2 \Delta_p \; , \\
H^+ \delta_3 \longleftrightarrow H^+ \delta_3 \; , & H^+ \Delta_p \longleftrightarrow  H^+ \Delta_p \; ,
\end{array}
\end{equation}
are all equal to $\lambda_{H \De}$.

\item[(XIII)]
The scattering amplitudes of the following processes
\begin{equation}
\begin{array}{rr}
\phi_2 \delta_3 \longleftrightarrow \phi_2 \delta_3 \; , & \phi_2 \Delta_p \longleftrightarrow  \phi_2 \Delta_p \; ,\\
G^0_H \delta_3 \longleftrightarrow G^0_H \delta_3 \; , &  G^0_H \Delta_p \longleftrightarrow  G^0_H \Delta_p \; , \\
G^p_H \delta_3 \longleftrightarrow G^p_H \delta_3 \; , &  G^p_H \Delta_p \longleftrightarrow  G^p_H \Delta_p \; ,
\end{array}
\end{equation}
are all equal to $\lambda_{\Phi \De}$.
\end{itemize}

To summarize:  For the above 13 groups of scattering processes, perturbative unitarity requires the following constraints
\begin{equation}
\begin{array}{rcc}
{\rm (I)} \Longrightarrow & \vert \lambda_i ( \mathcal M_1 ) \vert \leq 8 \pi \; , \;  \forall i = (1,\cdots , 10) \, ,& \\
{\rm (II)-(VIII)} \Longrightarrow &  \vert \lambda_H \vert  \leq  4 \pi \; , \vert \lambda_H^\prime \vert \leq 8 \sqrt 2 \pi \; ,
 \vert 2 \lambda_H \pm \lambda^\prime_H \vert \leq 8 \pi \; , \vert  \lambda_\Phi \vert  \leq  4 \pi \; , \vert \lambda_\De \vert  \leq   4 \pi \; , &\\
{\rm (IX),(X),(XI)} \Longrightarrow & \vert \lambda_{H\Phi} \vert  \leq  8 \pi \; , 
 \vert \widetilde \lambda_{H\Phi} \vert = \vert \lambda_{H\Phi} + \lambda_{H\Phi}^\prime \vert \leq  8 \pi \; , 
\vert \lambda^\prime_{H\Phi} \vert  \leq  8 \sqrt 2 \pi \; , & \\
{\rm (XII),(XIII)} \Longrightarrow & \vert \lambda_{H\Delta} \vert  \leq  8 \pi \; ,  \vert \lambda_{\Phi\Delta} \vert  \leq  8 \pi \; . &
\end{array}
\label{pertunitarity3}
\end{equation}

\subsection{Numerical results from vacuum stability and
perturbative unitarity}
\label{sec:VSPU}

In this section, we will present numerical results from the constraints of VS and PU. 
The VS constraints  correspond to Eqs.~\eqref{coposA}--\eqref{coposC} in Sec.~\ref{section:vac_stab}, 
i.e., the constraints on the copositivity of the matrix ${\bf Q}(\xi , \eta)$. 
On the other hand, the PU constraints  can be found in Sec.~\ref{section:pert_u} and are
summarized in Eq.~\eqref{pertunitarity3}.
For the two ratios $\xi$ and $\eta$ defined in Eqs.~\eqref{ratioxi} and \eqref{ratioeta}, 
we will use the endpoint values $\xi =0,1$ and $\eta = -1, 0$ in our analysis.
In the following, $\lambda_{H,\Phi,\De}$ and $\lambda^\prime_H$ are referred to as the diagonal couplings,
and $\lambda_{H\Phi,H\De,\Phi\De}$ and $\lambda^\prime_{H\Phi}$ as the off-diagonal couplings.

\begin{figure}
	\begin{minipage}[b]{0.475\textwidth}
   	\includegraphics[width=\textwidth]{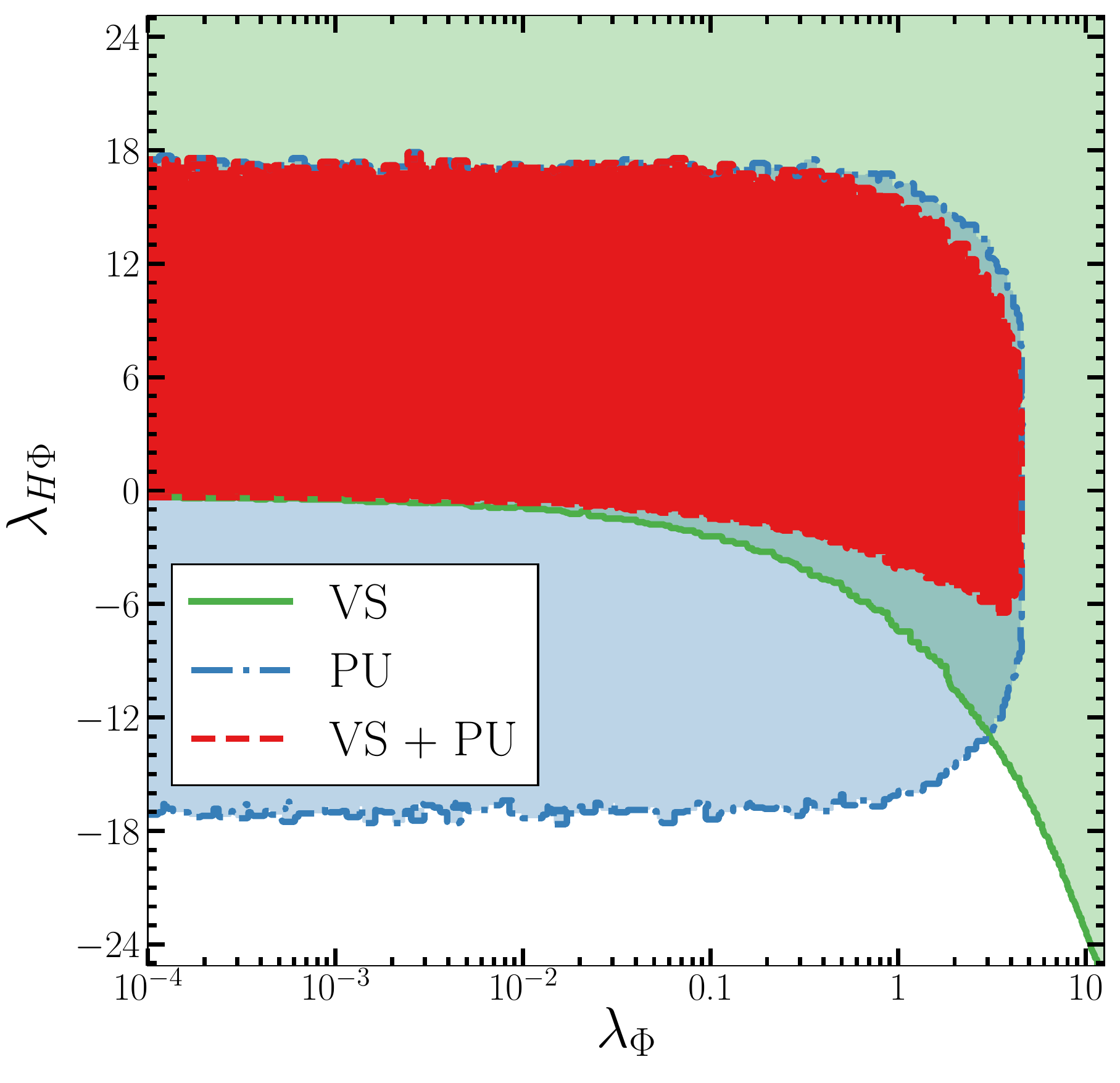}
   \end{minipage}
   \hfill
   \begin{minipage}[b]{0.475\textwidth}
	   \includegraphics[width=\textwidth]{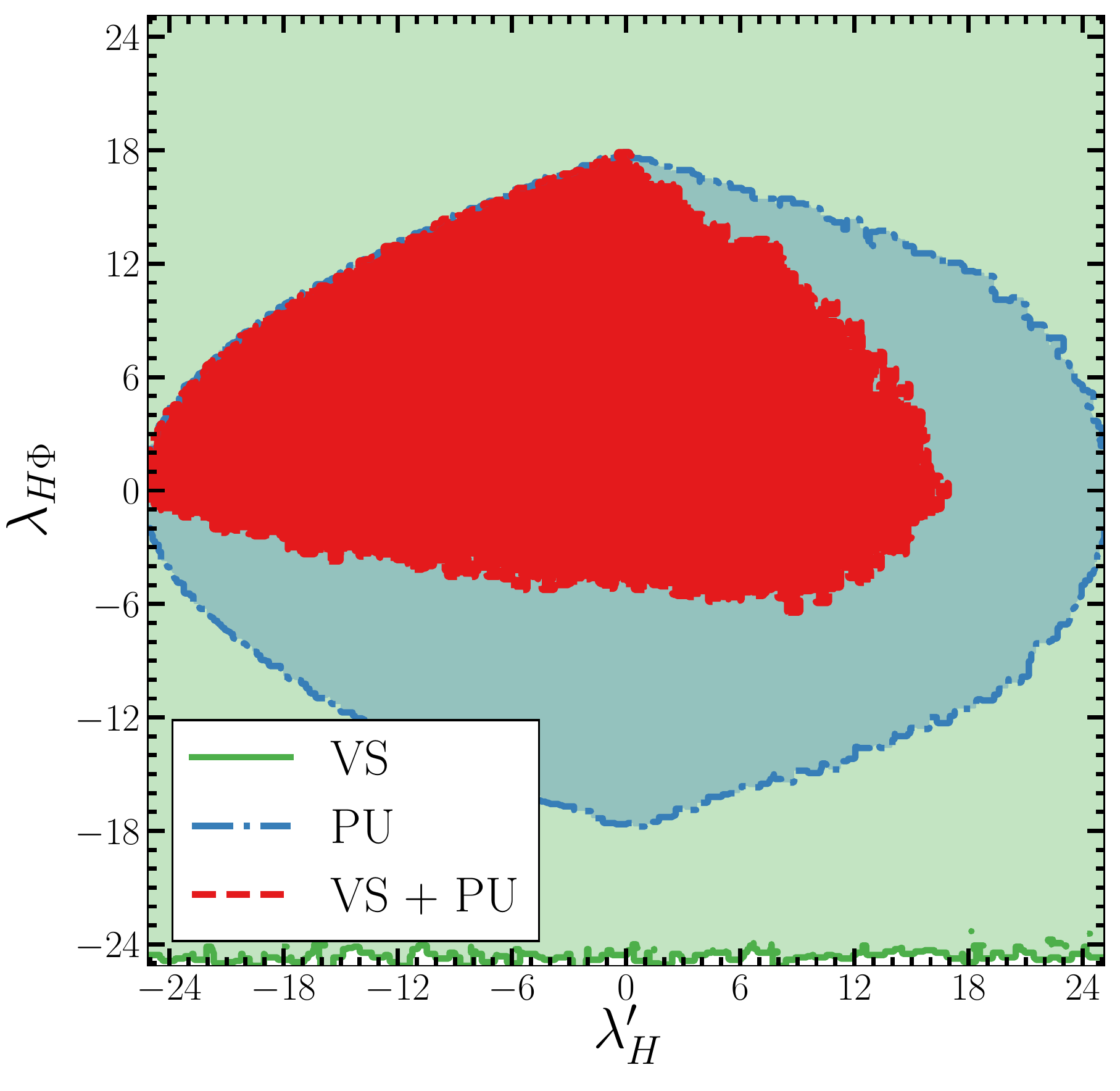}
   \end{minipage}

	\begin{minipage}[b]{0.475\textwidth}
	   \includegraphics[width=\textwidth]{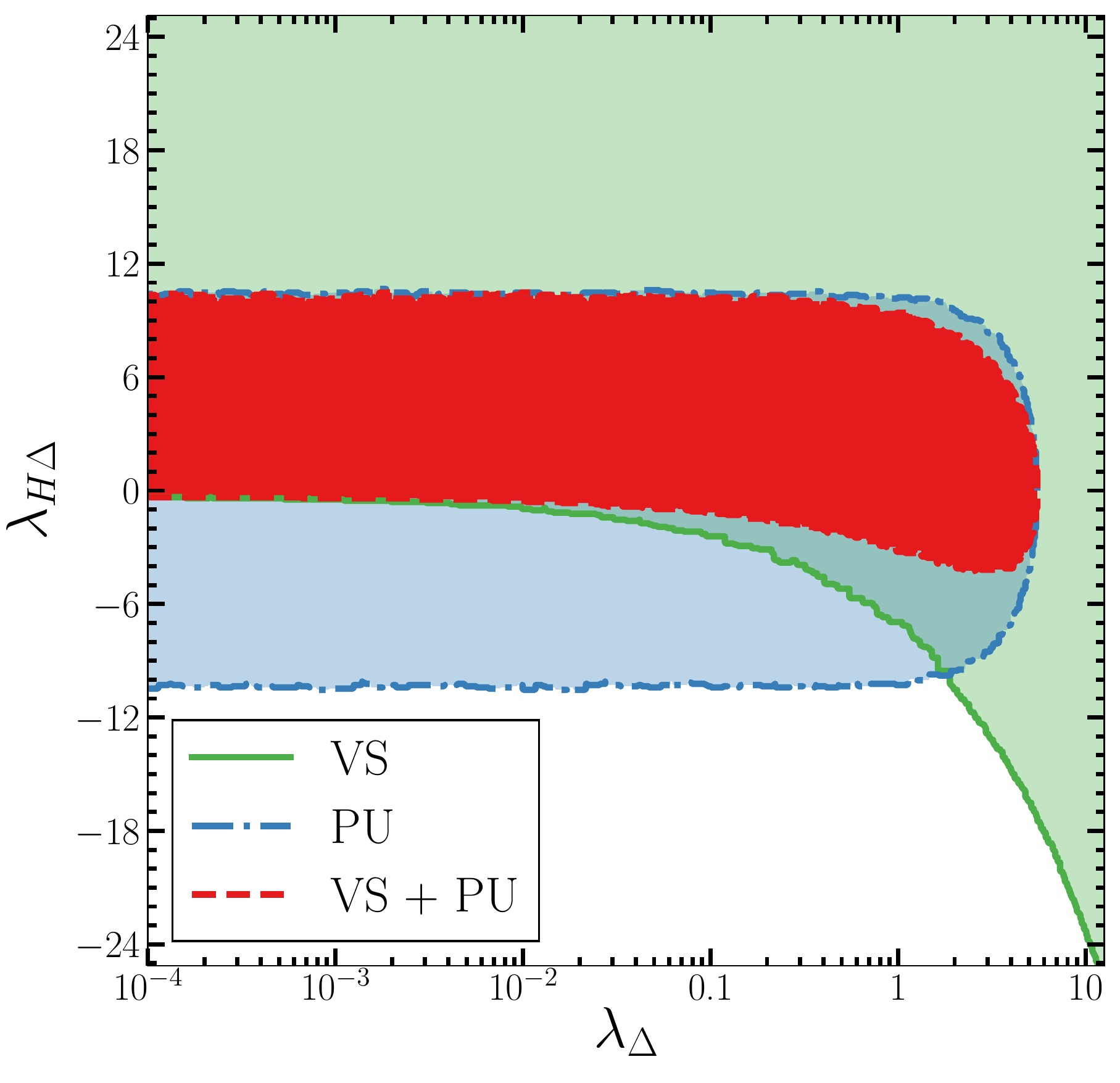}
   \end{minipage}
   \hfill
   \begin{minipage}[b]{0.475\textwidth}
	   \includegraphics[width=\textwidth]{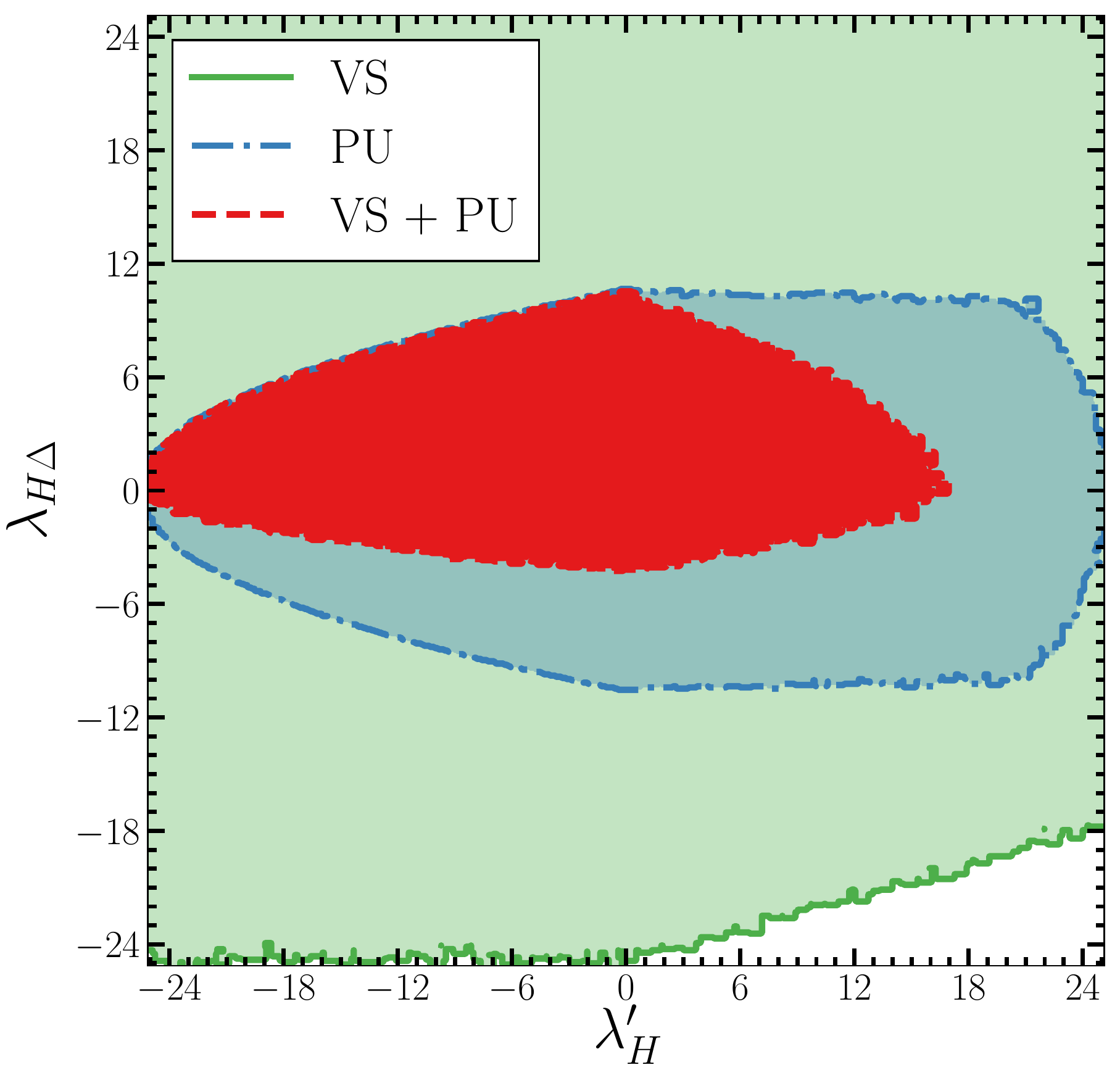}
   \end{minipage}
   \caption{
      \label{fig:lams}
      Allowed regions of the parameter space by the VS, PU and (VS+PU) constraints projected onto the 
      $(\lambda_\Phi, \lambda_{H\Phi})$, $(\lambda_\De, \lambda_{H\De})$,
      $(\lambda^\prime_H, \lambda_{H\Phi})$ and $(\lambda^\prime_H, \lambda_{H\De})$ planes.}
\end{figure}

In the upper- and lower-left panels of Fig.~\ref{fig:lams}, the allowed regions 
for diagonal couplings $\lambda_{\Phi}$ and $\lambda_{\Delta}$ versus
off-diagonal couplings $\lambda_{H\Phi}$ and $\lambda_{H\Delta}$ respectively are presented.
In fact, the allowed regions projected on to the $\la_H$-$\la_{H\Phi}$~($\la_H$-$\la_{H\De}$) plane
behave quite similarly to those of $\la_\Phi$-$\la_{H\Phi}$~($\la_\De$-$\la_{H\De}$).
The green and blue regions are the allowed regions by VS and PU respectively, 
while only the red regions are  allowed  when both VS and PU are considered. 
Clearly, the constraints from PU~[Eq.~\eqref{pertunitarity3}] alone will impose upper limits 
on all diagonal couplings $\lambda_{i}$, where $i=(H,\Phi,\Delta)$, as   shown
by the blue regions for the cases of $\lambda_\Phi$ and $\lambda_\De$.
On the other hand, the green regions from the constraints of VS Eqs.~\eqref{coposA}--\eqref{coposC} 
restrict the  off-diagonal
$\lambda_{H\Phi}$ and $\lambda_{H\Delta}$ to be greater than $-2$ for the diagonal 
$\lambda_{i}<0.1$.  
In contrast, for larger $\lambda_{i}$, 
the windows for negative values of $\lambda_{H\Phi}$ and $\lambda_{H\Delta}$ are widened.

The small wedgelike regions bounded by the green, blue and red contours 
with negative $\lambda_{H\Phi,H\Delta}$ and 
$0.1\lesssim\lambda_H,\lambda_\Phi\lesssim 4$
are allowed by either VS or PU alone.
However as both the VS and PU constraints are imposed, this region 
is excluded as the two constraints are in tension.
We found that this tension is mainly caused by a nontrivial combination of the
copositive conditions in Eq.~\eqref{coposB} and the unitarity constraints in Eq.~\eqref{pertunitarity3}.
Loosely speaking, if the off-diagonal $\lambda_{ij}$ is negative, 
then larger $\lambda_{i}$ and $\lambda_{j}$ are needed as indicated in Eq.~\eqref{coposB}.  
However, the PU conditions set limits on the size of the diagonal couplings
$\vert \lambda_{H,\Delta,\Phi} \vert$ which in turn implies lower limits on the off-diagonal $\lambda_{ij}$ 
via the VS constraint in Eq.~\eqref{coposB}.  
The last VS constraint of Eq.~\eqref{coposC} actually does not yield much more information
 since the only couplings therein that can be negative are 
$\widetilde \lambda_{H \Phi}(\xi)$, $\lambda_{H \De}$ and $\lambda_{\Phi\De}$, whose
lower limits are already constrained by Eq.~\eqref{coposB}.

In the upper- and lower-right panels of Fig.~\ref{fig:lams}, the allowed regions on the 
$(\lambda^\prime_{H},\lambda_{H\Phi})$ and $(\lambda^\prime_{H},\lambda_{H\Delta})$ plane are presented.
As in the left panels, the green boundaries are constrained by VS. 
However, unlike $\lambda_H$ constrained to be positive, 
a negative value of $\lambda^\prime_{H}$ is more favored. 
Recall that $\widetilde \lambda_H(\eta) \equiv \lambda_H + \eta \lambda^\prime_H$ with 
{$-0.25 \leq \eta \leq 0$.
$\widetilde \lambda_H(\eta) $ would turn negative if $\lambda_H^\prime$ becomes too positive. 
This violates one of the VS constraints, $\widetilde \lambda_H(\eta) \geq 0$ in Eq.~\eqref{coposA}. 
Similar to the two left panels, the small wedgelike regions surrounded by blue and red contours 
are excluded when the both VS and PU constraints are taken into account.

\begin{figure}
	\begin{minipage}[b]{0.475\textwidth}
	   \includegraphics[width=\textwidth]{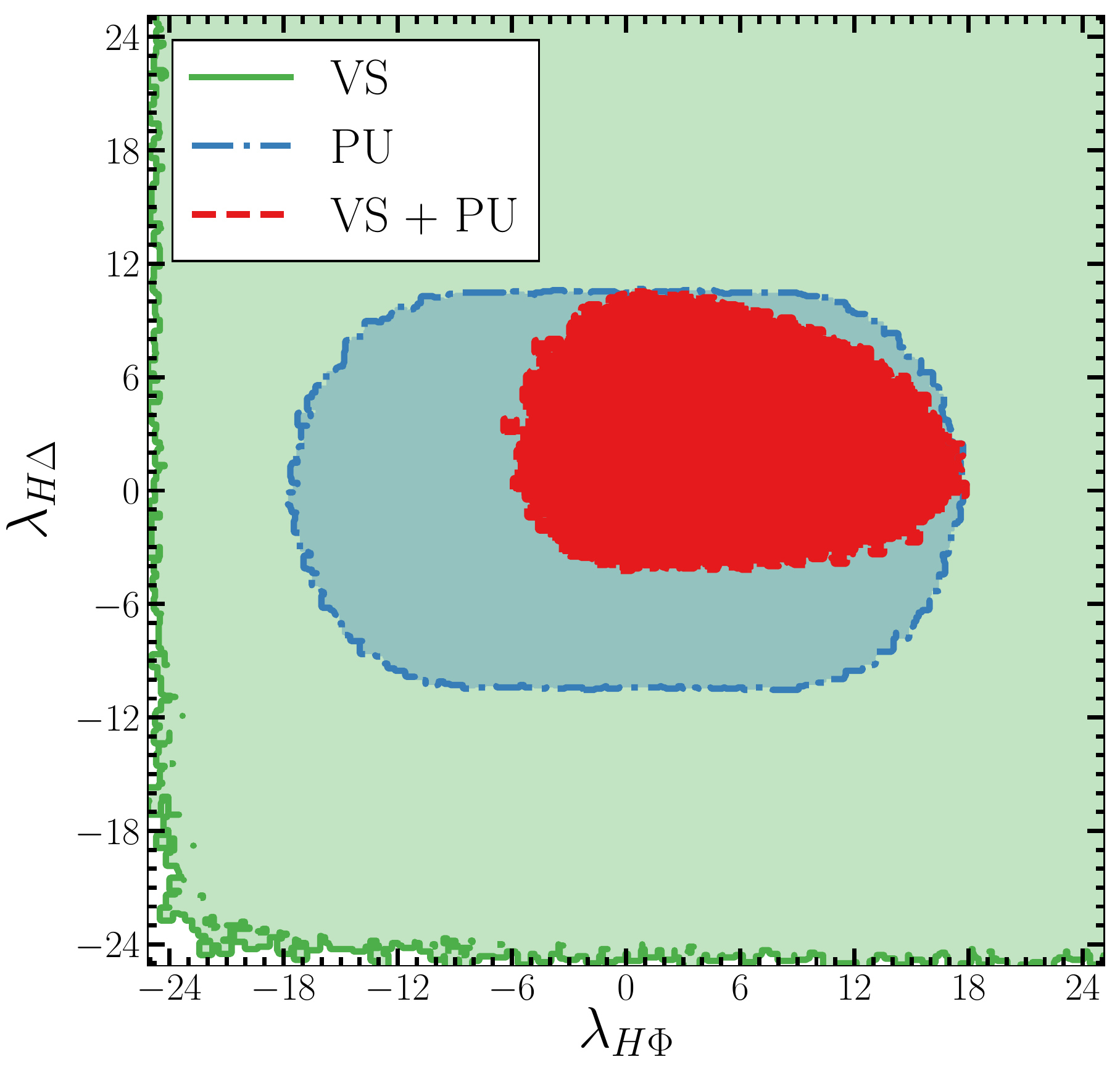}
	\end{minipage}
   \hfill
   \begin{minipage}[b]{0.475\textwidth}
   	\includegraphics[width=\textwidth]{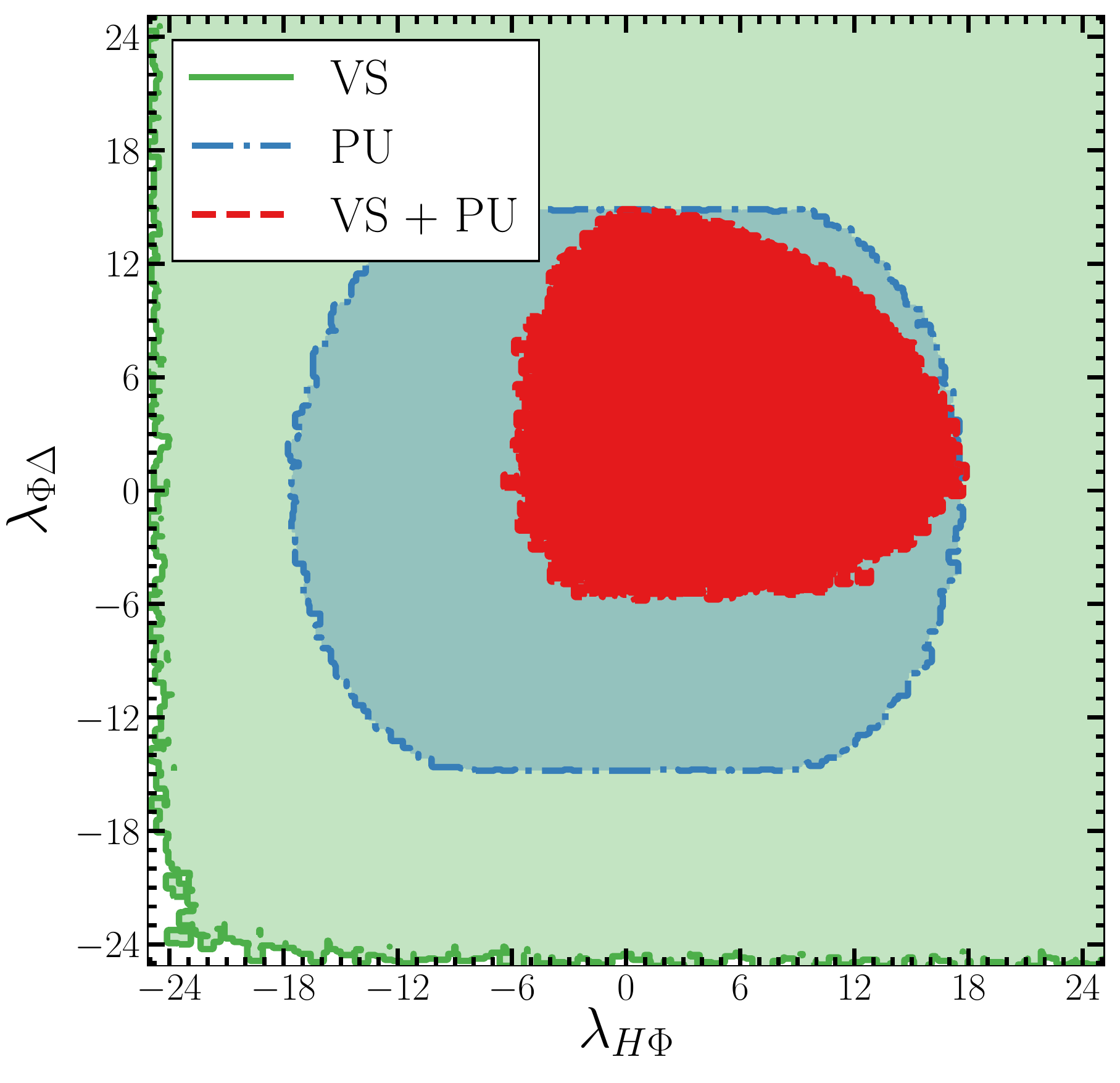}
   \end{minipage}

	\begin{minipage}[b]{0.475\textwidth}
	   \includegraphics[width=\textwidth]{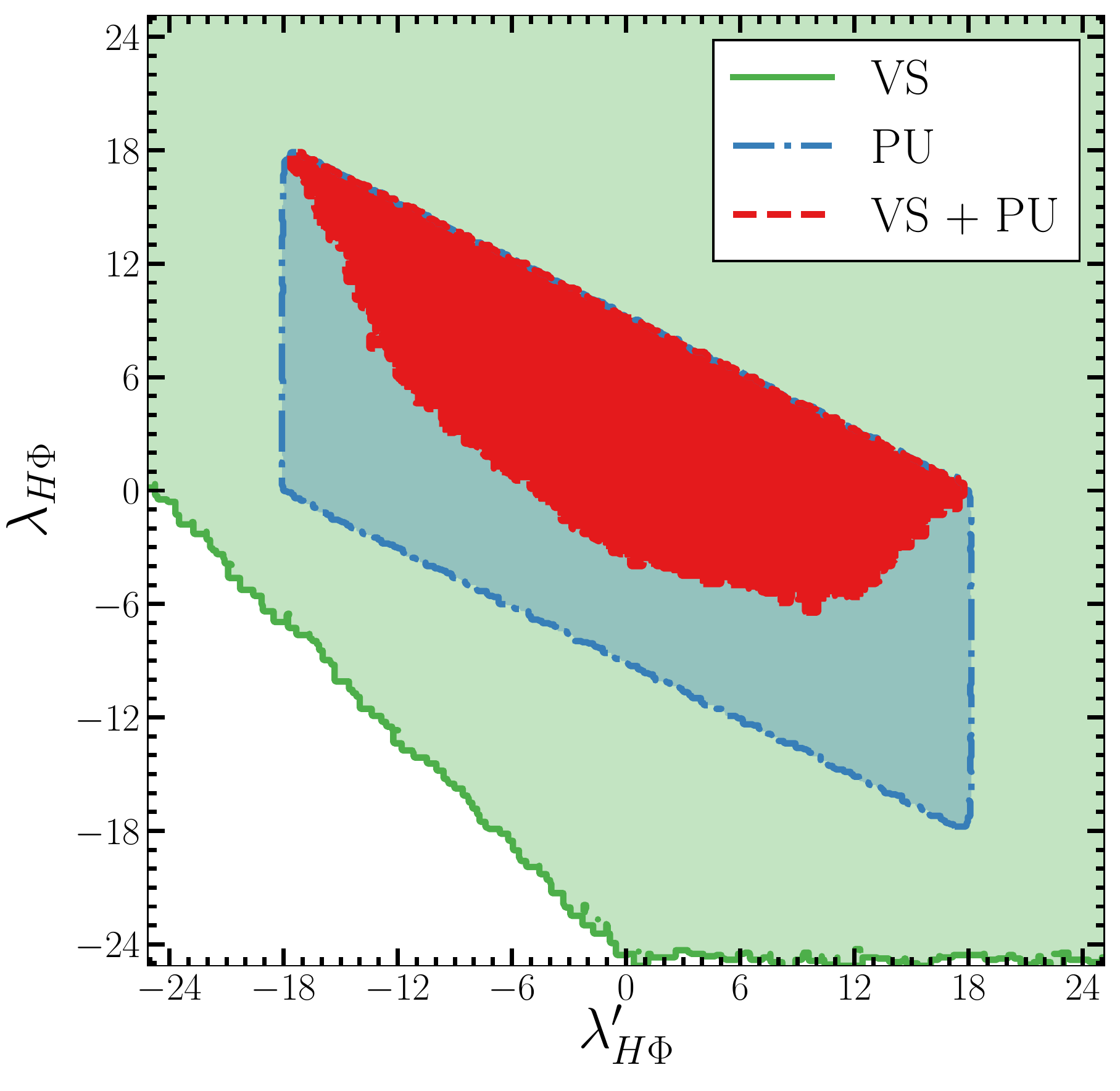}
   \end{minipage}
   \hfill
   \begin{minipage}[b]{0.475\textwidth}
   	\includegraphics[width=\textwidth]{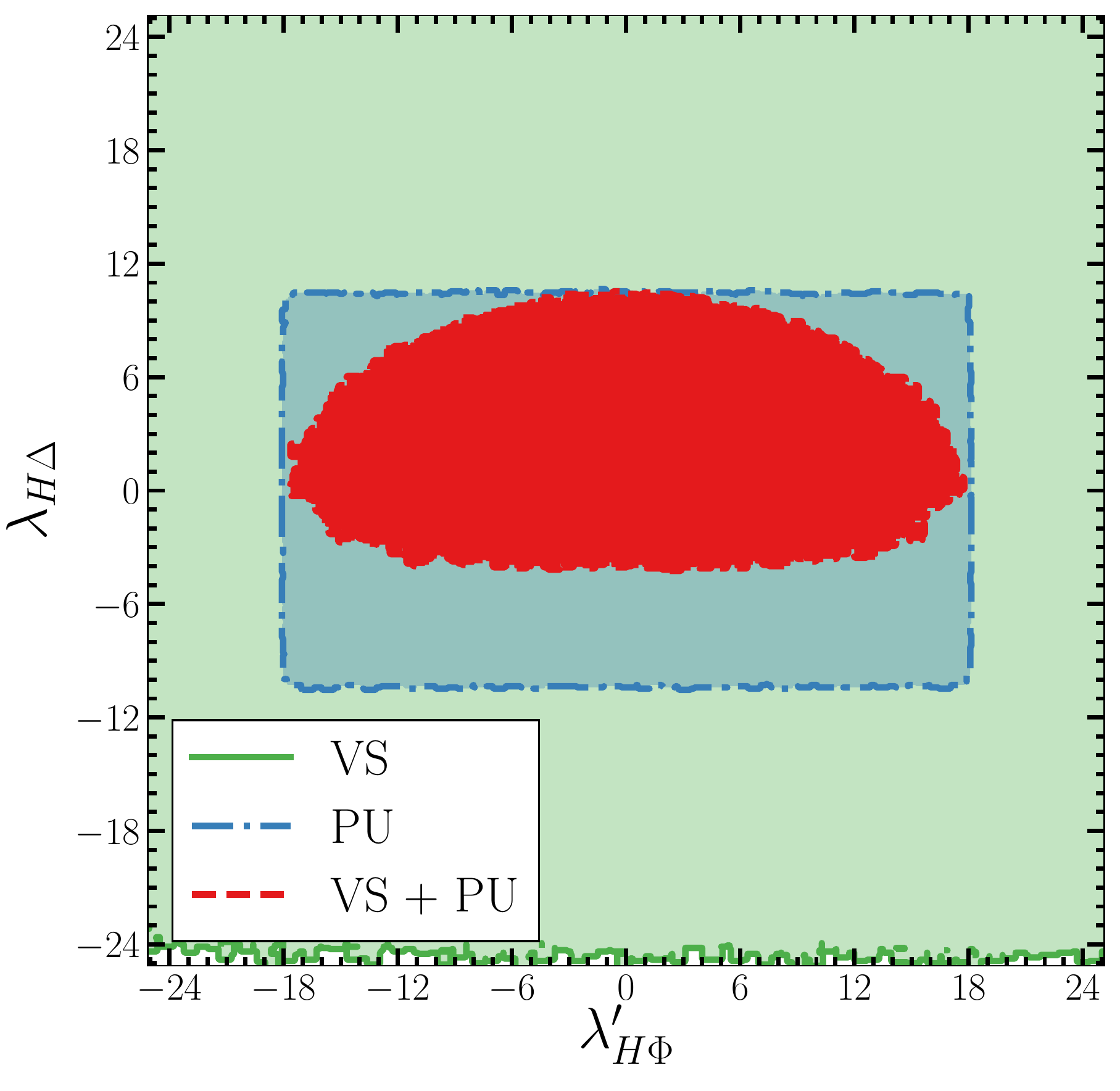}
   \end{minipage}
   \caption{
   \label{fig:lamij}   
   Allowed regions of the parameter space by the VS, PU and (VS+PU) constraints projected onto the planes of
      $(\lambda_{H\Phi}, \lambda_{H\De})$, $(\lambda^\prime_{H\Phi}, \lambda_{H\Phi})$,
      $(\lambda_{H\Phi}, \lambda_{\Phi\De})$ and $(\lambda^\prime_{H\Phi}, \lambda_{H\De})$.}
\end{figure}

In Fig.~\ref{fig:lamij}, we present the allowed regions for the off-diagonal terms $\lambda_{ij}$s. 
Obviously, smaller $\lambda_{ij}$s are always allowed by all the theoretical constraints. 
The limits of $\lambda_{ij}$s obtained from the PU conditions are $|\lambda_{H\Phi}|\lesssim 18$, 
$|\lambda_{H\Phi}^\prime|\lesssim 18$, $|\lambda_{\Phi\Delta}|\lesssim 15$, and $|\lambda_{H\Delta}|\lesssim 11$. 
Interestingly, $\lambda_{H\Delta}$ is slightly more stringent than others 
because it has relatively large coefficients 
inside the scattering matrix $\mathcal M_1$ in Eq.~\eqref{Mscattering1}.   

On the other hand, effects of the VS condition are hardly visible in the parameter space except for the $\lambda_{H\Phi}$-$\lambda_{H\Phi}^\prime$ plane. 
The constraints originate from the parameter ${\widetilde  \lambda_{H\Phi}}(\xi)$.
Recall that ${\widetilde  \lambda_{H\Phi}}(\xi) \equiv \lambda_{H\Phi} + \lphp \xi$ with $0 \leq  \xi \leq1 $.
From the first inequality of Eq.~\eqref{coposB}
${\widetilde  \lambda_{H\Phi}}(\xi) \geq - 2 \sqrt{\widetilde \lambda_H(\eta) \lambda_\Phi}$, 
it implies $\lambda_{H\Phi}$ and $\lambda^\prime_{H\Phi}$ 
cannot be too negative at the same time, leading to the blank bottom-left corner in the
bottom-left panel of Fig.~\ref{fig:lamij}.
As already explained in Fig.~\ref{fig:lams}, when we combine both constraints from VS and PU, 
the allowed region is further shrunk due to the tension between the copositive conditions Eq.~\eqref{coposB} 
and the unitarity limits Eq.~\eqref{pertunitarity3}.

\section{Higgs Phenomenology} \label{section:HiggsPheno}

In this section, we will consider the phenomenological constraints from the Higgs physics
at the LHC.

\subsection{Higgs diphoton decay} \label{section:h_gaga}

In G2HDM, as explained above the 125 GeV Higgs boson $h_1$ is a linear combination
of $h$, $\phi_2$ and $\delta_3$:
\begin{align}
h_1 = O_{11} h + O_{21} \phi_2 + O_{31} \delta_3 ,  
\end{align}
where $O_{ij}$ are the elements of the orthogonal matrix $O$ that diagonalizes the 
mass matrix $\mathcal M^2_0$ displayed in Eq.~\eqref{eq:OTM0sqO}. 
The mixing has an impact on both the $ggH$ production and  Higgs decay branching ratio into two photons.

Due to the narrow Higgs decay width, the Higgs production will be dominated by
the resonance region and thus the cross section $\sigma(pp \to h_1 \to \ga\ga)$ can be well approximated by~\cite{Franceschini:2015kwy}
\begin{align}
\sig\lee gg \to h_1 \to \ga\ga \rii = \frac{ \pi^2}{ 8 s \, m_{h_1} \, \Ga_{h_1} } f_{gg} \lee \frac{m_{h_1}}{\sqrt{s}} \rii
 \Ga\lee h_1 \to gg \rii \Ga\lee h_1 \to \ga\ga \rii ,
\end{align}
with the center of mass energy $\sqrt{s}=13$ TeV and the integral of the parton~(gluon in this case) distribution function product 
\begin{align}
 f_{gg} (\sqrt y) = \int^1_{y} \frac{d x}{x} g\lee x, \mu^2 \rii g\lee \frac{ y } {x}, \mu^2 \rii ,
\end{align}
evaluated at the scale of $\mu=m_{h_1}=125$ GeV. 
The corresponding signal strength for  $ggH$ production is
\begin{align}
    \mu^{\gamma\gamma}_{\text{ggH}}=  \frac{ \Ga^{\text{SM}}_{h} }{ \Ga_{h_1} }  \frac{\Ga\lee h_1 \to gg \rii \Ga\lee h_1 \to \ga\ga \rii }
 {\Ga^{\text{SM}}\lee h \to gg \rii \Ga^{\text{SM}}\lee h \to \ga\ga \rii },
 \label{eq:muggH}
\end{align}   
where the superscript SM refers to the SM Higgs boson $h$. In G2HDM, $\Ga\lee h_1 \to gg \rii$ receives additional contributions from the new colored heavy fermions while $\Ga\lee h_1 \to \ga\ga \rii$
has extra contributions from both the new charged heavy fermions and the charged Higgs $H^\pm$. 
As a result, one has~\cite{Huang:2015wts,Gunion:1989we,Djouadi:2005gi,Djouadi:2005gj,Chen:2013vi}
\begin{align}
\Gamma \, (h_1\to \gamma\gamma) =& \frac{G_{F}\, \alpha^{2}\,m_{h_1}^{3}  O_{11}^2  }
{128\,\sqrt{2}\,\pi^{3}} 
\left| 
A_1(\tau_{W^\pm}) + 
\sum_{f} N_{C} Q_f^2 A_{1/2}(\tau_f) 
\right. \nn\\
 & \left. \;\;\; + \; \mathcal{C}_h
\frac{ \widetilde{\lambda}_H v^2}{m_{H^\pm}^2} A_0(\tau_{H^\pm})
+ \frac{O_{21}}{O_{11}} \frac{v}{v_\Phi} \sum_{F}    N_{C} Q_F^2 A_{1/2}(\tau_F) 
\right|^2  \; ,
\label{eq:hgagaH2}
\end{align}
with $N_C$ being the number of color and
\begin{align}
	\label{eq:Chfactor}
\mathcal{C}_h = 1 + \frac{O_{21}}{O_{11}} \frac{ (\la_{H\Phi} + \la'_{H\Phi}) v_{\Phi} }{ 2\widetilde{\lambda}_H v} - \frac{O_{31}}{O_{11}} \frac{ 2 \la_{H\De} v_{\De} + M_{H\De} }{ 4\widetilde{\lambda}_H v}
 \; ,
\end{align}
where $\widetilde \lambda_H =  \lambda_H - \lambda_H^\prime/2$.
The symbol $f$ refers to the SM fermions while $F$ denotes the heavy fermions.
The form factors for spins $0$, $\frac{1}{2}$ and 1 particles are given by
\begin{eqnarray}
A_{0}(\tau) & =  & - [\tau -f(\tau)]\, \tau^{-2} \, ,\nonumber \\
A_{1/2}(\tau) & = & 2 [\tau +(\tau -1)f(\tau)]\, \tau^{-2}  \, , \nonumber \\   
A_1(\tau) & = & - [2\tau^2 +3\tau+3(2\tau -1)f(\tau)]\, \tau^{-2} \, ,
\label{eq:A0+Af+Aw}
\end{eqnarray}
with the function $f(\tau)$ defined as
\begin{eqnarray}
f(\tau)=\left\{
\begin{array}{ll}  \displaystyle
\arcsin^2\sqrt{\tau} & {\rm , \;\; for} \; \tau\leq 1 \; ; \\
\displaystyle -\frac{1}{4}\left[ \log\frac{1+\sqrt{1-\tau^{-1}}}
{1-\sqrt{1-\tau^{-1}}}-i\pi \right]^2 \hspace{0.5cm} & {\rm , \; \; for} \; \tau>1 \; .
\end{array} \right.
\label{eq:ftau}
\end{eqnarray}
The parameters $\tau_i= m_{h_1}^2/4m_i^2$ with $i=H^\pm,f,F,W^\pm$ are related to the
corresponding masses of the particles in the loops. 
Other symbols not defined in Eq.~\eqref{eq:hgagaH2} are self-explanatory.
As is well known in the SM, even if the $W$ gauge boson or top quark becomes infinitely heavy~($\tau_i \to 0$), 
they do not decouple in the triangle loop and 
have still finite contributions due to $A_{1/2}(0)=4/3$ and $A_1(0)= -7$.
Moreover, $h_{\rm SM} \to \gamma \gamma$ is dominated by the
$W^\pm$ loop contribution, while the top quark contribution is subdominant and has 
destructive interference with that of $W^\pm$. 
In G2HDM, the extra contributions from charged Higgs and new heavy charged fermions  
to the diphoton channel can be either constructive or destructive interferences with the SM
ones, depending on the signs of $\mathcal C_h$ and $O_{21}/O_{11}$ respectively. 
However, due to an extra factor of $1/m^2_{H^\pm}$ in front of 
$A_0(\tau_{H^\pm})$ and the small ratio $v / v_\Phi$ in front of 
$A_{1/2}(\tau_{F})$ in Eq.~\eqref{eq:hgagaH2}, 
both the contributions from the very heavy charged Higgs 
and extra fermions are not significant in the diphoton channel.
They are effectively decoupled in the large mass limit in G2HDM.

On the other hand, the partial decay width of $h_1$ into two gluons mediated by the SM quarks and the new 
colored fermions is~\cite{Gunion:1989we,Djouadi:2005gi,Djouadi:2005gj}
\begin{eqnarray}
\Gamma \, (h_1\to g g) = \frac{ \alpha^{2}_s\,  m^3_{h_1}  O_{11}^2 }
{72 \, v^2  \,\pi^{3}} 
\left| 
\sum_{f}  \frac{3}{4} A_{1/2}(\tau_f) 
+
\frac{O_{21}}{O_{11}} \frac{v}{v_\Phi} \sum_{F}  \frac{3}{4} A_{1/2}(\tau_F) 
\right|^2 \; .
\label{eq:hggH2}
\end{eqnarray}
Depending on the sign of $O_{21}/O_{11}$, the contributions 
from the new heavy quarks in G2HDM 
can increase or decrease the branching fraction.
Note also the suppression factor of the small ratio $v / v_\Phi$ 
for the contributions from new heavy quarks.

Finally, the SM prediction for the branching ratios of $h \to gg$ and $h \to \ga\ga$
can be easily obtained from the previous formulas 
by setting $O_{11}=1$ and $O_{21}=O_{31}=0$ and neglecting the contributions from the
new heavy fermions and $H^\pm$.

\subsection{Numerical results from Higgs physics}

In this subsection, we will discuss constraints on the scalar quartic
couplings $\lambda$s from Higgs physics only~(marked in orange) 
and from Higgs physics~(HP) plus the aforementioned theoretical conditions~(marked in magenta). 
We also colored in red the theoretical constraints alone  as before. 
Here, we include the HP experimental constraints in the $2\sigma$ range from 
the ATLAS experiment~\cite{Aaboud:2018xdt}:
\begin{enumerate}[(i)]
\item Higgs mass $m_h=125.09\pm 0.24\gev$, and 
\item The signal strength of  the Higgs boson produced  through gluon-gluon fusion decaying to two photons, 
    $\mu^{\gamma\gamma}_{ggH}=0.81^{+0.19}_{-0.18}$.
\end{enumerate}
In addition, other theoretical criteria 
are also implemented: (i) the charged Higgs $H^\pm$ and the extra gauge boson $W^{\prime (p,m)}$ 
must be heavier than the DM particle $D$, (ii) any point with negative scalar mass squared is discarded, and 
(iii) the vacuum must be a global minimum.
Since $\lambda_H^\prime$ contributes to the charged Higgs mass 
and the $h_1 H^+H^-$ coupling through mixing effects 
but not to the neutral Higgs mass spectra, 
its constraint from the current LHC Higgs data is coming from 
the charged Higgs running inside the triangle diagram of diphoton channel, 
which is quite loose, in particular when the charged Higgs mass is getting very heavy.

To deduce the constraints from HP, we will adopt the allowed ranges for all the $\lambda$-parameters 
from the theoretical VS and PU constraints in Eqs.~\eqref{coposA}, \eqref{coposB}, \eqref{coposC} 
and~\eqref{pertunitarity3}. Additional
parameters needed to be considered in this subsection are $M_{H\Delta}$, $M_{\Phi\Delta}$
and the VEVs ($v$, $v_\Phi$, $v_\Delta$).
We will fix $v=246$ GeV as the SM Higgs VEV  but 
$v_\Phi$ will be set to $10$ TeV  in order to make the new gauge bosons heavy
and thus satisfy the bounds from  LHC  high-mass resonances searches~\cite{Aaboud:2016cth,CMS:2016abv}. 
The third VEV $v_\Delta$ will take values in the range
$\{0.5,20\}$ TeV and the two parameters $M_{H\Delta}$ and
$M_{\Phi\Delta}$ will be varied in the range $\{-1,1\}$ TeV.
Once these parameters have been chosen, the minimization conditions
Eqs.~\eqref{vevv}, \eqref{vevphi} and \eqref{vevdelta} will determine the numerical
value of the $\mu^2_i$ parameters. 

As mentioned in Sec.~\ref{subsection:higgspotential},
we require $\mu^2_\Delta > 0$ to break $SU(2)_H$ which induces $SU(2)_L$ breaking.
To verify this, we examine the sign of $\mu^2_H$ given the previous parameter choices.
It turns out that for those points which satisfy all the constraints, 40\% of them have positive $\mu^2_H$.
It implies that $SU(2)_H$ breaking in some cases can indeed
trigger $SU(2)_L$ breaking even in the presence of positive $\mu^2_H$.

To ensure the new heavy fermion masses will not contradict our assumption that
$D$ is the DM candidate, we will take them to be 1 TeV heavier than
$D$. However, our results are not sensitive to the precise values of the new
heavy fermion masses as long as they are much heavier than the SM Higgs mass.

\begin{figure}
	\begin{minipage}[b]{0.44\textwidth}
   	\includegraphics[width=\textwidth]{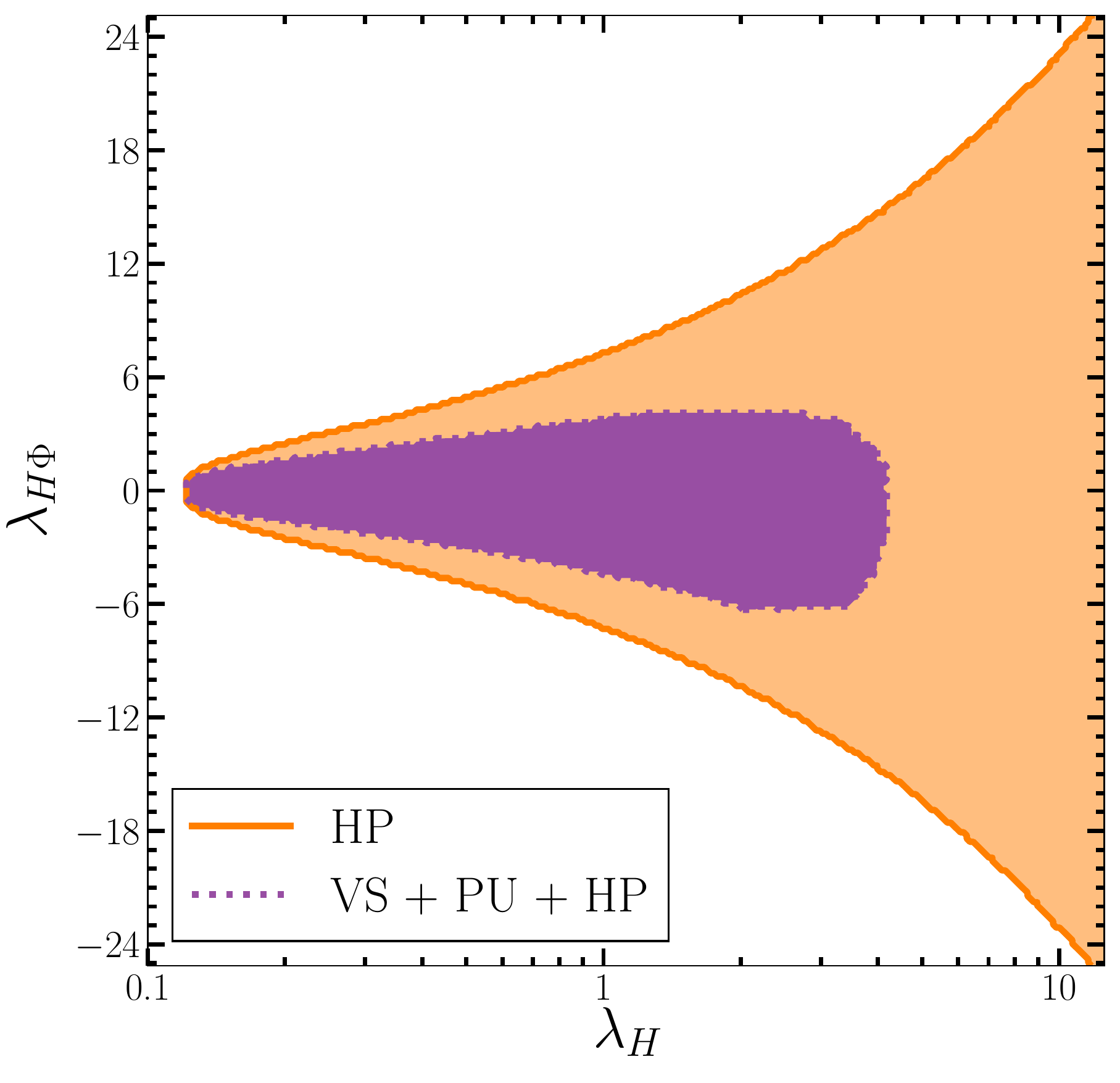}
   \end{minipage}
   \hfill
   \begin{minipage}[b]{0.44\textwidth}
	   \includegraphics[width=\textwidth]{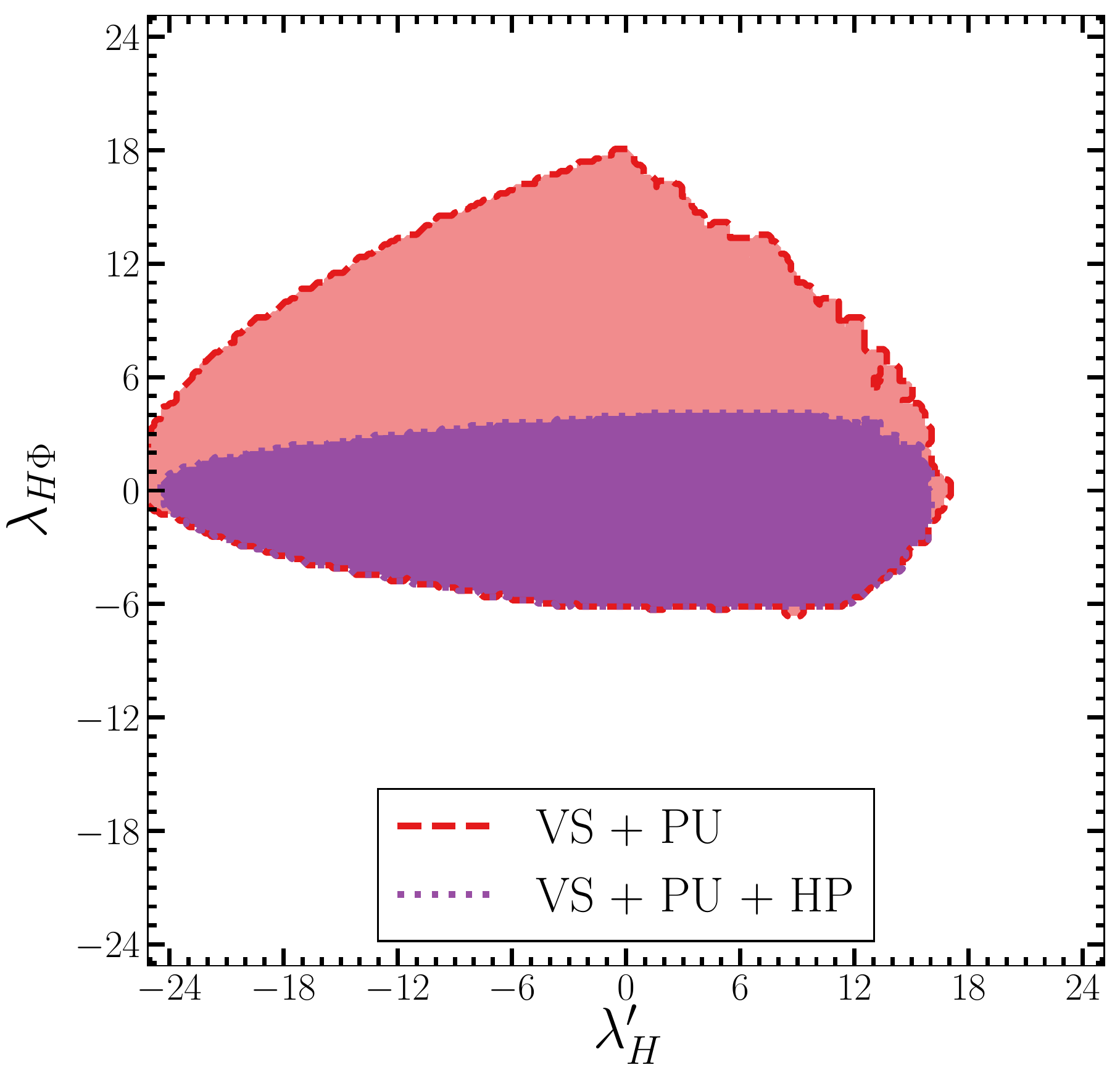}
   \end{minipage}

	\begin{minipage}[b]{0.44\textwidth}
	   \includegraphics[width=\textwidth]{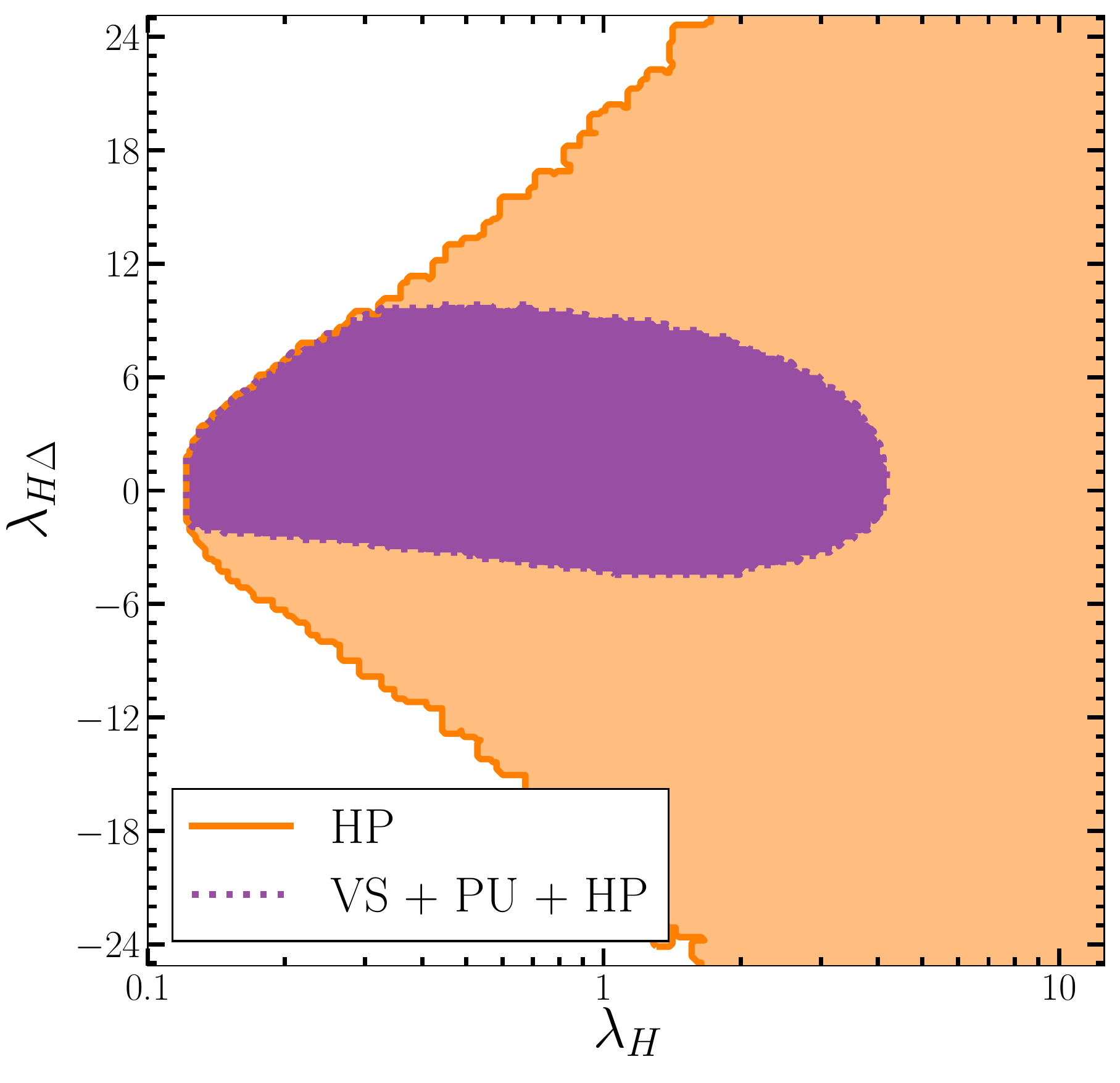}
   \end{minipage}
   \hfill
   \begin{minipage}[b]{0.44\textwidth}
	   \includegraphics[width=\textwidth]{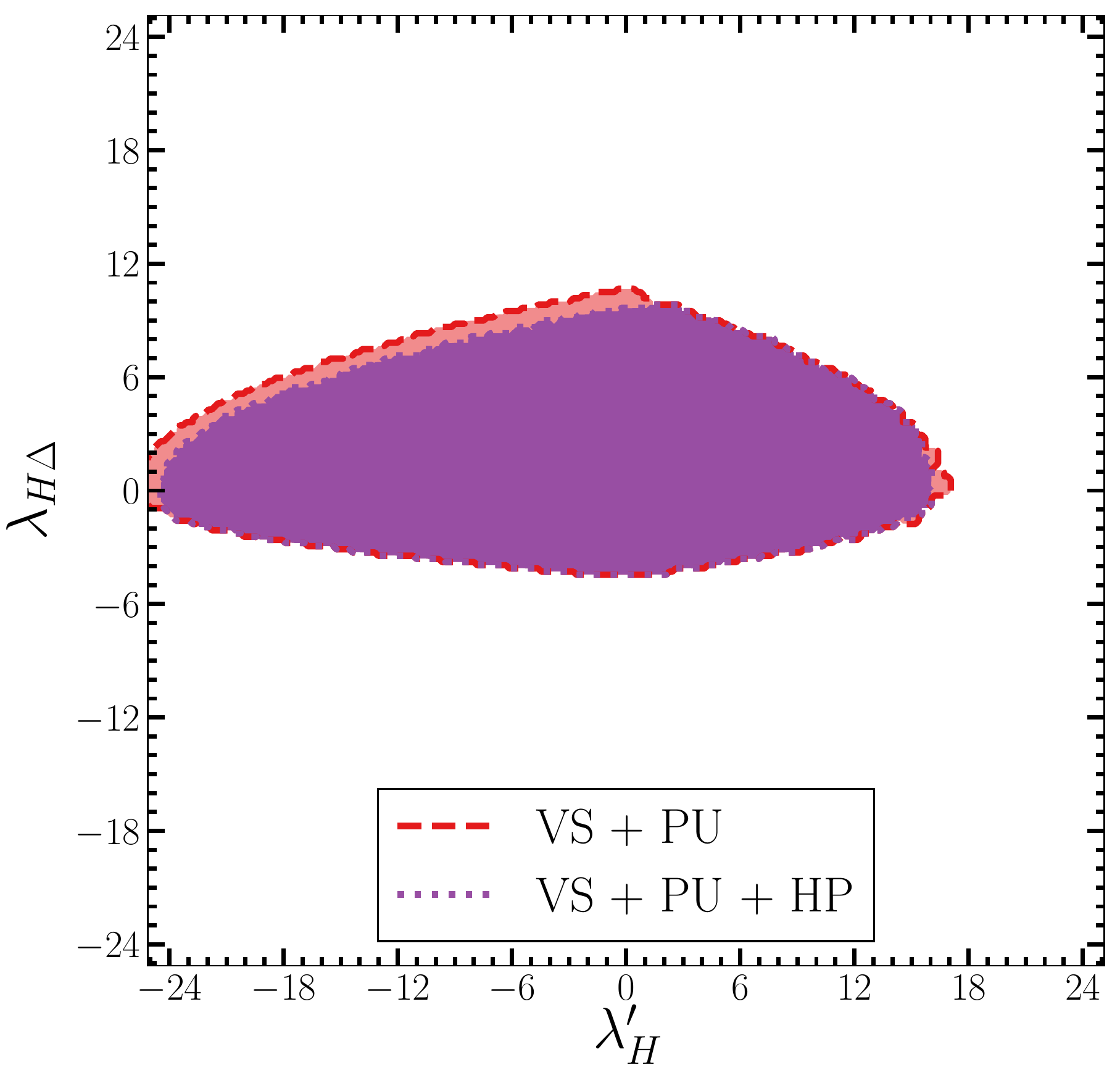}
   \end{minipage}

	\begin{minipage}[b]{0.44\textwidth}
	   \includegraphics[width=\textwidth]{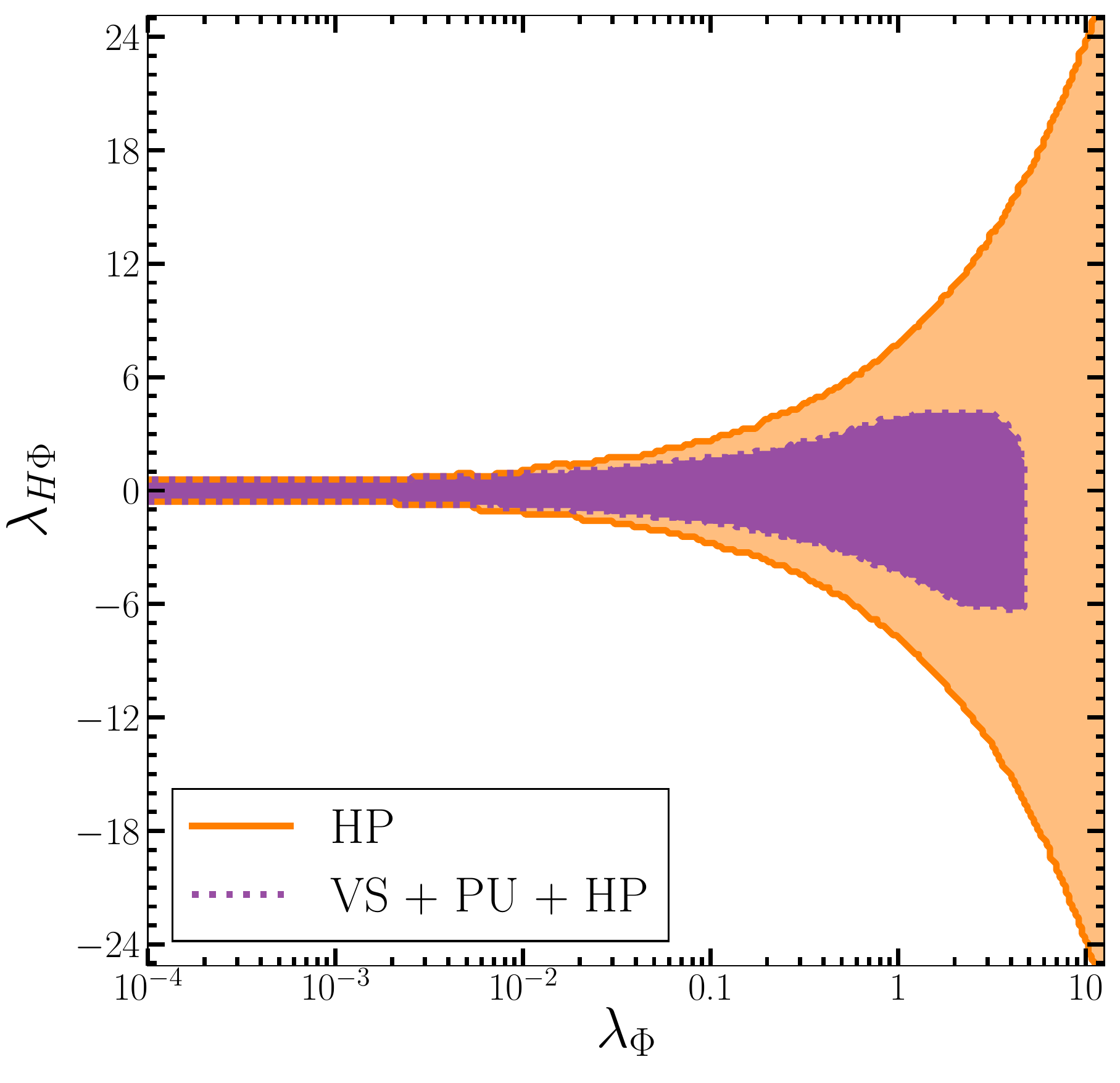}
   \end{minipage}
   \hfill
   \begin{minipage}[b]{0.44\textwidth}
	   \includegraphics[width=\textwidth]{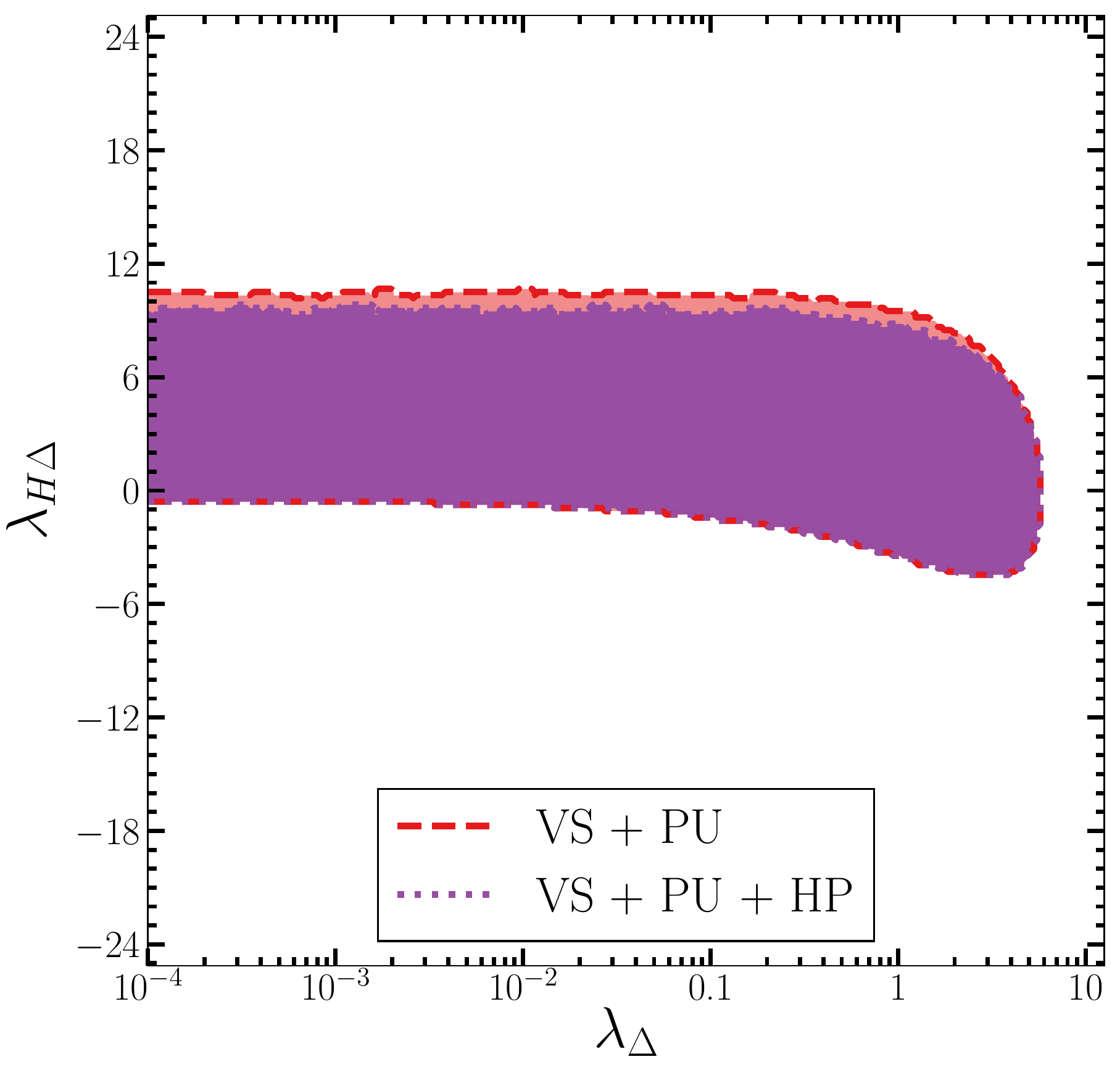}
   \end{minipage}
   \caption{
      \label{fig:diag_vs_non_diag}
	  Allowed regions of the parameter space by the HP, (VS+PU) and (VS+PU+HP) constraints projected onto the planes of
	  $(\lambda_H,\lambda_{H\Phi})$, $(\lambda_H,\lambda_{H\De})$, $(\lambda_\Phi,\lambda_{H\Phi})$, 
	  $(\lambda^\prime_H,\lambda_{H\Phi})$, $(\lambda^\prime_H,\lambda_{H\De})$ 
	  and $(\lambda_\De,\lambda_{H\De})$.}
\end{figure}

In the left columns of Fig.~\ref{fig:diag_vs_non_diag}, from top to bottom, we
project out the allowed regions of $\lambda_H$, $\lambda_\Phi$,
$\lambda_{H\De}$ and $\lambda_{H\Phi}$ on the $(\lambda_H , \lambda_{H\Phi})$,
$(\lambda_H , \lambda_{H\De})$ and $(\lambda_\Phi, \lambda_{H\Phi})$ planes
respectively.  The HP constraints are shown in orange while the combined
(VS+PU+HP) constraints are shown in magenta.
We notice that the allowed regions for Higgs physics constraints are symmetric under
$\lambda_{H\Phi,H\Delta} \to - \lambda_{H\Phi,H\Delta}$ in these three plots.
That is because the Higgs mass $m_{h_1}$ and the mixing with other scalars are invariant
under $\lambda_{H\Phi} \to - \lambda_{H\Phi}$ and/or $\lambda_{H\De} \to - \lambda_{H\De}$ with
 $M_{H\De} \to - M_{H\De}$ as can be seen from Eq.~\eqref{eq:scalarbosonmassmatrix}.
For values of $\lambda_{H}$ close to the SM one $m_h^2/2 v^2 = 0.13$, the 125
GeV Higgs observed at the LHC is very much SM-like, and one expects the mixing
effects controlled by $\lambda_{H\Phi,H\Delta} $ should be minuscule.  This
fact is clearly reflected in the top and middle plots that $\lambda_{H\Phi}$
and $\lambda_{H\De}$ are confined to be small when $\lambda_H \to 0.13$.

The Higgs data constraints alone sets a lower limit for $\lambda_H \geq 0.13$
but not for $\lambda_{\Phi}$. The lower limit stems from the fact that in the
limit of $ v \ll v_{\Phi}$ and $v_\De$, the mixing among $h$, $\phi_2$ and
$\delta_3$ can only {\it decrease} the lightest eigenstate mass $m_{h_1}$
\footnote{
    Mixing effects that lead to decreasing the lightest eigenstate mass can
    also occur in the fermion case. One famous example is the Type-I seesaw
    mechanism, in which the contribution from the heavy right-handed neutrino
    to the light neutrino mass is actually {\it negative} which can be made
    positive via field redefinition.
}
which can be only compensated by a larger $\la_H$~($\geq 0.13$).  To increase
$m_{h_1}$, one would need $\la_H \geq 0.13$.  By including the constraints
from VS and PU conditions, one obtains $0.13 \leq \lambda_H \leq 4.0$
and $0 \leq  \lambda_\Phi \leq 4.19$.  The allowed range of
$\lambda_{H\Delta}$ is wider than that of $\lambda_{H\Phi}$ is simply because
the value of $v_\Phi$ is fixed but $v_\Delta$ is varied in our setup.       In
general the theoretical (VS+PU) constraints (in particular the PU constraints)
are much stringent than the Higgs data constraint. 

Similarly, in the right column of Fig.~\ref{fig:diag_vs_non_diag}, we plot the allowed regions
from top to bottom for the $(\lambda^\prime_H , \lambda_{H\Phi})$, 
$(\lambda^\prime_H , \lambda_{H\De})$ and 
$(\lambda_\De, \lambda_{H\De})$ planes. 
Since as indicated in the left panels  
current LHC Higgs data is more sensitive to the parameters 
$\lambda_H$, $\lambda_{H\Phi}$ and to a lesser extent $\lambda_{H\De}$ but not others, 
we will not show the Higgs data constraints in all the subsequent plots. 
Instead we present only the theoretical (VS + PU) constraints in red and the
combined (VS+PU+HP) constraints in magenta.  It is clear that combining the
Higgs data constraints with the theoretical constraints can further reduce the
allowed regions for these parameters ---most significantly on the
$(\lambda^\prime_H , \lambda_{H\Phi})$ plane.  Allowed regions of other
two-dimensional planes of diagonal versus off-diagonal parameters will be
summarized in the next section. 

\begin{figure}
	\begin{minipage}[b]{0.44\textwidth}
   	\includegraphics[width=\textwidth]{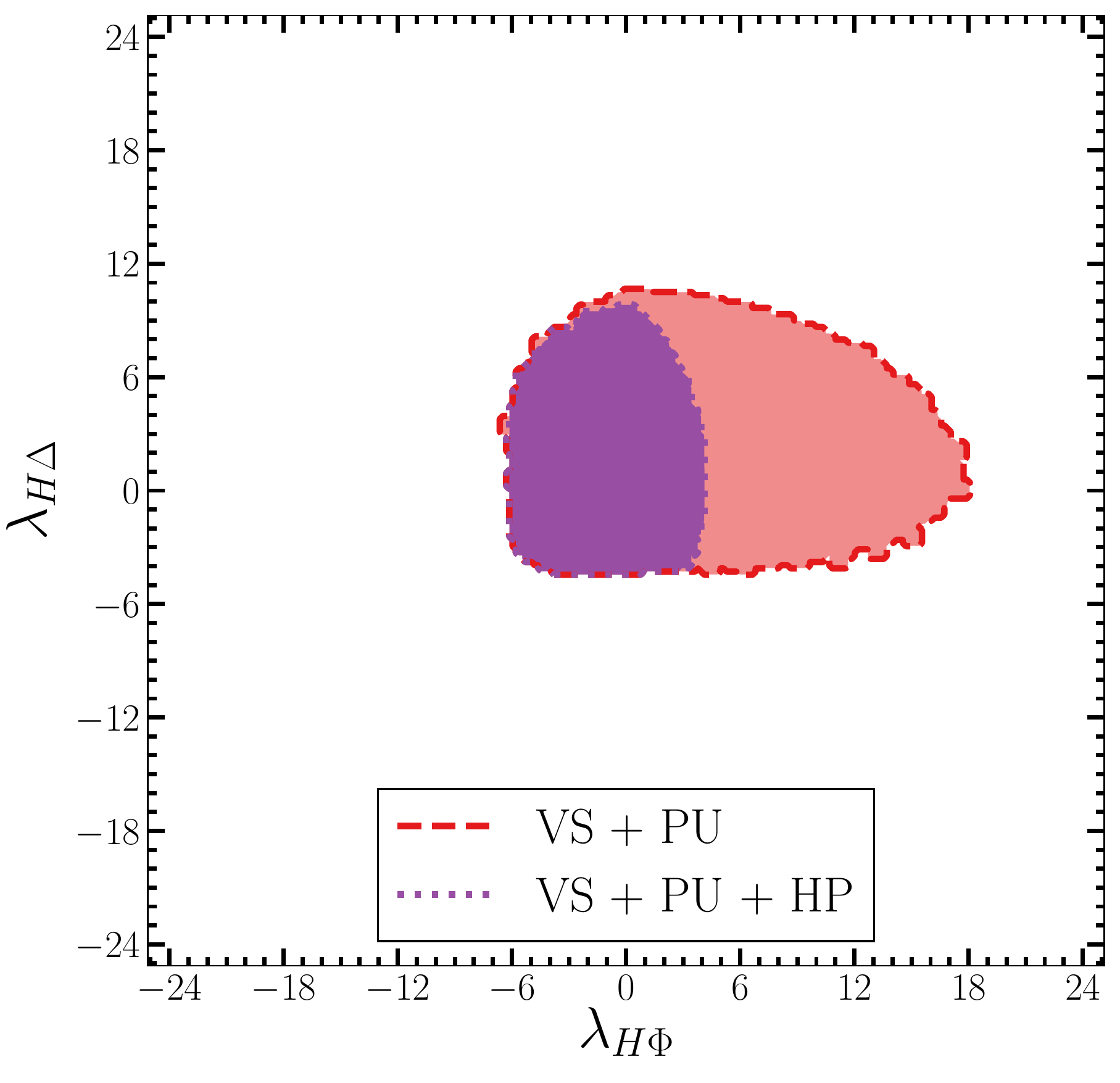}
   \end{minipage}
   \hfill
   \begin{minipage}[b]{0.44\textwidth}
	   \includegraphics[width=\textwidth]{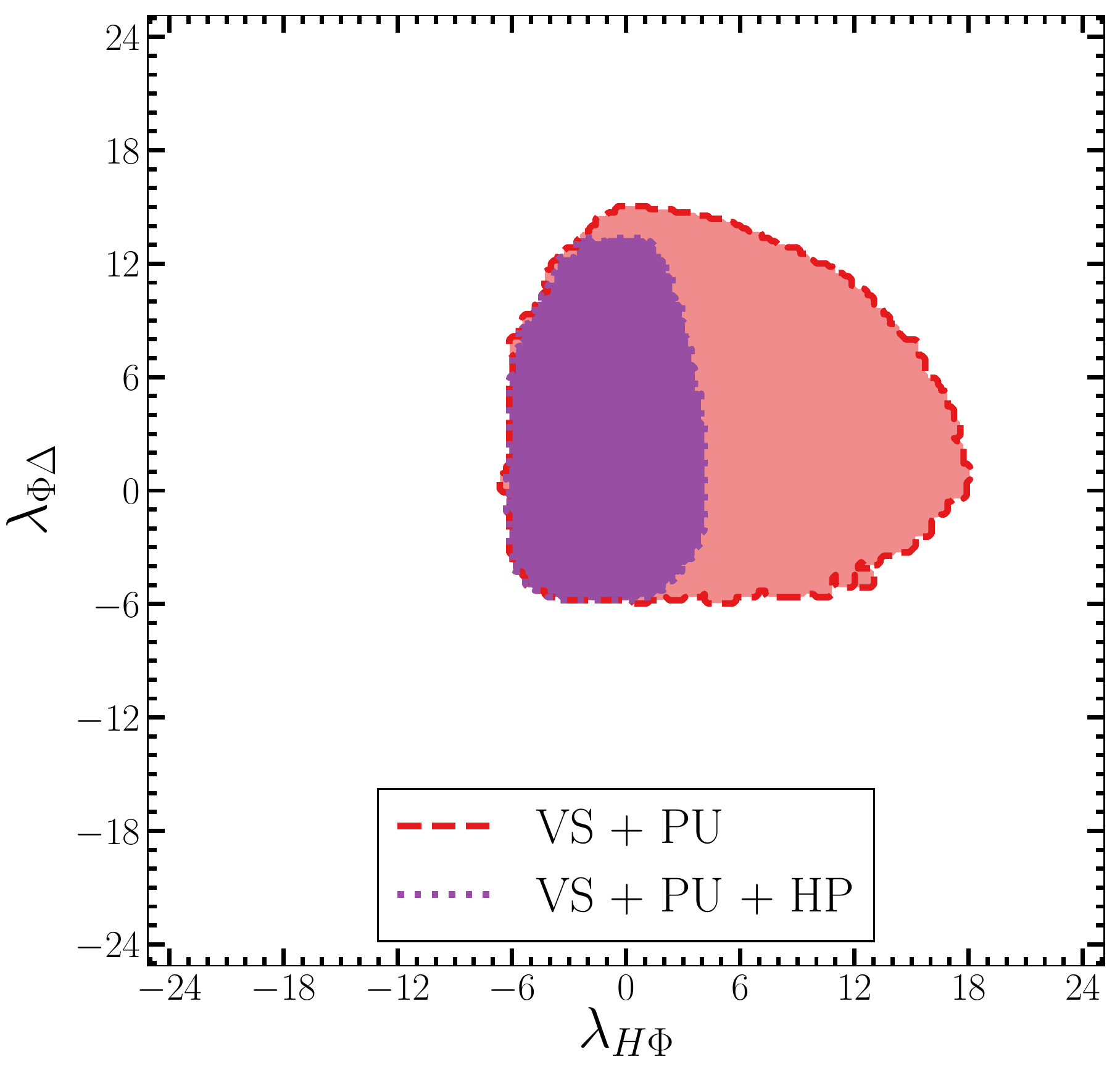}
   \end{minipage}

	\begin{minipage}[b]{0.44\textwidth}
	   \includegraphics[width=\textwidth]{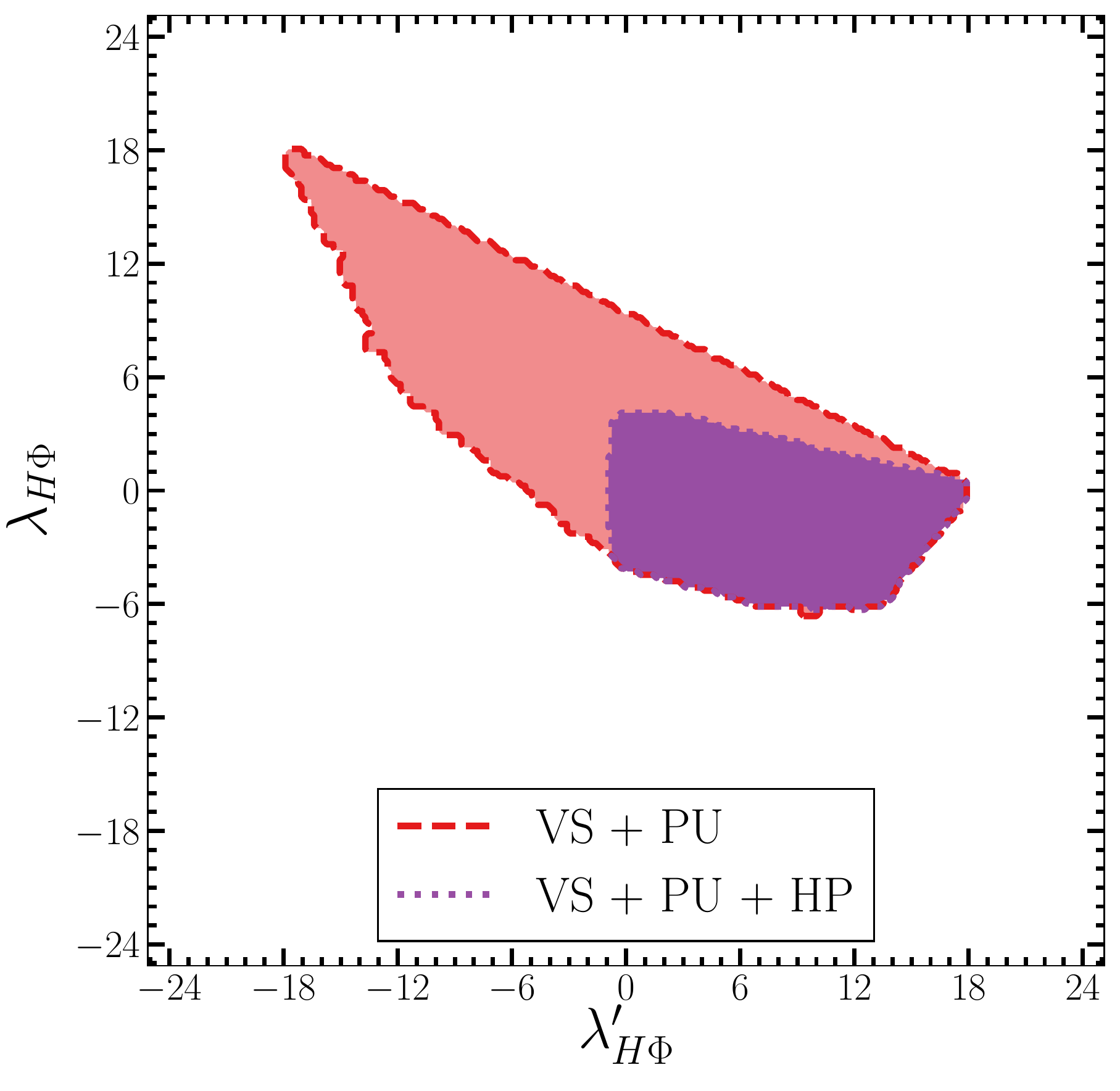}
   \end{minipage}
   \hfill
   \begin{minipage}[b]{0.44\textwidth}
	   \includegraphics[width=\textwidth]{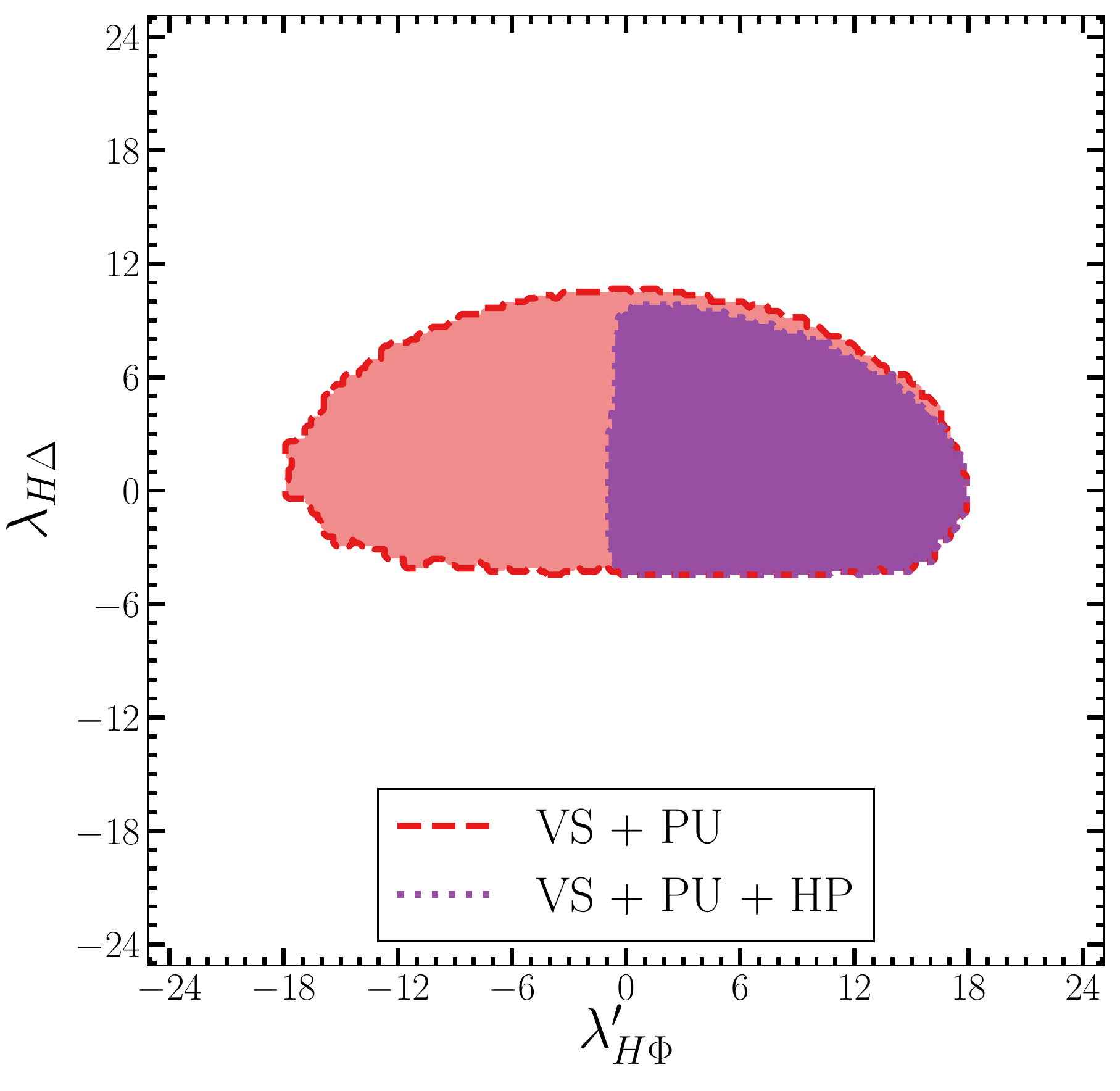}
   \end{minipage}

	\begin{minipage}[b]{0.44\textwidth}
	   \includegraphics[width=\textwidth]{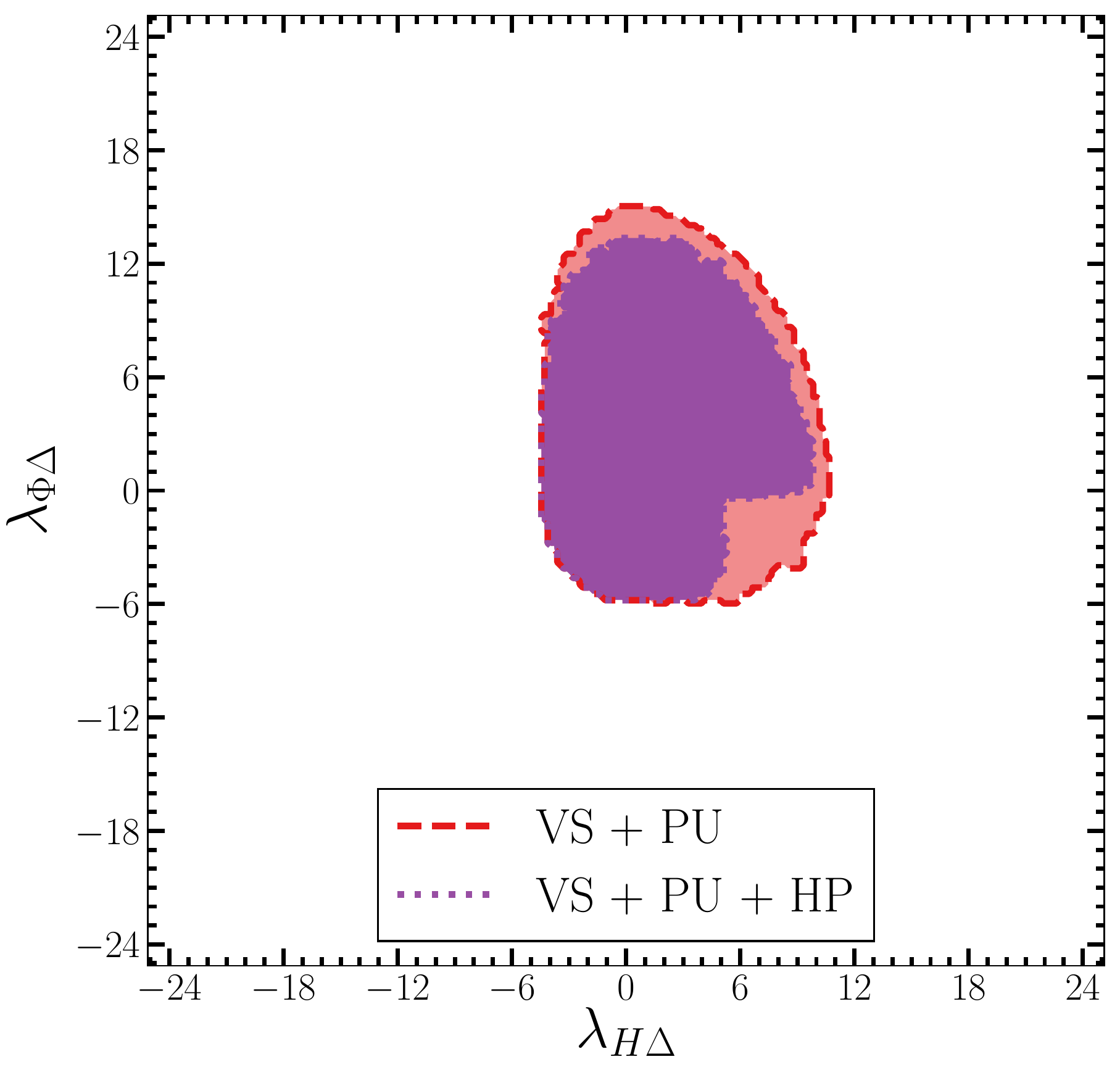}
   \end{minipage}
   \hfill
   \begin{minipage}[b]{0.44\textwidth}
	   \includegraphics[width=\textwidth]{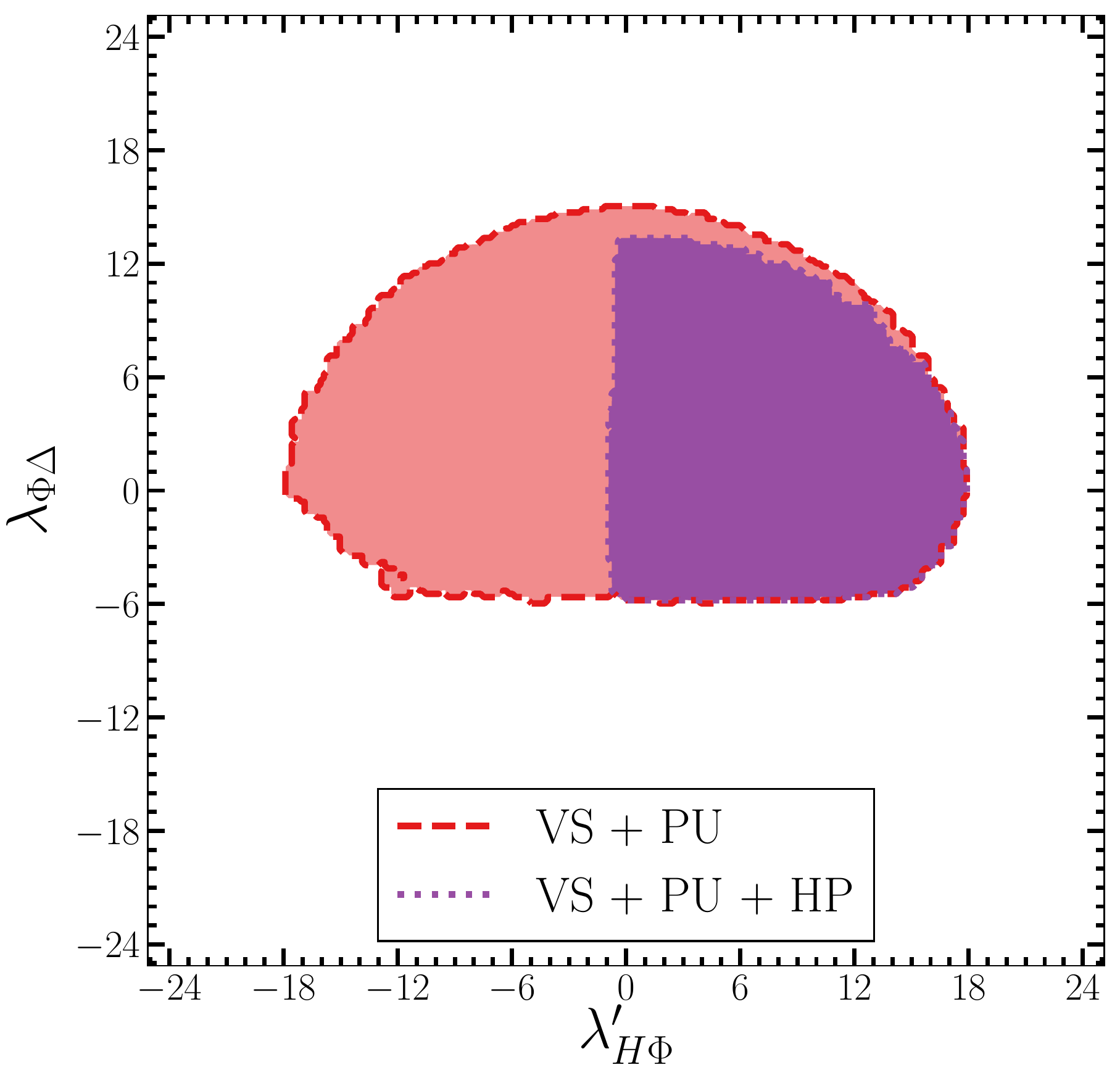}
   \end{minipage}
   \caption{
      \label{fig:non_diag}
	  Allowed regions of the parameter space by the (VS+PU) and (VS+PU+HP) constraints projected onto the
	  $(\lambda_{H\Phi},\lambda_{H\De})$,
      $(\lambda^\prime_{H\Phi},\lambda_{H\Phi})$, 
	  $(\lambda_{H\De},\lambda_{\Phi\De})$, 
	  $(\lambda_{H\Phi},\lambda_{\Phi\De})$,
      $(\lambda^\prime_{H\Phi},\lambda_{H\De})$ 
	  and $(\lambda^\prime_{H\Phi},\lambda_{\Phi\De})$ planes.}
\end{figure}

Next, we present the impacts of HP constraints on some of purely
 off-diagonal parameter planes  $(\lambda_{ij}, \lambda_{kl})$ 
in Fig.~\ref{fig:non_diag}.
Comparing the two parameters $\lambda_{H\Phi}$ 
and $\lambda_{H\Phi}^\prime$, their favored regions by the Higgs data behave 
quite differently as they are constrained in different ways.
The exclusion of the region $\lambda_{H\Phi}\gtrsim 4$ 
is owing to the SM Higgs mass, whereas $\lambda_{H\Phi}^\prime$ is forced to be greater 
than $-1$, in light of  prohibition of tachyonic modes for any of the scalars.
Overall, $\lambda_{H\Phi}$ is more constrained 
by the Higgs physics data than $\lambda_{\Phi\Delta}$, $\lambda_{H\Delta}$ and $\lambda^\prime_{H\Phi}$.
Allowed regions on the other planes of off-diagonal versus off-diagonal parameters will be summarized in the next section. 

In addition to the above bounds on $\lambda$s, we find that the mass parameter $M_{\Phi\Delta}$ 
has a lower limit $\sim -0.56$ GeV
while the mass parameter $M_{H\Delta}$ and the VEV $v_\Delta$ remain unconstrained 
within their ranges of scanning. The  lower bound on $M_{\Phi\Delta}$ arises  
from the observation that the $B$ term defined in Eq.~\eqref{ABC} must be negative in order to attain
positive $M_D$ and $M_{\widetilde \Delta}$.
Note that since $v_\Delta$ and $v_\Phi$ are much larger than $v$, 
$B$ can then be approximated as 
$-2 \left( 4 v_\De m^2_{H^\pm} + M_{\Phi \Delta} \left( 4 v_\Delta^2 + v_\Phi^2 \right) \right)$. 
This means that $M_{\Phi\Delta}$ cannot be too negative. 
However  the lower bound, $M_{\Phi\Delta} \gtrsim -0.58 $ TeV, also depends
on the chosen range  of  $-1 \leq (M_{H\Delta}/\mbox{TeV}) \leq 1$ in the scans  
as it may help to cancel a negative $M_{\Phi\Delta}$.
If one enlarges the scanning range of $M_{H\Delta}$, the lower bound
becomes more negative.

Before leaving this section, we discuss the impact of the measured Higgs diphoton signal strength $\mu^{\gamma\gamma}_{ggH}$ 
from the LHC.
Note that a similar analysis was presented in the previous
work~\cite{Huang:2015wts}.  We here improve our previous analysis by including
effects of the new heavy fermion loops on both  the Higgs production and the
decay into two photons.  In addition, we have also checked the Higgs to
    $\tau^+\tau^-$ decay channel which has a larger uncertainty~(a factor of 2 or so) than the diphoton channel~\cite{Sirunyan:2017khh}, as shown in Fig.
    \ref{fig:tautau}. Note that $\mu^{\tau\tau}_{ggH}$ is equal to
    $\Gamma(h_1\to gg)/\Gamma^{\text{SM}}(h\to gg)$, since the
    $h_1\to\tau^+\tau^-$ decay width differs from the SM one by $O^2_{11}$
    cancelling the total decay width ratio
    $\Gamma_h^\text{SM}/\Gamma_{h_1}\approx O^{-2}_{11}$. One can expect
	a similar behavior in other decay channels such as 
    $b\bar{b}$, $W^+W^-$ and $ZZ$, which also have bigger uncertainties. We would like to come
    back to this point for a more complete analysis in the future, when more data are collected
    at the LHC.

\begin{figure}
       \includegraphics[width=0.5\textwidth]{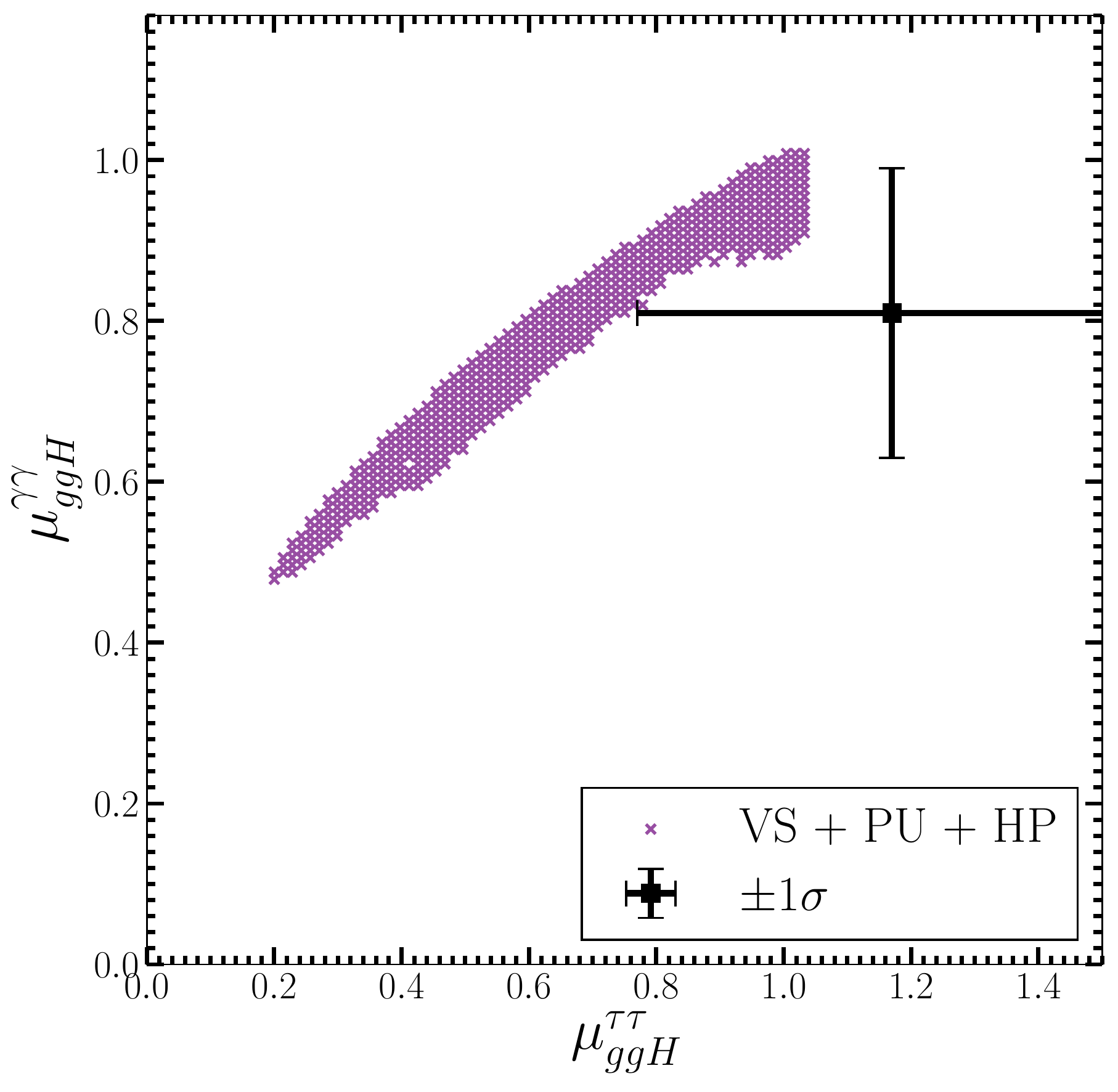}
       \caption{\label{fig:tautau} Comparison between signal strengths for
           the $\tau^+\tau^-$ and $\gamma\gamma$. The scatter plot represents
           the prediction from G2HDM and the error bar represents the
           $1\sigma$ region for the current measured signal
           strengths.
        }
\end{figure}

In Fig.~\ref{fig:scattermuggh}, we show the scan results of 
$\mu_{gg H}$ versus $O_{11}^2$ (top-left panel) and 
$ O_{21}/O_{11} $ (top-right panel), as well as the relative (bottom-left
panel) and absolute (bottom-right panel) contributions from new
physics to $\mu_{gg H}$. The relative contribution refers to the contribution from either the charged 
Higgs or heavy fermions to $\mu_{ggh}$ divided by the two contributions combined, 
while the absolute contribution denotes the individual amplitude in
Eq.~\eqref{eq:hgagaH2} without the common coefficient in front.
The Higgs constraints alone are labeled by orange dots, while the combined 
(VS+PU+HP) constraints are labeled by magenta crosses.
Our results indicate that our model has no problem having $\mu_{gg H}$  
around the experimental central value of $0.81$ with the $1 \sigma$ and $2 \sigma$ errors
marked by the dark and light gray regions respectively.
One can see that there are also points accumulating near the SM value
$\mu^{\gamma\gamma}_{ggH}=1$, which is barely inside the $1\sigma$ range.

\begin{figure}
    \begin{minipage}[b]{0.44\textwidth}
        \includegraphics[width=\textwidth]{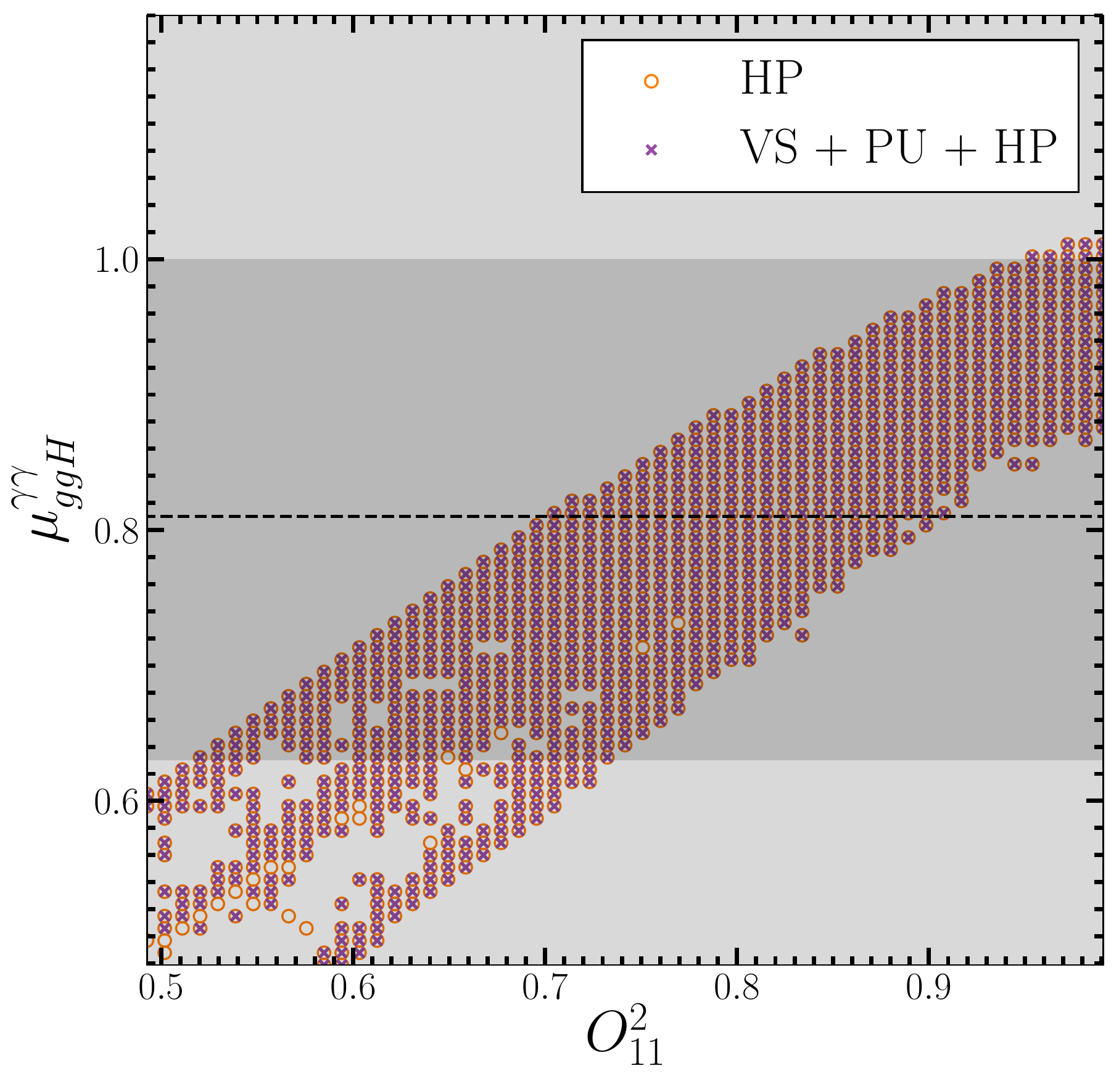}
    \end{minipage}
    \hfill
    \begin{minipage}[b]{0.44\textwidth}
         \includegraphics[width=\textwidth]{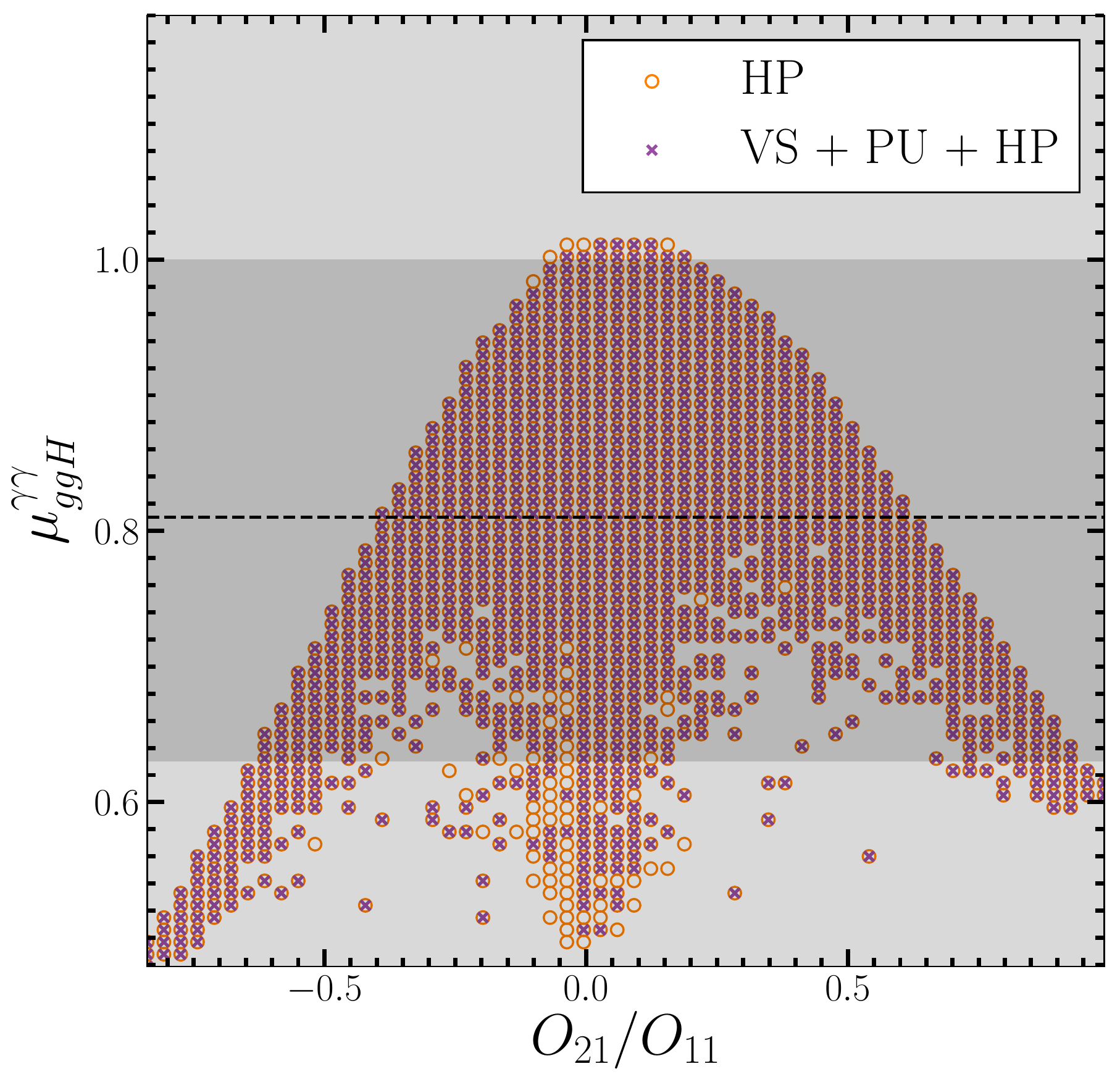}
    \end{minipage}

    \begin{minipage}[b]{0.44\textwidth}
        \includegraphics[width=\textwidth]{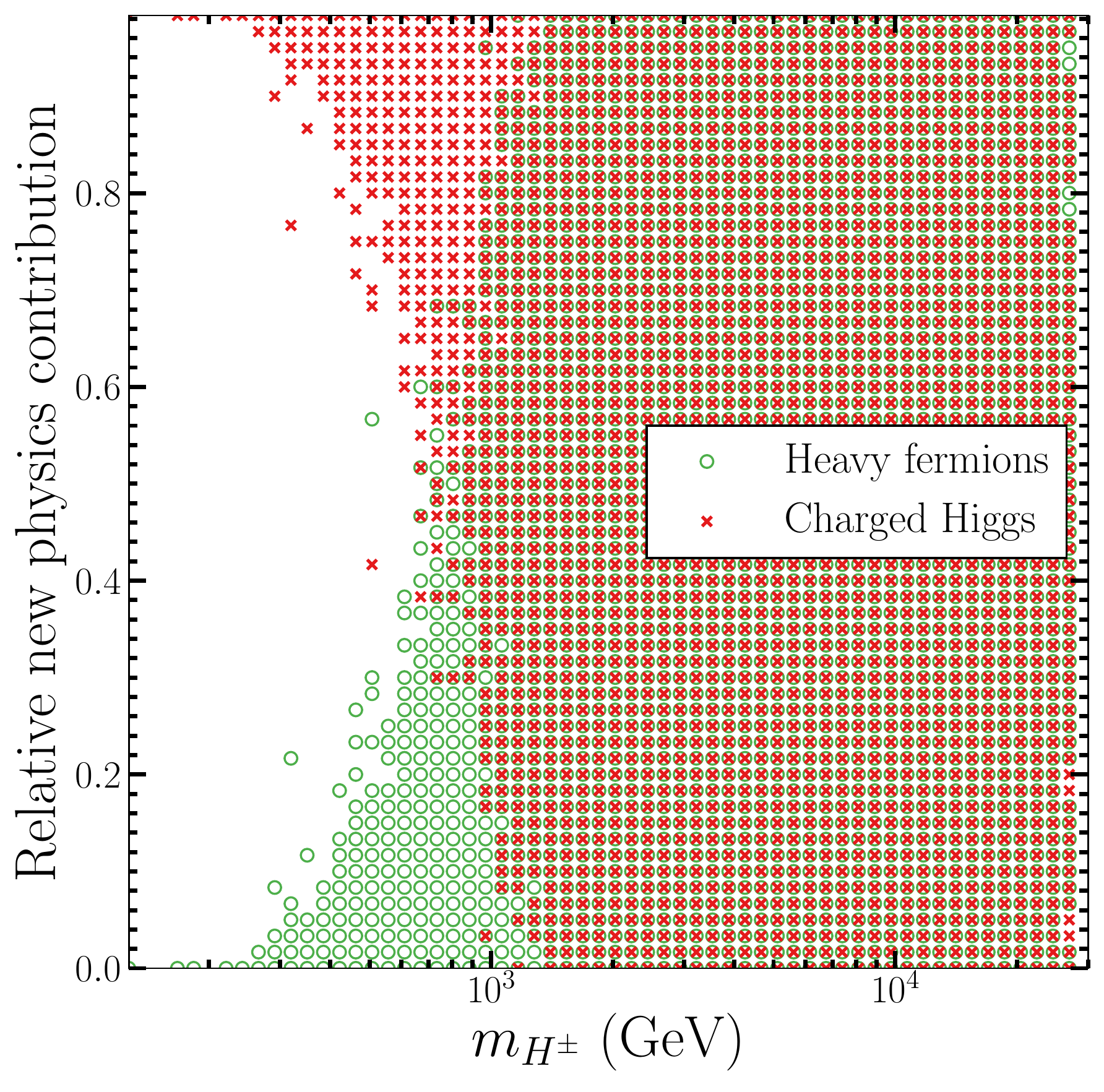}
    \end{minipage}
    \hfill
    \begin{minipage}[b]{0.44\textwidth}
        \includegraphics[width=\textwidth]{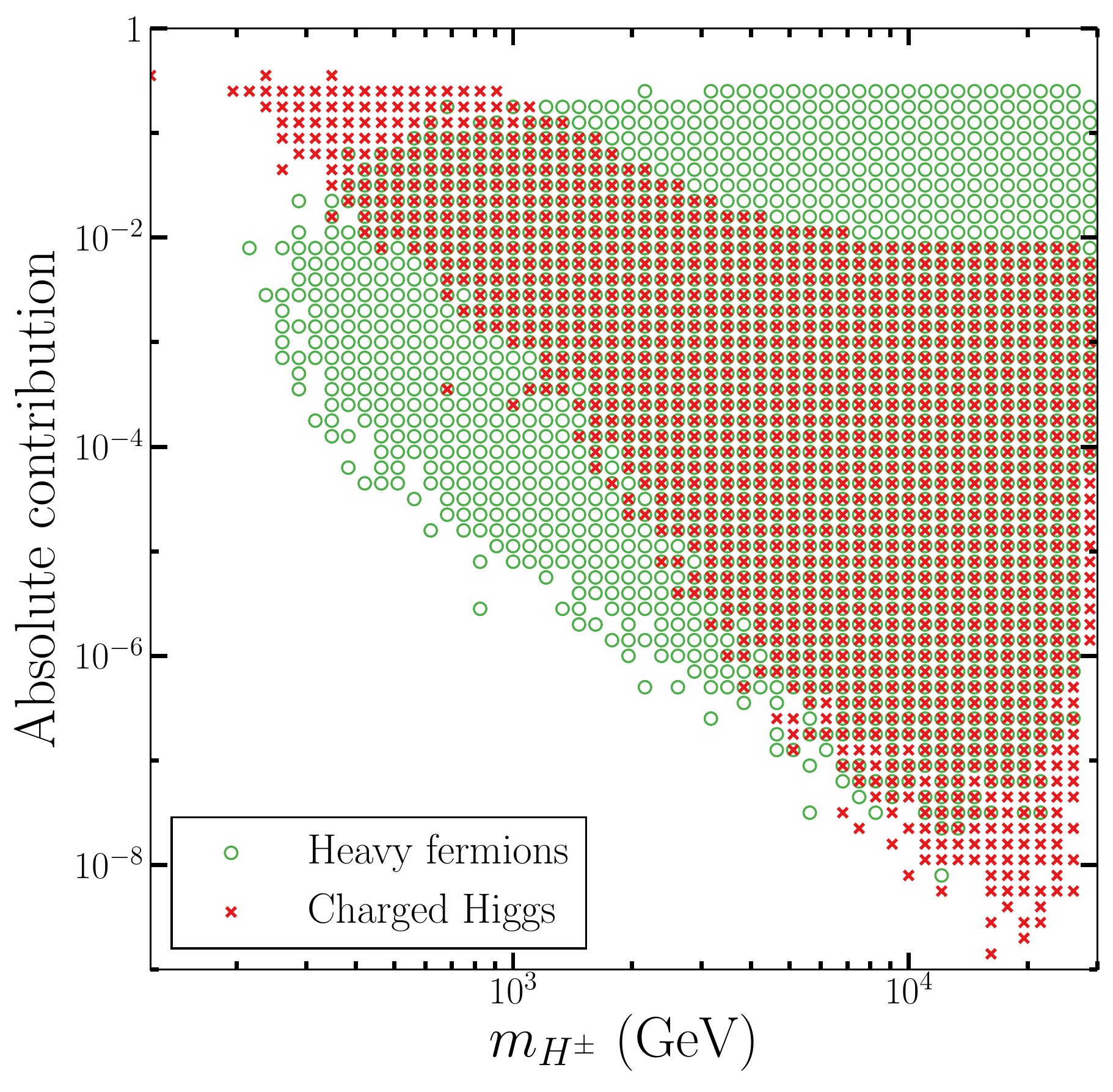}
    \end{minipage}

    \caption{
      \label{fig:scattermuggh}
      The correlations of the signal strength $\mu^{\gamma\gamma}_{ggH}$ with the squared
      mixing matrix element $O^2_{11}$ (upper-left) and the ratio
      $O_{21}/O_{11}$ (upper-right) is shown. The dashed line indicates the
      most recent ATLAS result for the central value of $\mu^{\gamma\gamma}_{ggH} =
      0.81$~\cite{Aaboud:2018xdt}.  The light and dark gray bands indicate the
  2$\sigma$ and 1$\sigma$ errors in $\mu^{\gamma\gamma}_{ggH}$  respectively. Below, the
  relative (lower-left) and absolute (lower-right) contribution to $\mu^{\gamma\gamma}_{ggH}$
  from new physics and its correlation with the mass of the charged Higgs.}
\end{figure}

In the top-left panel of Fig.~\ref{fig:scattermuggh}, for $O^2_{11}<0.75$ we can
see that the points allowed by all constraints (magenta cross) split into three
branches for small $O^2_{11}$ , depending on values of
$O_{21}/O_{11}$. The upper, middle and lower branches correspond to $O_{21}/O_{11}> 0.5$, 
$ -0.5 < O_{21}/O_{11} < 0.5$ and $O_{21}/O_{11}< -0.5$, respectively, 
as can be visually identified from the top-right panel of Fig.~\ref{fig:scattermuggh}.
While this ratio of $O_{21}/O_{11}$ appears several places in our $\mu^{\gamma\gamma}_{ggH}$ calculation,
this shape is mostly controlled by the heavy colored fermion contributions to the
partial decay width of $h_1$ decays into two gluons in Eq.~\eqref{eq:hggH2}. 
This two gluon partial width 
is multiplied by the diphoton partial width, which also
contains some less significant effects of this ratio through 
the heavy charged fermions as well as the charged Higgs contributions,
communicating the effects of this ratio to the whole signal strength
$\mu^{\gamma\gamma}_{ggH}$ in Eq.~\eqref{eq:muggH}.

The deviation from the SM predicted $\mu^{\gamma\gamma}_{ggH}=1$ has three sources:
the overall factor $O^2_{11}$ in Eqs.~\eqref{eq:hgagaH2} and \eqref{eq:hggH2}, 
and the extra loop contributions from the new heavy fermions and charged Higgs.
Figure~\ref{fig:scattermuggh} shows that the combined effects from these three factors
tend to decrease the signal strength: $\mu^{\gamma\gamma}_{ggH} \lsim 1$. 
In the end, any enhancement is not strong enough to
make $\mu^{\gamma\gamma}_{ggH} > 1$ as
it has to overcome the overall factor $O^2_{11}$.
An important reason for
$\mu^{\gamma\gamma}_{ggH} \lsim 1$ is that the new heavy fermion contributions
(in additional to the top quark loop) 
cancel the dominant contribution of the  SM $W^\pm$ loop for $h_1 \to \ga \ga$.

Furthermore, it might seem strange to find that values of
$\mu^{\gamma\gamma}_{ggH}$ do not fall sharply to the SM value of 1
as $O^2_{11}$ grows.
However, upon closer inspection of Eq.~\eqref{eq:hgagaH2}, we notice that 
the charged Higgs, whether it is heavy or light, will still contribute to the diphoton channel 
even if $O_{11} = 1$. This is because $O_{21}$ and  $O_{31}$ will be vanishing and $\mathcal C_h$ in 
Eq.~\eqref{eq:Chfactor}  becomes unity.
A relatively light charged Higgs mass can indeed move $\mu^{\gamma\gamma}_{ggH}$ away from the SM value
even if the mixing is small, {\it i.e.} $O^2_{11}\approx 1$.
When all the constraints are taken into account, the contribution
from the charged Higgs dominates over heavy fermion ones completely for
$m_{H^\pm} \lesssim 400$ GeV, but they become comparable
around $m_{H^\pm} \gtrsim 10^{3}$ GeV, as demonstrated by the
lower-right panel of Fig.~\ref{fig:scattermuggh}.  It is important to note
that while the lower panels of Fig.~\ref{fig:scattermuggh} apparently show an
upper bound for the mass of the charged Higgs, this is actually the maximum
value reached with our choice of scan range of $v_\Delta$ and has no
theoretical implication on the upper limit of the charged Higgs mass.

To further illustrate the role of new physics contributions from 
the charged Higgs and heavy fermions in G2HDM,
we compare the new contributions with the SM contributions 
in the loops of $\Gamma(h_1\to\gamma\gamma)$ 
in Fig.~\ref{fig:scatterloops}.
At the top-left panel the ratio of
charged Higgs loop to SM loops is shown, while at the top-right panel we
have the ratio of total contributions from new physics loops
(charged Higgs plus new heavy fermions) to SM loops. At the two bottom
panels we show the ratio of new heavy fermions to SM loops projected on
the common heavy fermion mass (left) and charged Higgs mass (right).

\begin{figure}
    \begin{minipage}[b]{0.44\textwidth}
        \includegraphics[width=\textwidth]{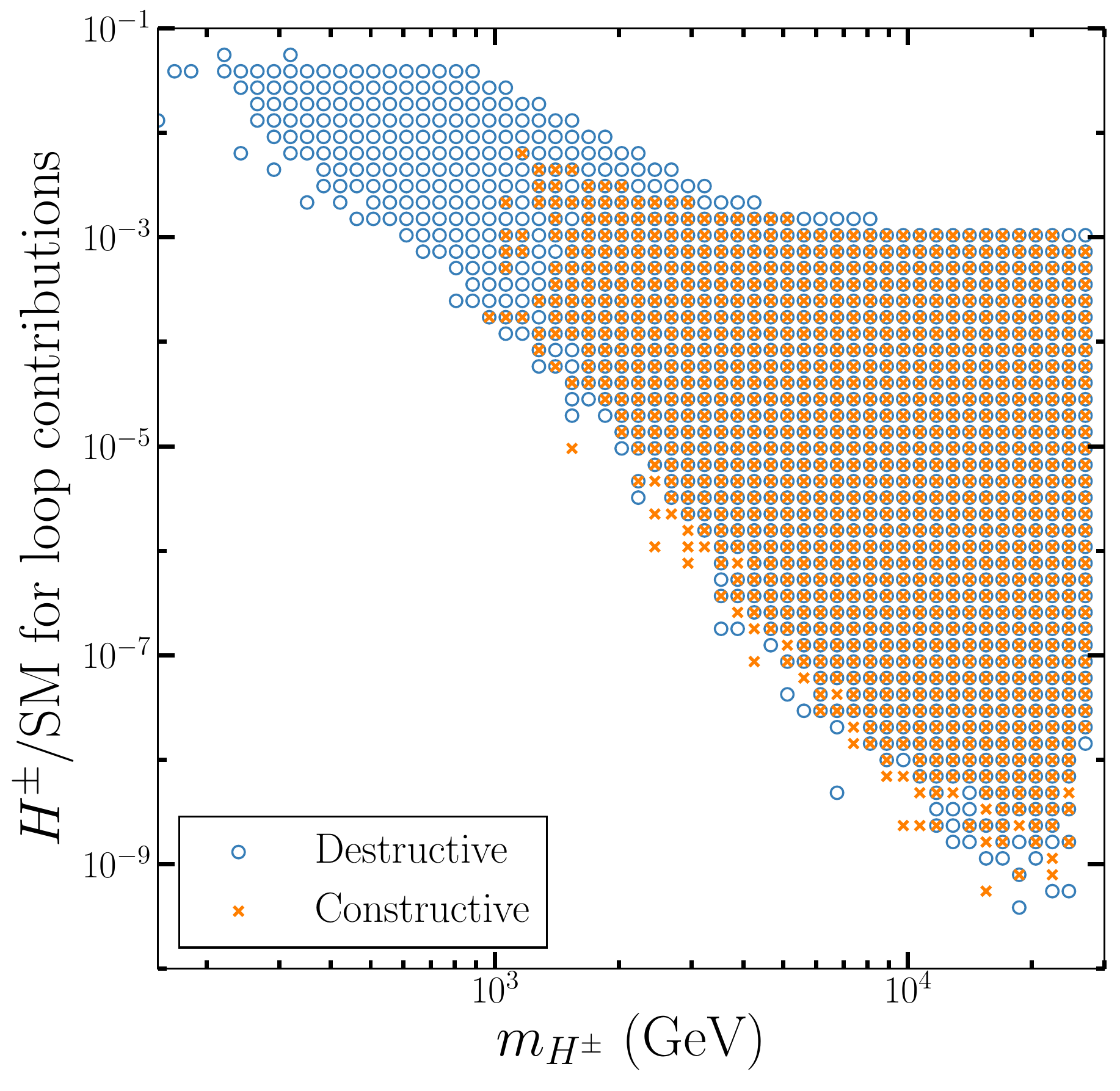}
    \end{minipage}
    \hfill
    \begin{minipage}[b]{0.44\textwidth}
        \includegraphics[width=\textwidth]{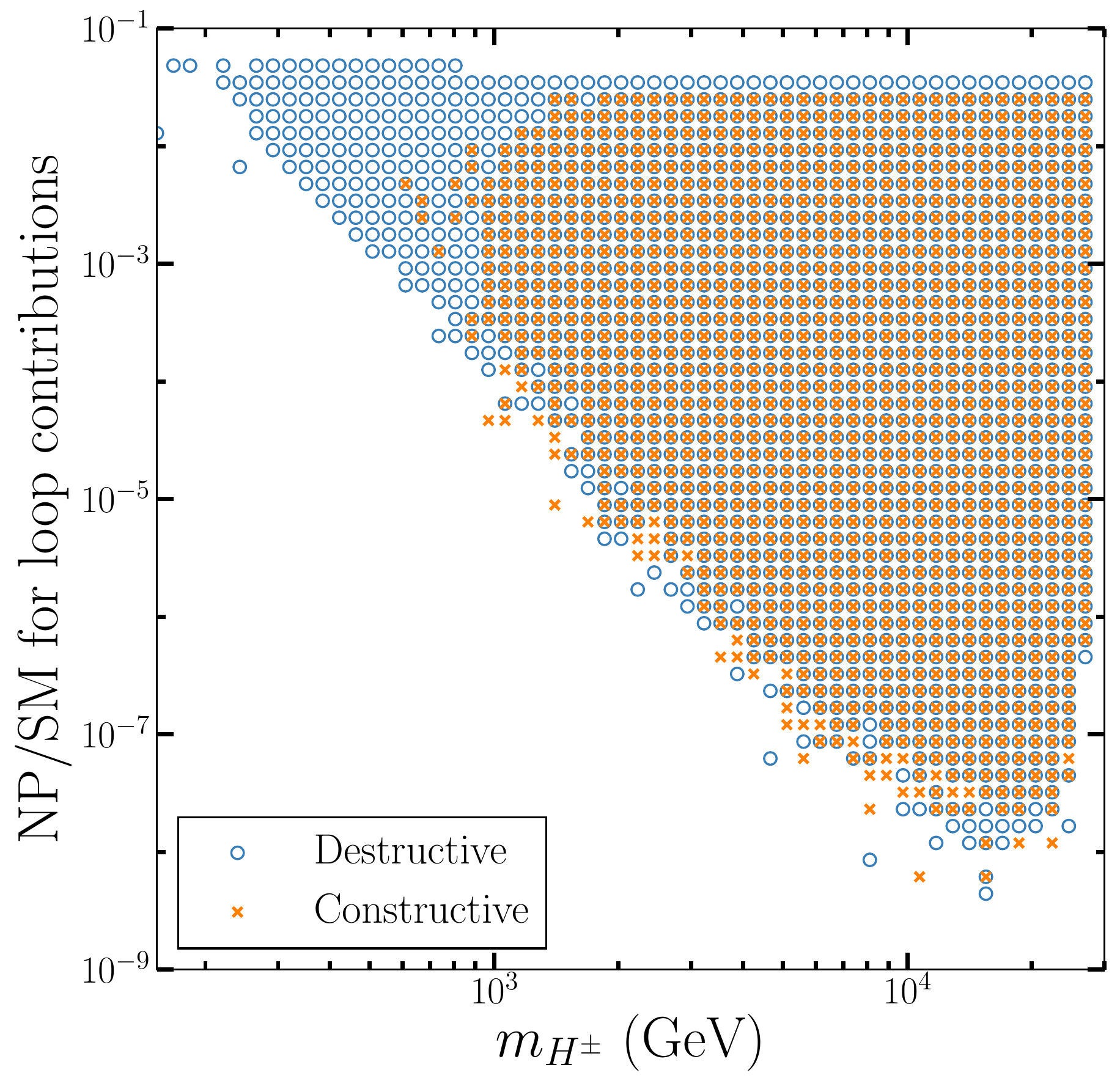}
    \end{minipage}

    \begin{minipage}[b]{0.44\textwidth}
        \includegraphics[width=\textwidth]{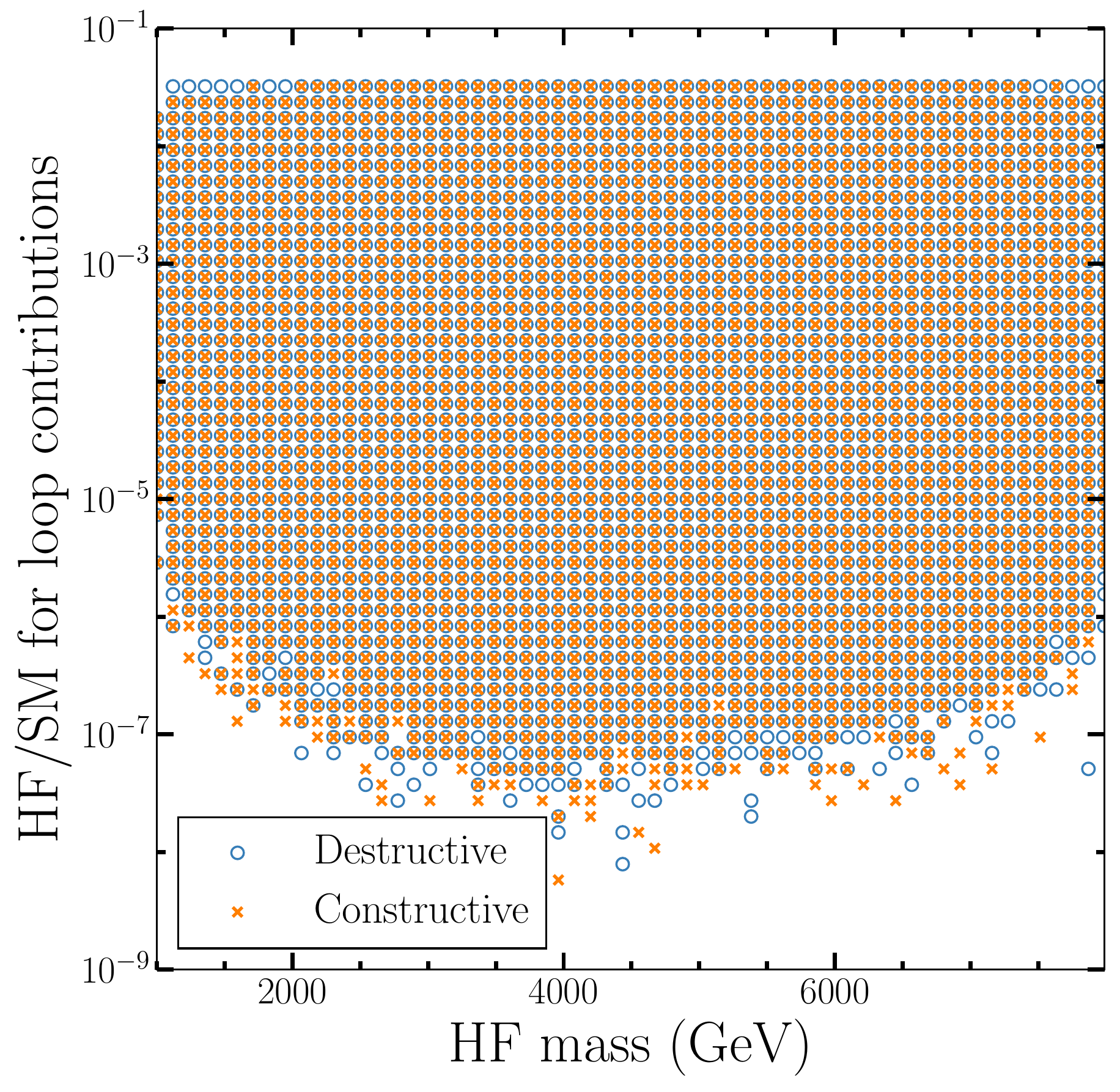}
    \end{minipage}
    \hfill
    \begin{minipage}[b]{0.44\textwidth}
        \includegraphics[width=\textwidth]{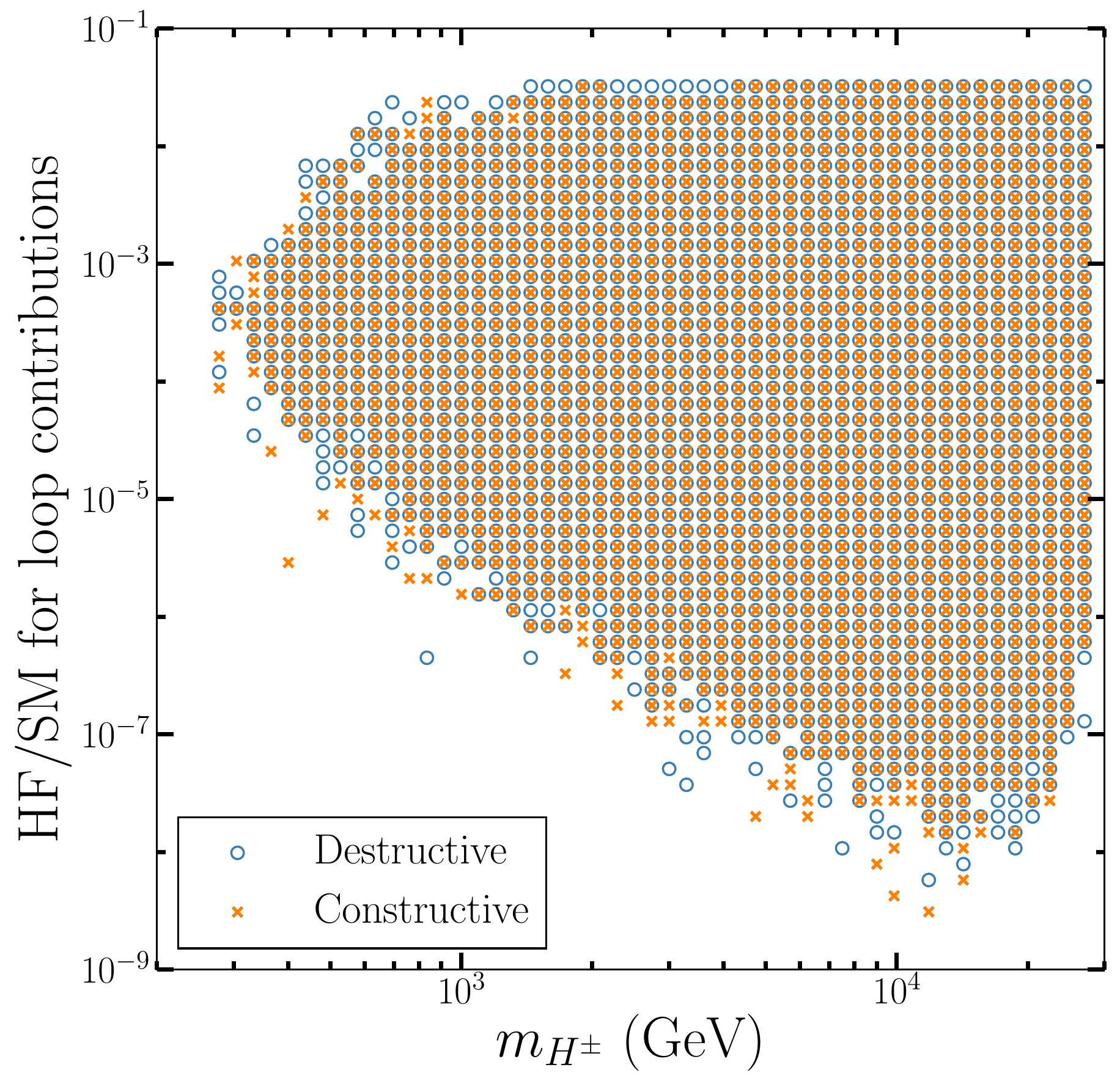}
    \end{minipage}

    \caption{
        \label{fig:scatterloops}
        Comparisons between different contributions to the loops in 
        $h_1\to\gamma\gamma$ decay.  
        HF is the abbreviation for all heavy fermions 
        while NP is for both charged Higgs 
        and all heavy fermions in G2HDM.
      }
\end{figure}

\begin{figure}
	   \includegraphics[width=0.5\textwidth]{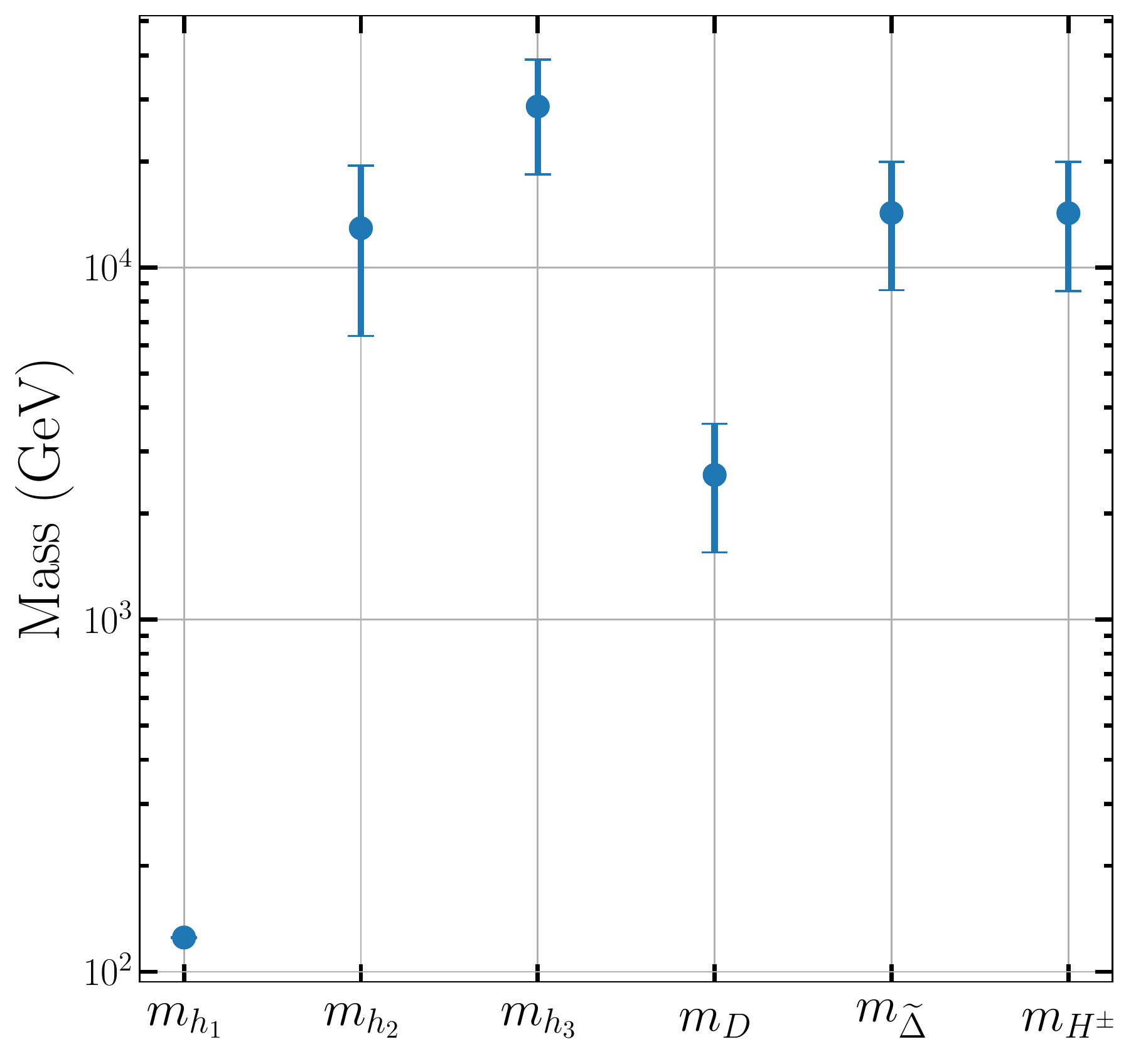}
   \caption{
      \label{fig:spectrumbars}
      Typical scalar masses in the final allowed parameter space due to the (VS+PU+HP) constraints. 
      The dot indicates the average mass and the error bar corresponds to the $\pm
      1\sigma$ deviation.
  }
\end{figure}

Lastly, with the final allowed parameter space determined, we compute
    the mean values and the standard deviations for all the scalar masses
    using the allowed points. We thus obtain distributions for the masses of
    all the scalars in G2HDM. The resulting mass spreads are shown in
Fig.~\ref{fig:spectrumbars}, with dots indicating their mean values and error
bars for the $\pm 1 \sigma$ deviation.  One can easily see that while the
DM candidate $D$ in G2HDM has a mass around $1 - 3$ TeV, all other
scalars are much heavier ($\gtrsim 10$ TeV).  Note that, while we have
constrained all the quartic couplings in the scalar potential, the cubic
couplings $M_{H\Delta}$ and $M_{\Phi\Delta}$ remain untamed in a very wide
range.  By choosing suitable ranges for these parameters, e.g. by
tuning them to achieve $4AC \ll B^2$ in Eq.~\eqref{darkmattermass}, one can
give $D$ a much lighter mass.

\section{Summary} \label{section:conclusions}

\begin{figure}
	   \includegraphics[width=\textwidth]{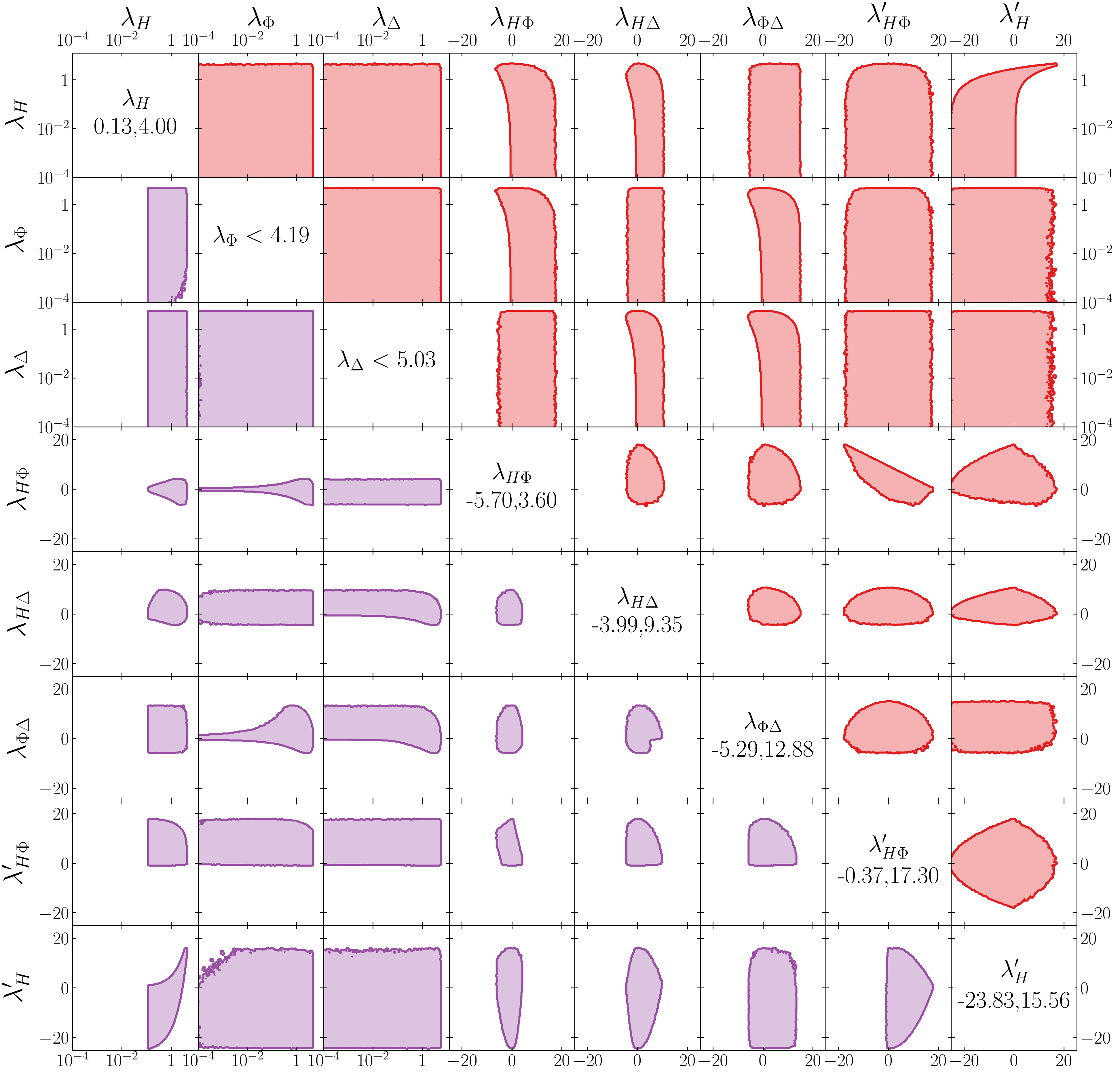}
   \caption{
      \label{fig:matrixplot}
	  A summary of the parameter space allowed by the theoretical and
	  phenomenological constraints. The red regions show the results from
	  the theoretical constraints (VS+PU) of Sec.~\ref{section:th_constraints}. The
	  magenta regions are constrained by Higgs physics 
	  as well as the theoretical constraints (HP+VS+PU), as discussed in
	  Sec.~\ref{section:HiggsPheno}. 
  }
\end{figure}

G2HDM was proposed to promote the discrete $Z_2$ symmetry, that ensures the stability of DM 
in IHDM, to a local gauge symmetry by embedding the two Higgs doublets 
into the fundamental representation of the new $SU(2)_H$ gauge group.
In this work we have studied the theoretical constraints from vacuum stability  and perturbative unitarity 
imposed on the scalar potential of G2HDM at tree level. The 125 GeV Higgs physics data
from the LHC are also taken into account, in particular the Higgs mass and its diphoton signal strength.
We have improved the G2HDM scalar potential by including 
two new couplings $\lambda_H^\prime$ and $\lambda_{H\Phi}^\prime$, which were missing 
in our previous work. Although these two new couplings
do not alter the minimization conditions of the scalar potential, their impacts on the scalar 
mass spectrum and effects on the VS, PU and HP constraints are analyzed in detail.
We have recomputed the diphoton signal strength for the 125 GeV Higgs. 
We have included the contributions of the new heavy fermions in G2HDM
and demonstrated that their effects can be significant in the diphoton channel.
The charged Higgs contribution is also found to be significant if its mass is 
in the range of 100 to 300 GeV. Overall the diphoton signal strength is found to be
$\lsim 1$ in G2HDM. In particular  G2HDM can naturally accommodate 
the current ATLAS central value of 0.81 for the diphoton signal strength from gluon-gluon fusion production.
We note that the corresponding central value from CMS is 1.10~\cite{Sirunyan:2018ouh}
which is not favorable in G2HDM with the present numerical setup in this work.
However we are quoting the LHC Run II data from ATLAS and CMS 
in our analysis, while the combined results from both experiments 
are not available yet.

In Fig.~\ref{fig:matrixplot}, we summarize the allowed regions of the
multidimensional parameter 
space projected on all the two-dimensional planes comprised of the  four diagonal couplings
$\lambda_{H, \Phi, \De}$, $\lambda_{H}^\prime$ and the four off-diagonal couplings
$\lambda_{H\Phi,H\De,\Phi\De}$, $\lambda^\prime_{H\Phi}$. The upper red triangular block
corresponds to (VS+PU) constraints, while the lower magenta triangular block corresponds to the
(VS+PU+HP) constraints. The diagonal panels indicate the allowed ranges of the eight couplings. 
One can see that among the eight couplings only two of them $\lambda_H$ and $\lambda_{H\Phi}$ are 
significantly constrained by (VS+PU+HP). 
Some of the couplings like $\lambda_{H}^\prime$, $\lambda_{H\Phi}^\prime$ and 
$\lambda_{\Phi\De}$ are loosely constrained as shown clearly in Fig.~\ref{fig:matrixplot}.

We emphasize that our analysis in this work is based upon tree-level VS and PU.
One loop corrections to the scalar potential in G2HDM 
is important as it has been demonstrated in~\cite{Staub:2017ktc,Krauss:2017xpj}
that such loop corrections may lead to larger parameter space for perturbativity constraints in BSM.
Perturbative unitarity constraints can also be improved by including the cubic scalar couplings
$M_{H\De}$ and $M_{\Phi\De}$.
Recently, cubic scalar couplings are found to be significantly improved by the unitarity constraints in various models 
beyond the SM if the center-of-mass energy is not taken asymptotically infinite~\cite{Goodsell:2018tti,Goodsell:2018fex,Krauss:2018orw}.
Further improvement of our current analysis can be carried out by including renormalization group 
running effects for all the quartic couplings as well as the two cubic couplings.
These issues are interesting but beyond the scope of this work.
We would like to come back to these issues in the future.

G2HDM is an interesting variant of the popular IHDM with a neutral component in the second Higgs doublet 
as a DM candidate. It would be interesting to include DM constraints from cosmology, 
direct and indirect detections as well as collider searches in our analysis. This work 
is in progress~\cite{G2HDM-DMConstraints} and we will report it elsewhere.

\newpage 

\section*{Acknowledgments}
The analysis presented here was done using the resources of the high-performance 
T3 Cluster at the Institute of Physics at Academia Sinica.
This work was supported in part by the Ministry of Science and Technology (MoST) of Taiwan under
Grant No. 104-2112-M-001-001-MY3 (TCY),
the Moroccan Ministry of Higher Education and Scientific Research MESRSFC and CNRST:
``Projet dans les domaines prioritaires de la recherche scientifique et du d'eveloppement technologique": 
PPR/2015/6 (AA),
and the Independent Research Fund Denmark, grant number 
DFF 6108-00623 (WCH). 
The CP3-Origins centre is partially funded by the Danish National Research Foundation, grant number DNRF90.


\begin{thebibliography}{99}

\bibitem{Branco:2011iw}
  G.~C.~Branco, P.~M.~Ferreira, L.~Lavoura, M.~N.~Rebelo, M.~Sher, and J.~P.~Silva,
  ``Theory and phenomenology of two-Higgs-doublet models,''
  Phys.\ Rep.\  {\bf 516}, 1 (2012)
  [arXiv:1106.0034 [hep-ph]].



\bibitem{Deshpande:1977rw} 
  N.~G.~Deshpande and E.~Ma,
  ``Pattern of symmetry breaking with two Higgs doublets,''
  Phys.\ Rev.\ D {\bf 18}, 2574 (1978).



\bibitem{Ma:2006km} 
  E.~Ma,
  ``Verifiable radiative seesaw mechanism of neutrino mass and dark matter,''
  Phys.\ Rev.\ D {\bf 73}, 077301 (2006)
  [hep-ph/0601225].



\bibitem{Barbieri:2006dq} 
  R.~Barbieri, L.~J.~Hall, and V.~S.~Rychkov,
  ``Improved naturalness with a heavy Higgs: An alternative road to LHC physics,''
  Phys.\ Rev.\ D {\bf 74}, 015007 (2006)
  [hep-ph/0603188].



\bibitem{LopezHonorez:2006gr} 
  L.~Lopez Honorez, E.~Nezri, J.~F.~Oliver, and M.~H.~G.~Tytgat,
  ``The inert doublet model: An archetype for dark matter,''
  JCAP {\bf 02}, 028 (2007)
  [hep-ph/0612275].



\bibitem{Arhrib:2013ela} 
  A.~Arhrib, Y.~L.~S.~Tsai, Q.~Yuan, and T.~C.~Yuan,
  ``An updated analysis of inert Higgs doublet model in light of the recent results from LUX, PLANCK, AMS-02 and LHC,''
  JCAP {\bf 06} (2014) 030
  [arXiv:1310.0358 [hep-ph]].



\bibitem{Ilnicka:2015jba} 
  A.~Ilnicka, M.~Krawczyk, and T.~Robens,
  ``Inert doublet model in light of LHC Run I and astrophysical data,''
  Phys.\ Rev.\ D {\bf 93}, no. 5, 055026 (2016)
  [arXiv:1508.01671 [hep-ph]].



\bibitem{Belyaev:2016lok} 
  A.~Belyaev, G.~Cacciapaglia, I.~P.~Ivanov, F.~Rojas-Abatte, and M.~Thomas,
  ``Anatomy of the inert two Higgs doublet model in the light of the LHC and non-LHC dark matter searches,''
  Phys.\ Rev.\ D {\bf 97}, no. 3, 035011 (2018)
  [arXiv:1612.00511 [hep-ph]].



\bibitem{Eiteneuer:2017hoh} 
  B.~Eiteneuer, A.~Goudelis, and J.~Heisig,
  ``The inert doublet model in the light of Fermi-LAT gamma-ray data: a global fit analysis,''
  Eur.\ Phys.\ J.\ C {\bf 77}, no. 9, 624 (2017)
  [arXiv:1705.01458 [hep-ph]].



\bibitem{Kephart:2015oaa} 
  T.~W.~Kephart and T.~C.~Yuan,
  ``Origins of inert Higgs doublets,''
  Nucl.\ Phys.\ {\bf B906}, 549 (2016)
  [arXiv:1508.00673 [hep-ph]].



\bibitem{Krauss:1988zc} 
  L.~M.~Krauss and F.~Wilczek,
  ``Discrete Gauge Symmetry in Continuum Theories,''
  Phys.\ Rev.\ Lett.\  {\bf 62}, 1221 (1989).



\bibitem{Kallosh:1995hi} 
  R.~Kallosh, A.~D.~Linde, D.~A.~Linde, and L.~Susskind,
  ``Gravity and global symmetries,''
  Phys.\ Rev.\ D {\bf 52}, 912 (1995)
  [hep-th/9502069].



\bibitem{Huang:2015wts} 
  W.~C.~Huang, Y.~L.~S.~Tsai, and T.~C.~Yuan,
  ``G2HDM : Gauged two Higgs doublet model,''
  JHEP {\bf 04} (2016) 019
  [arXiv:1512.00229 [hep-ph]].



\bibitem{Huang:2015rkj} 
  W.~C.~Huang, Y.~L.~S.~Tsai, and T.~C.~Yuan,
  ``Gauged two Higgs doublet model confronts the LHC 750 GeV diphoton anomaly,''
  Nucl.\ Phys.\ {\bf B909}, 122 (2016)
  [arXiv:1512.07268 [hep-ph]].



\bibitem{Huang:2017bto} 
  W.~C.~Huang, H.~Ishida, C.~T.~Lu, Y.~L.~S.~Tsai, and T.~C.~Yuan,
  ``Signals of new gauge bosons in gauged two Higgs doublet model,''
  Eur.\ Phys.\ J.\ C {\bf 78}, 613 (2018)
  [arXiv:1708.02355 [hep-ph]].


\bibitem{Kanemura:1999xf} 
  S.~Kanemura, T.~Kasai, and Y.~Okada,
  ``Mass bounds of the lightest $CP$ even Higgs boson in the two Higgs doublet model,''
  Phys.\ Lett.\ B {\bf 471}, 182 (1999)
  [hep-ph/9903289].



\bibitem{Arhrib:2000is} 
  A.~Arhrib,
  ``Unitarity constraints on scalar parameters of the standard and two Higgs doublets model,''
  [hep-ph/0012353].



\bibitem{Akeroyd:2000wc} 
  A.~G.~Akeroyd, A.~Arhrib, and E.~M.~Naimi,
  ``Note on tree level unitarity in the general two Higgs doublet model,''
  Phys.\ Lett.\ B {\bf 490}, 119 (2000)
  [hep-ph/0006035].



\bibitem{Arhrib:2012ia} 
  A.~Arhrib, R.~Benbrik, and N.~Gaur,
  ``$H\to \gamma \gamma$ in inert Higgs doublet model,''
  Phys.\ Rev.\ D {\bf 85}, 095021 (2012)
  [arXiv:1201.2644 [hep-ph]].



\bibitem{Hung:2006ap} 
  P.~Q.~Hung,
  ``A Model of electroweak-scale right-handed neutrino mass,''
  Phys.\ Lett.\ B {\bf 649}, 275 (2007)
  [hep-ph/0612004].


\bibitem{Ko:2012hd} 
  P.~Ko, Y.~Omura, and C.~Yu,
  ``A resolution of the flavor problem of two Higgs doublet models with an
  extra $U(1)_H$ symmetry for Higgs flavor,''
  Phys.\ Lett.\ B {\bf 717}, 202 (2012)
  [arXiv:1204.4588 [hep-ph]].
  
  

\bibitem{Campos:2017dgc} 
  M.~D.~Campos, D.~Cogollo, M.~Lindner, T.~Melo, F.~S.~Queiroz, and W.~Rodejohann,
  ``Neutrino masses and absence of flavor changing interactions in the 2HDM from gauge principles,''
  JHEP {\bf 08} (2017) 092
  [arXiv:1705.05388 [hep-ph]].
  
  
  
\bibitem{Kors:2004dx} 
  B.~Kors and P.~Nath,
  ``A Stueckelberg extension of the standard model,''
  Phys.\ Lett.\ B {\bf 586}, 366 (2004)
  [hep-ph/0402047].



\bibitem{Kors:2005uz} 
  B.~Kors and P.~Nath,
  ``Aspects of the Stueckelberg extension,''
  JHEP {\bf 07} (2005) 069
  [hep-ph/0503208].



\bibitem{Feldman:2007wj} 
  D.~Feldman, Z.~Liu, and P.~Nath,
  ``The Stueckelberg Z-prime extension with kinetic mixing and milli-charged dark matter from the hidden sector,''
  Phys.\ Rev.\ D {\bf 75}, 115001 (2007)
  [hep-ph/0702123 [HEP-PH]].



\bibitem{Cheung:2007ut} 
  K.~Cheung and T.~C.~Yuan,
  ``Hidden fermion as milli-charged dark matter in Stueckelberg Z- prime model,''
  JHEP {\bf 03} (2007) 120
  [hep-ph/0701107].



\bibitem{ElKaffas:2006gdt} 
  A.~W.~El Kaffas, W.~Khater, O.~M.~Ogreid, and P.~Osland,
  ``Consistency of the two Higgs doublet model and CP violation in top production at the LHC,''
  Nucl.\ Phys.\ {\bf B775}, 45 (2007)
  [hep-ph/0605142].



\bibitem{Arhrib:2011uy} 
  A.~Arhrib, R.~Benbrik, M.~Chabab, G.~Moultaka, M.~C.~Peyranere, L.~Rahili, and J.~Ramadan,
  ``The Higgs potential in the type II seesaw model,''
  Phys.\ Rev.\ D {\bf 84}, 095005 (2011)
  [arXiv:1105.1925 [hep-ph]].



\bibitem{Kannike:2012pe} 
  K.~Kannike,
  ``Vacuum stability conditions from copositivity criteria,''
  Eur.\ Phys.\ J.\ C {\bf 72}, 2093 (2012)
  [arXiv:1205.3781 [hep-ph]].



\bibitem{Kannike:2016fmd} 
  K.~Kannike,
  ``Vacuum stability of a general scalar potential of a few fields,''
  Eur.\ Phys.\ J.\ C {\bf 76}, no. 6, 324 (2016)
  Erratum: [Eur.\ Phys.\ J.\ C {\bf 78}, no. 5, 355 (2018)]
  [arXiv:1603.02680 [hep-ph]].


\bibitem{Klimenko:1984qx} 
  K.~G.~Klimenko,
  ``Conditions for certain Higgs potentials to be bounded below,''
  Theor.\ Math.\ Phys.\  {\bf 62}, 58 (1985)
  [Teor.\ Mat.\ Fiz.\  {\bf 62}, 87 (1985)].
  
  
\bibitem{Bonilla:2015eha} 
  C.~Bonilla, R.~M.~Fonseca, and J.~W.~F.~Valle,
  ``Consistency of the triplet seesaw model revisited,''
  Phys.\ Rev.\ D {\bf 92}, no. 7, 075028 (2015)
  [arXiv:1508.02323 [hep-ph]].



\bibitem{Arhrib:2015dez} 
  A.~Arhrib, C.~B\oe hm, E.~Ma, and T.~C.~Yuan,
  ``Radiative model of neutrino mass with neutrino interacting MeV dark matter,''
  JCAP {\bf 04} (2016) 049
  [arXiv:1512.08796 [hep-ph]].



\bibitem{Franceschini:2015kwy} 
  R.~Franceschini, G.~F.~Giudice, J.~F.~Kamenik, M.~McCullough, A.~Pomarol, R.~Rattazzi, M.~Redi, F.~Riva, A.~Strumia, and R.~Torre,
  ``What is the $\gamma \gamma$ resonance at 750 GeV?,''
  JHEP {\bf 03} (2016) 144
  [arXiv:1512.04933 [hep-ph]].



\bibitem{Gunion:1989we} 
  J.~F.~Gunion, H.~E.~Haber, G.~L.~Kane, and S.~Dawson,
  ``The Higgs hunter's guide,''
  Front.\ Phys.\  {\bf 80}, 1 (2000).



\bibitem{Djouadi:2005gi} 
  A.~Djouadi,
  ``The Anatomy of electro-weak symmetry breaking. I: The Higgs boson in the standard model,''
  Phys.\ Rep.\  {\bf 457}, 1 (2008)
  [hep-ph/0503172].



\bibitem{Djouadi:2005gj} 
  A.~Djouadi,
  ``The Anatomy of electro-weak symmetry breaking. II. The Higgs bosons in the minimal supersymmetric model,''
  Phys.\ Rep.\  {\bf 459}, 1 (2008)
  [hep-ph/0503173].



\bibitem{Chen:2013vi} 
  C.~S.~Chen, C.~Q.~Geng, D.~Huang, and L.~H.~Tsai,
  ``New scalar contributions to $h\to Z\gamma$,''
  Phys.\ Rev.\ D {\bf 87}, 075019 (2013)
  [arXiv:1301.4694 [hep-ph]].



\bibitem{Aaboud:2018xdt} 
  M.~Aaboud {\it et al.} [ATLAS Collaboration],
  ``Measurements of Higgs boson properties in the diphoton decay channel with 36 fb$^{-1}$ of $pp$ collision data at $\sqrt{s} = 13$ TeV with the ATLAS detector,''
  Phys.\ Rev.\ D {\bf 98}, 052005 (2018)
  [arXiv:1802.04146 [hep-ex]].


\bibitem{Aaboud:2016cth} 
  M.~Aaboud {\it et al.} [ATLAS Collaboration],
  ``Search for high-mass new phenomena in the dilepton final state using proton-proton collisions at $\sqrt{s}=13$ TeV with the ATLAS detector,''
  Phys.\ Lett.\ B {\bf 761}, 372 (2016)
  [arXiv:1607.03669 [hep-ex]].



\bibitem{CMS:2016abv} 
  CMS Collaboration,
  ``Search for a high-mass resonance decaying into a dilepton final state in 13 fb$^{-1}$ of pp collisions at $\sqrt{s}=13~\mathrm{TeV}$,''
  CERN Report No. CMS-PAS-EXO-16-031.



\bibitem{Sirunyan:2017khh} 
  A.~M.~Sirunyan {\it et al.} [CMS Collaboration],
  ``Observation of the Higgs boson decay to a pair of $\tau$ leptons with the CMS detector,''
  Phys.\ Lett.\ B {\bf 779}, 283 (2018)
  [arXiv:1708.00373 [hep-ex]].


\bibitem{Sirunyan:2018ouh} 
  A.~M.~Sirunyan {\it et al.} [CMS Collaboration],
  ``Measurements of Higgs boson properties in the diphoton decay channel in pp collisions at $\sqrt{s} =$ 13 TeV,''
  [arXiv:1804.02716 [hep-ex]].


\bibitem{Staub:2017ktc} 
  F.~Staub,
  ``Reopen parameter regions in Two-Higgs Doublet Models,''
  Phys.\ Lett.\ B {\bf 776}, 407 (2018)
  [arXiv:1705.03677 [hep-ph]].


\bibitem{Krauss:2017xpj} 
  M.~E.~Krauss and F.~Staub,
  ``Perturbativity constraints in BSM models,''
  Eur.\ Phys.\ J.\ C {\bf 78}, no. 3, 185 (2018)
  [arXiv:1709.03501 [hep-ph]].


\bibitem{Goodsell:2018tti} 
  M.~D.~Goodsell and F.~Staub,
  ``Unitarity constraints on general scalar couplings with SARAH,''
  Eur.\ Phys.\ J.\ C {\bf 78}, no. 8, 649 (2018)
  [arXiv:1805.07306 [hep-ph]].


\bibitem{Goodsell:2018fex} 
  M.~D.~Goodsell and F.~Staub,
  ``Improved unitarity constraints in two-Higgs-doublet-models,''
  arXiv:1805.07310 [hep-ph].


\bibitem{Krauss:2018orw} 
  M.~E.~Krauss and F.~Staub,
  ``Unitarity constraints in triplet extensions beyond the large s limit,''
  Phys.\ Rev.\ D {\bf 98}, no. 1, 015041 (2018)
  [arXiv:1805.07309 [hep-ph]].


\bibitem{G2HDM-DMConstraints}
C.-R. Chen, W.-C. Huang, Y.-X. Lin, C. S. Nugroho, 
R. Ramos, Y. L. S. Tsai, and T.-C. Yuan
(to be published)

\end{thebibliography}
\end{document}